\documentclass[11pt]{article}

\pdfoutput=1
\usepackage [latin1]{inputenc}
\usepackage{ifpdf}
\usepackage{amsmath}
\usepackage{graphicx}
\usepackage{array}
\usepackage{indentfirst}
\usepackage{amssymb}
\usepackage{cite}
\usepackage{color}
\usepackage{subfigure}
\usepackage[colorlinks,linkcolor=blue,citecolor=blue,urlcolor=blue,hyperindex]{hyperref}
\usepackage{indentfirst}
\usepackage{amsmath}
\usepackage{amssymb}
\usepackage{latexsym}
\usepackage{amsmath}
\usepackage{color}
\usepackage{lineno}
\usepackage{graphicx}
\usepackage{float}
\usepackage{subfigure}
\usepackage{geometry}
\usepackage{graphics}
\usepackage{subfigure}
\usepackage{graphicx}
\usepackage{multirow}
\usepackage{booktabs}
\usepackage{hyperref}
\usepackage{cite}

\begin{document}

\title{Effects of dark matter on shadows and rings of Brane-World black holes illuminated by various accretions}

\date{}
\maketitle

\begin{center}
\author{Xiao-Xiong Zeng}$^{a,b}$,
\author{Ke-Jian He}$^{c,}$ and
\author{Guo-Ping Li}$^{d,}$$\footnote{\texttt{Corresponding author:gpliphys@yeah.net}}$
\vskip 0.25in
$^{a}$\it{State Key Laboratory of Mountain Bridge and Tunnel Engineering, Chongqing Jiaotong University, Chongqing 400074, China, People's Republic of China}\\
$^{b}$\it{Department of Mechanics, Chongqing Jiaotong University, Chongqing 400074, China, People's Republic of China }\\
$^{c}$\it{College Of Physics, Chongqing University, Chongqing 401331, People's Republic of China}\\
$^{d}$\it{Physics and Space College, China West Normal University, Nanchong 637000, People's Republic of China}

\end{center}
\vskip 0.6in
{\abstract
{In this study, by taking the accretions into account, the observed shadows and rings cast by the Brane-World black hole were numerically investigated when the observer was located at the cosmological horizon. The results showed that the radius $r_p$ of the photon sphere increased with the cosmological parameter $\alpha$ and dark matter parameter $\beta$, while the impact parameter $b_p$ decreased with $\alpha$ and increased with $\beta$.
For thin disk accretion, the total observed intensity is mainly composed of direct emission. Simultaneously, the lensing ring and photon ring have only small and negligible contributions, respectively. We also found that shadows and rings exhibit different and exciting features when the disk is located at different positions.
For static and infalling spherical accretions, it is evident that the size of shadows is always the same for both accretions. This implies that shadows are only related to space-time geometry in this case. The luminosity of the shadow and photon sphere is closely associated with the Doppler effect and the emissivity per unit volume $j({\nu_e})$.
In addition, the influence of dark matter and cosmological constant on the observed intensity of shadows and rings is carefully emphasized throughout this paper.
Finally, we obtained the burring images of shadows and rings using the nominal resolution of the Event Horizon Telescope. We also studied the upper limits of the X-clod dark matter parameter $\beta$ using the data of the shadow of M87.\\

}
}

\thispagestyle{empty}
\newpage
\setcounter{page}{1}

\section{Introduction}\label{sec1}
As a particular object in our universe, a black hole possesses a powerful gravitational field, creating a significantly curved space-time. For a long time, black holes have attracted theoretical and experimental interest. The Laser Interferometer Gravitational-Wave Observatory first observed the gravitational wave event (GW150914) in 2015, providing direct evidence of a black hole.
The international collaboration of Event Horizon Telescope (EHT) in 2019 declared that the first shadow image of a supermassive black hole of M87 was captured using a long-baseline interferometer experiment\cite{Akiyama2019L1,Akiyama2019L2,Akiyama2019L3,Akiyama2019L4,Akiyama2019L5,Akiyama2019L6}. In 2021, when the magnetic fields and the plasma properties were taken into account, the polarized image was also observed in \cite{Akiyama2021L12,Akiyama2021L13}. As a milestone in the history of a black hole, this fact is scientifically significant. For instance, it provides strong evidence for general relativity, but it can also reveal the information of black holes, the accretion process for various matters, and radiation mechanism. The image of M87 clearly shows a dark interior inside and a compact asymmetric bright region outside, related to black hole shadows and photon rings.

In general, the size and shape of black hole shadow closely depend on the nature of space-time and the observer¡¯s location. As a bound photon orbit, the photon ring describes the size of the shadow in a black hole image. In \cite{Lumi}, the photon ring of Schwarzschild black hole is 3$\sqrt{3}M$, where $M$ is the mass of the black hole.
For the Kerr black hole, the results show that the shape of the shadow is a black disk when the observer is located at the axis of rotation. However, as the location of the observer is fixed to the equatorial plane, the black disk will evolve into a D-shaped shadow\cite{Kerr1, Kerr2}. In recent years, Wei {et al.} investigated the shadows of Einstein-Maxwell-Dilaton-Axion and noncommutative rotating black holes and obtained exciting results\cite{Wei1, Wei2}. In 2016, Huang et al. studied the shadow of a phantom black hole as photons coupled to the Weyl tensor. They found that the coupling photons would propagate along different paths when different polarization directions are considered; thereby, one shadow would change to two shadows in this case\cite{Huang1}.
The Portuguese scholars (Cunha et al.) discussed black hole shadows and space-time instabilities. They showed that shadow can represent a cuspy shadow at some certain parameters\cite{Cunha}. Wang et al. also obtained the D-shape and cuspy shadows in the Konoplya-Zhidenko black hole\cite{csb77}. In addition, one discussed the influence of other parameters on the size and shape of shadow and promoted a series of exploratory studies on constraining black hole parameters with shadow\cite{Shadow22,Shadow17,Shadow13,Shadow19,Shadow18,Shadow3,Shadow20,Shadow16,Shadow4,Shadow21,Shadow7,Shadow5,Shadow8,Shadow6,Shadow10,Shadow9, Shadow11,Shadow12,Shadow113,csb7,csb3,csb2,csb5,csb4,csb1,GMY5,GMY1,GMY2,GMY4,Soo,add1,add3,add4,add5,add6,add7,add10,add11,add12,add13,add15, add16,add17,add18,wsw1,wsw2,wsw3,chen1,Zilong}.
It is generally believed that there are various matters around a black hole on the base of the first black hole image. Given this, by considering the case that a simple model of spherically symmetric accretion surrounded the Schwarzschild black hole, a recent study \cite{Gammie} shows that shadow and photon sphere are only related to the geometry of space-time, and their sizes are scarcely influenced by the accretion. Then, Gan et al. studied the shadow and photon sphere of a hairy black hole in this case and obtained the optical appearance of shadow with the inclusion of two-photon spheres existed\cite{can1}. In 2019, when a black hole was surrounded by an optically and geometrically thin disk, Wald et al. found that shadow is composed of a dark disk and a bright ring outside of it as the observer is located at the North Pole, where the bright ring includes the direct emission, lensing ring, and photon ring\cite{Wald}. Importantly, the size and intensity of the bright ring are closely related to the location of the thin disk\cite{Wald}.
In this case, the optically observed appearance of a thin-shell wormhole and the polarized image of a black hole are carefully addressed later, where some exciting results are obtained\cite{guoadd,guoadd1}. Considering the case that the spherically symmetric accretion is infalling, the shadow and photon sphere of a black hole are also investigated in later studies\cite{ZXX1,ZXX2,lgp1}.
At present, more attention has been given to studying shadow and photon sphere\cite{Li1,Li2,Li3,Li33,Li4,GMY3,guoadd3}. However, it is still necessary to further research the shadow of black holes with the inclusion of various models of accretions and other possible factors.

On the other hand, since the discovery of dark matter by Swiss astronomer Zwicky, evidence supporting the existence of dark matter in our universe has continued to grow \cite{dm1,dm2,dm3,dm4,dm5}. Dark matter has received much attention both in theoretical physics and modern astronomy. In 2019, the CRESST experiment, as a direct dark matter search, extracted an upper limit on the dark matter-nucleus scattering cross-section\cite{dm7}. Based on \cite{dm8}, although one can assume dark matter is a form of elementary particles, primordial black holes can also be part of it. The effects of X-cold dark matter on optical properties have been carefully discussed by\cite{dm9}. With the inclusion of X-cold dark matter and the generalized uncertainty principle, Liu et al. studied the thermodynamics and phase transition of black holes\cite{dm6}. However, the influence of dark matter on shadows and rings of black holes illuminated by various accretions remains unclear. Combined with these facts, in this paper, we investigate shadows and rings of black holes illuminated by various accretions, where the influence of dark matter and the cosmological constant are emphasized. As black holes are illuminated by thin disk accretions, static spherical accretion, and infalling spherical accretion, we will carefully study the observed intensity of the Brane-World black hole seen by an observer located at the asymptotic infinity and the cosmological horizon of a black hole and discuss the effects of dark matter and cosmological constant on the observed appearance of a black hole.

The organization of this paper is as follows: In Sec.\ref{sec2}, we will first study the effective potential and photon orbits of a Brane-World black hole. In Sec.\ref{sec3}, the shadows and rings of this black hole surrounded by thin disk accretion are carefully studied. Sec.\ref{Sec4} is devoted to obtaining the observed appearance of a black hole with the inclusion of the static and infalling spherical accretions. Finally, Sec.\ref{Sec5} ends up with some discussion and conclusions.

\section{The effective potential and photon orbits of brane-World black hole}\label{sec2}
\begin{figure}[!h]\label{adfigure2}
\centering
\subfigure[$\beta = 0.3$]{
\includegraphics[scale=0.40]{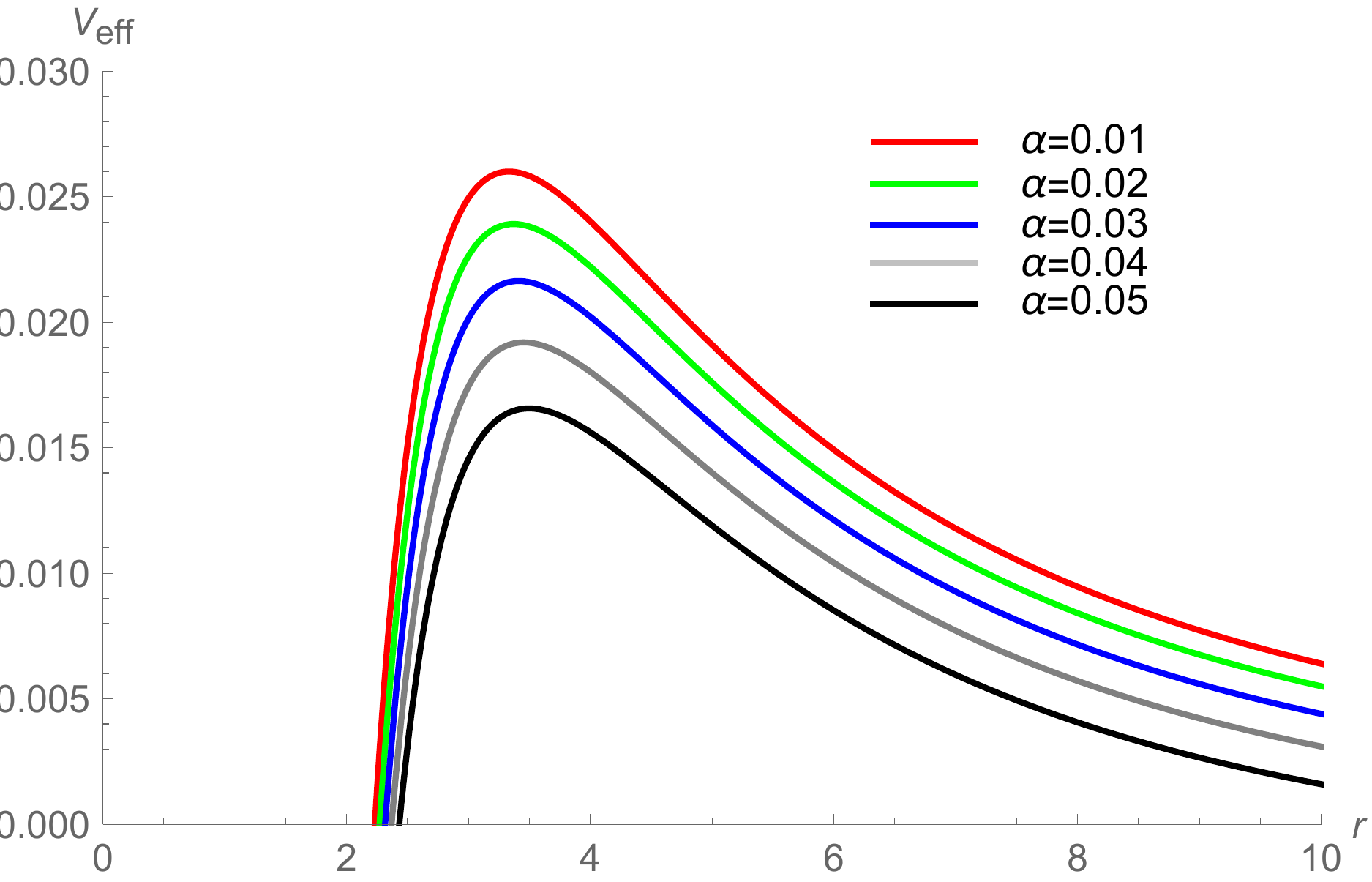}}
\subfigure[$\alpha =0.003$]{
\includegraphics[scale=0.35]{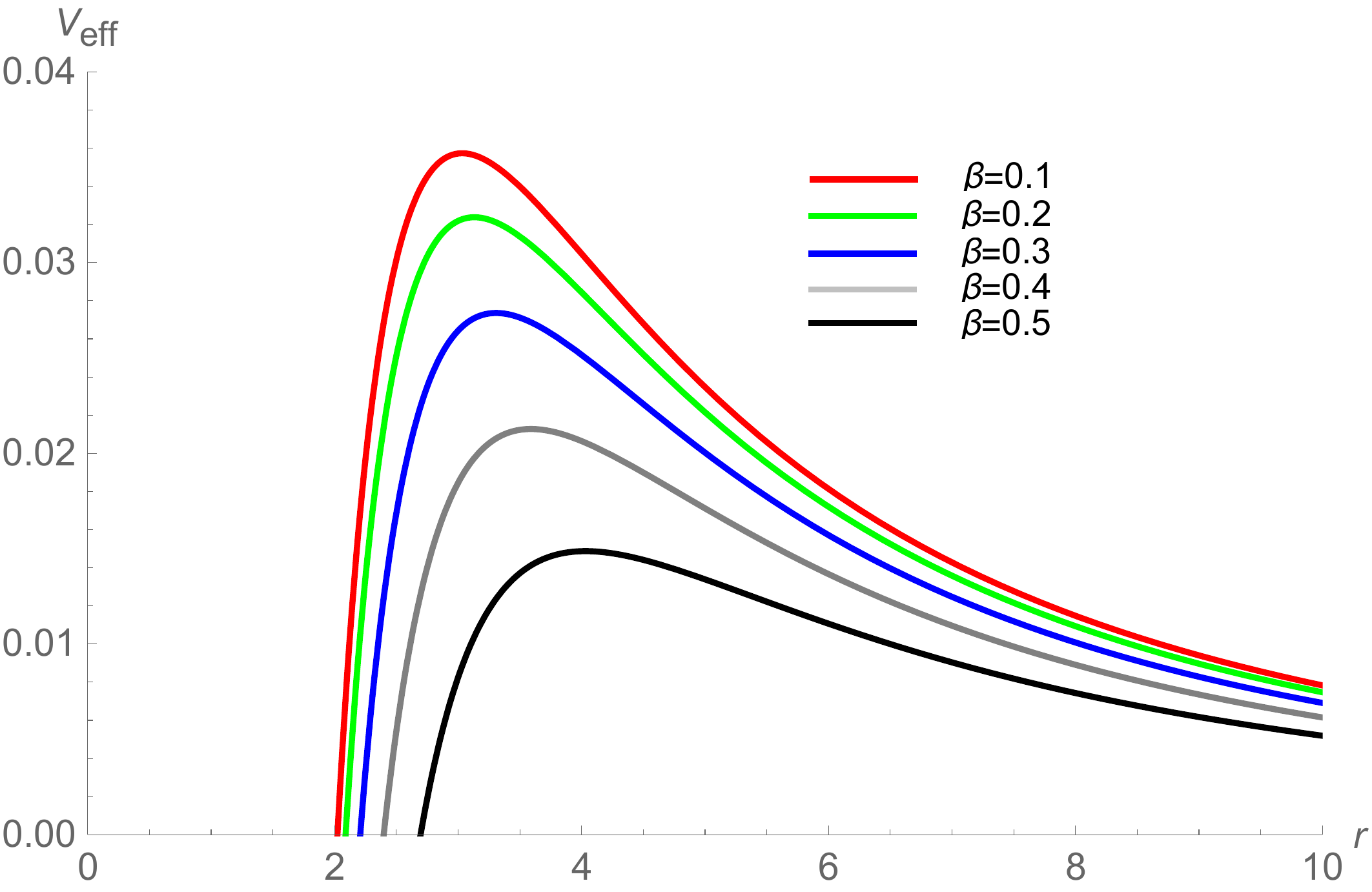}}
\caption{Effective potential vs. radius $r$. The (a) represents the effective potential when $\beta$ fixed, and the (b) for a fixed $\alpha$.}\label{fig1}
\end{figure}

As mentioned in \cite{lm1}, the total action for space-time $(\mathcal{M},\mathcal{G}_{AB})$ with boundary $(\Sigma,g_{\mu\nu})$ can be expressed as:
\begin{align}\label{eq1}
\mathcal{S}= \frac{1}{2 \alpha^*} \int_{\mathcal{M}}\rm{d}^m X\sqrt{\mathcal{G}}(\mathcal{R}-2\Lambda^{(b)})+\int_{\Sigma}\rm{d}^4 \sqrt{-g}(\mathcal{L}_{surface}+\mathcal{L}_{m}).
\end{align}
In this case, the Einstein equations in the bulk space read as follows:
\begin{align}\label{eq2}
G_{AB} + \Lambda^{(b)} \mathcal{G}_{AB} = \alpha^* \left( T_{AB} + 1/2 \cdot \mathcal{V} \cdot \mathcal{G}_{AB} \right).
\end{align}
where $\alpha^*$ is related to the fundamental scale of energy in the bulk space, and $\Lambda^{(b)}$ represents the cosmological constant of the bulk. $T_{AB}$ and $\mathcal{V}$ are the energy-momentum tensor of the matter and the confining potential, respectively.
When some conditions are considered for vacuum field equations, the induced effective Einstein field equation in the original brane can be written as:
\begin{align}\label{eq3}
G_{\mu \nu} = -\Lambda g_{\mu\nu} + Q_{\mu \nu} - \varepsilon_{\mu \nu},
\end{align}
with
\begin{align}\label{eq4}
 Q_{\mu \nu} = ( K K_{\mu \nu} - K_{\mu \alpha} K^\alpha _\nu) + \frac{1}{2} ( K_{\alpha \beta } K^{\alpha \beta }- K^2)g_{\mu \nu}.
\end{align}
$K_{\mu \nu}$, $\varepsilon_{\mu \nu}$ and $Q_{\mu \nu}$ are the extrinsic curvature, local effects from free bulk gravitational field, and energy-momentum tensor related to the X-cold dark matter, respectively. Based on Eq.(\ref{eq3}), a class of spherical symmetric Brane-World black hole vacuum solutions with the inclusion of dark matter was obtained by Heydari-Fard and Razmi\cite{lm1}, which is:
\begin{align}\label{eq5}
ds^2=A(r) d t^2 - \frac{1}{B(r)} d r^2 - r^2 (d \theta ^2 + {\sin^2\theta} d \phi ^2),
\end{align}
with
\begin{align}\label{eq6}
A(r) = B(r) = 1 - \frac{2M}{r} - \alpha^2 r^2 - 2 \alpha \beta r - \beta^2.
\end{align}
The parameter $\alpha$ is a constant that is related to a cosmological constant, and $\beta$ is the parameter of the X-cold dark matter. As $\beta \rightarrow 0$, this space-time will naturally reduce to a de Sitter black hole. The cosmological horizon $r_c$ and event horizon $r_h$ can be obtained by solving Equation $B(r)=0$. Next, we will investigate the null geodesic of the Brane-World black hole. In this space-time, it is easy to get the Lagrangian $\mathcal{L}$ of a particle, which is:
\begin{align}\label{eq7}
\mathcal{L}=-\frac{1}{2}g_{\mu \nu }\dot{x}^{\mu }\dot{x}^{\nu }=\frac{1}{2}\left[A(r)\dot{t}^2-\frac{\dot{r}^2}{B(r)} - r^2 \dot{\theta}^2 - r^2 \sin^2\theta \dot{\phi }^2\right],
\end{align}
$\dot{x}^{\mu }$ is the four-velocity of the photon, i.e., $\dot{x}^{\mu } = \frac{\partial {x}^{\mu }}{\partial \lambda}$, where $\lambda$ denotes the affine parameter.
In general, we always assume that the {motion} of a particle is located at the equatorial plane, i.e., $\theta=\pi/2$. This also implies the four-velocity $\dot{\theta}=0$. Meanwhile, two Killing fields, $\partial_t$ and $\partial_\phi$, correspond to the two conserved quantities in this space-time. They are the energy $E$ and the orbital angular momentum $L$,
\begin{align}\label{eq8}
E = \frac{\partial \mathcal{L}}{\partial \dot{t}} = A(r)\dot{t} , \qquad L = -\frac{\partial \mathcal{L}}{\partial \dot{\phi}} = r^2 \dot{\phi}.
\end{align}
Meanwhile, we have $g_{\mu \nu }\dot{x}^{\mu }\dot{x}^{\nu }= 0$ for null geodesic. Combined with the above facts, the four-velocity of time, azimuthal and radial components are:
\begin{align}
&\dot{t}=\frac{1}{b_{c} \left[ 1 - \frac{2M}{r} - \alpha^2 r^2 - 2 \alpha \beta r - \beta^2 \right] }, \label{eq100} \\ \quad
&\dot{\phi} = \pm \frac{1}{r^2} \label{eq101} , \\ \quad
&\dot{r}^2= \frac{1}{b_{ c}^2} - \frac{1}{r^2}\left[ 1 - \frac{2M}{r} - \alpha^2 r^2 - 2 \alpha \beta r - \beta^2 \right].\label{eq102}
\end{align}
In the above equation, the impact parameter $b_c$ is defined as $b_c = \frac{|L|}{E}$, and the affine parameter $\lambda$ is redefined as $\lambda/|L|$. The sign $\pm$ corresponds to the counterclockwise and clockwise direction of the light rays, respectively. By rewriting Eq.(\ref{eq102}) as $\dot{r}^2+V_{{eff}}=\frac{1}{b_{c}^2}$, the effective potential $V_{{eff}}$ is defined as:
\begin{align}\label{eq9}
V_{{ eff}}=\frac{1}{r^2}\left[ 1 - \frac{2M}{r} - \alpha^2 r^2 - 2 \alpha \beta r - \beta^2 \right].
\end{align}
Due to the photon sphere orbit conditions $\dot{r}=0$ and $\ddot{r}=0$, the effective potential should satisfy the condition:
\begin{align}\label{eq10}
V_{{eff}}=\frac{1}{b_{ c}^2}, \quad V_{{ eff}}'=0.
\end{align}
In a four-dimensional space-time, it can also be reexpressed as $r_p{}^2=b_{ p}^2 A(r), 2 b_p^2 A(r)^2=r_p^3 A'(r)$, where $r_p$ is the radius of the photon sphere, and $b_p$ is an impact parameter for the photon sphere. The numerical results of the event horizon $r_h$, the cosmological horizon $r_c$, the radius of the photon sphere $r_p$, and an impact parameter for the photon sphere $b_p$ for different values of $\alpha$ and $\beta$ are shown in Tables 1 and 2. We also plotted the effective potential vs. radius $r$ in Figure \ref{fig1}. The results show that the effective potential $V_{{eff}}$ decreased with parameters $\alpha$ and $\beta$.
\begin{center}
{\footnotesize{\bf Table 1.} The event horizon $r_h$, the cosmological horizon $r_c$, the radius $r_p$ and impact parameter $b_{p}$ of photon sphere for different values of $\beta$, where $M = 1$ and $\alpha=0.003$\footnote{Here, the values of $M$ are always set to $1$ for the following numerical results. }.\\
\vspace{1mm}
\begin{tabular}{ccccccccccc}
\hline &{$\beta=0.1$} &{$\beta=0.2$} &{$\beta=0.3$} &{$\beta=0.4$} &{$\beta=0.5$} &{$\beta=0.6$}        \\ \hline
{$r_{h}$}   &{2.02276}     &{2.08887}      &{2.20755}     &{2.39752}     &{2.69597}     &{3.18242}             \\
{$r_{c}$}   &{298.883}     &{265.408}      &{231.893}     &{198.315}     &{164.636}     &{130.775}             \\
{$r_{p}$}   &{3.03309}     &{3.13113}      &{3.30752}     &{3.58984}     &{4.03252}      &{4.75098}              \\
{$b_{p}$}   &{5.29033}     &{5.55767}      &{6.04609}     &{6.85628}     &{8.20085}     &{10.5779}             \\
\hline
\end{tabular}}\label{T1}
\end{center}
\vspace{1mm}
\begin{center}
{\footnotesize{\bf Table 2.} The event horizon $r_h$, the cosmological horizon $r_c$, the radius $r_p$ and impact parameter $b_{p}$ of photon sphere for different values of $\alpha$. where $\beta=0.3$.\\
\vspace{1mm}
\begin{tabular}{ccccccccccc}
\hline &{$\alpha=0.001$} &{$\alpha=0.002$} &{$\alpha=0.003$} &{$\alpha=0.004$} &{$\alpha=0.005$} &{$\alpha=0.006$}  \\ \hline
{$r_{h}$}   &{2.20101}     &{2.20426}      &{2.20755}     &{2.21088}     &{2.21426}     &{2.21769}           \\
{$r_{c}$}   &{698.567}     &{348.563}      &{231.893}     &{173.555}     &{138.551}     &{115.214}           \\
{$r_{p}$}   &{3.30029}     &{3.30390}      &{3.30752}     &{3.31116}     &{3.31482}     &{3.31849}           \\
{$b_{p}$}   &{6.00548}     &{6.02558}      &{6.04609}     &{6.06700}     &{6.08834}     &{6.11011}           \\
\hline
\end{tabular}}
\end{center}\label{T22}

In order to clearly present the trajectories of light ray, the motion equation of photon can be reexpressed as the following form:
\begin{align}\label{eq15}
\frac{{dr}}{d \phi }=\pm r^2\sqrt{\frac{1}{b_{ c}^2} - \frac{1}{r^2}\left[ 1 - \frac{2M}{r} - \alpha^2 r^2 - 2 \alpha \beta r - \beta^2 \right]}.
\end{align}

By rearranging $u=1/r$, we will have:
\begin{align}\label{eq16}
\mathcal{H}(u,b)\equiv \frac{d u}{d \phi }=\sqrt{\frac{1}{b_{ c}^2} - u^2\left[ 1 - {2Mu} - \frac{\alpha^2}{ u^2} - \frac{2 \alpha \beta}{ u} - \beta^2 \right]}.
\end{align}

\begin{figure}[!h]\label{figure2}
\centering
\subfigure[$\alpha = 0.001, \beta =0.3$]{
\includegraphics[scale=0.355]{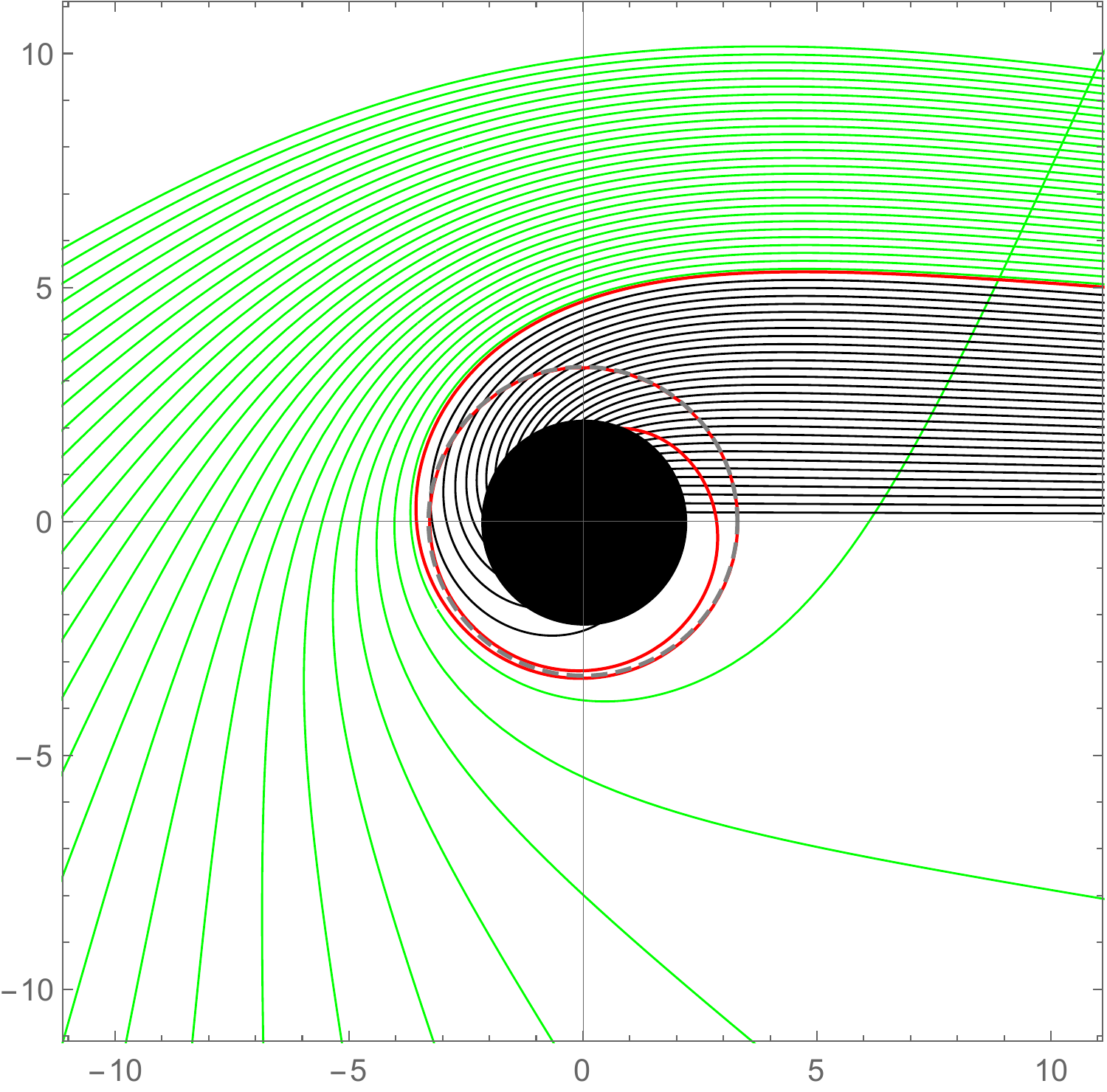}}
\subfigure[$\alpha = 0.003, \beta =0.3$]{
\includegraphics[scale=0.355]{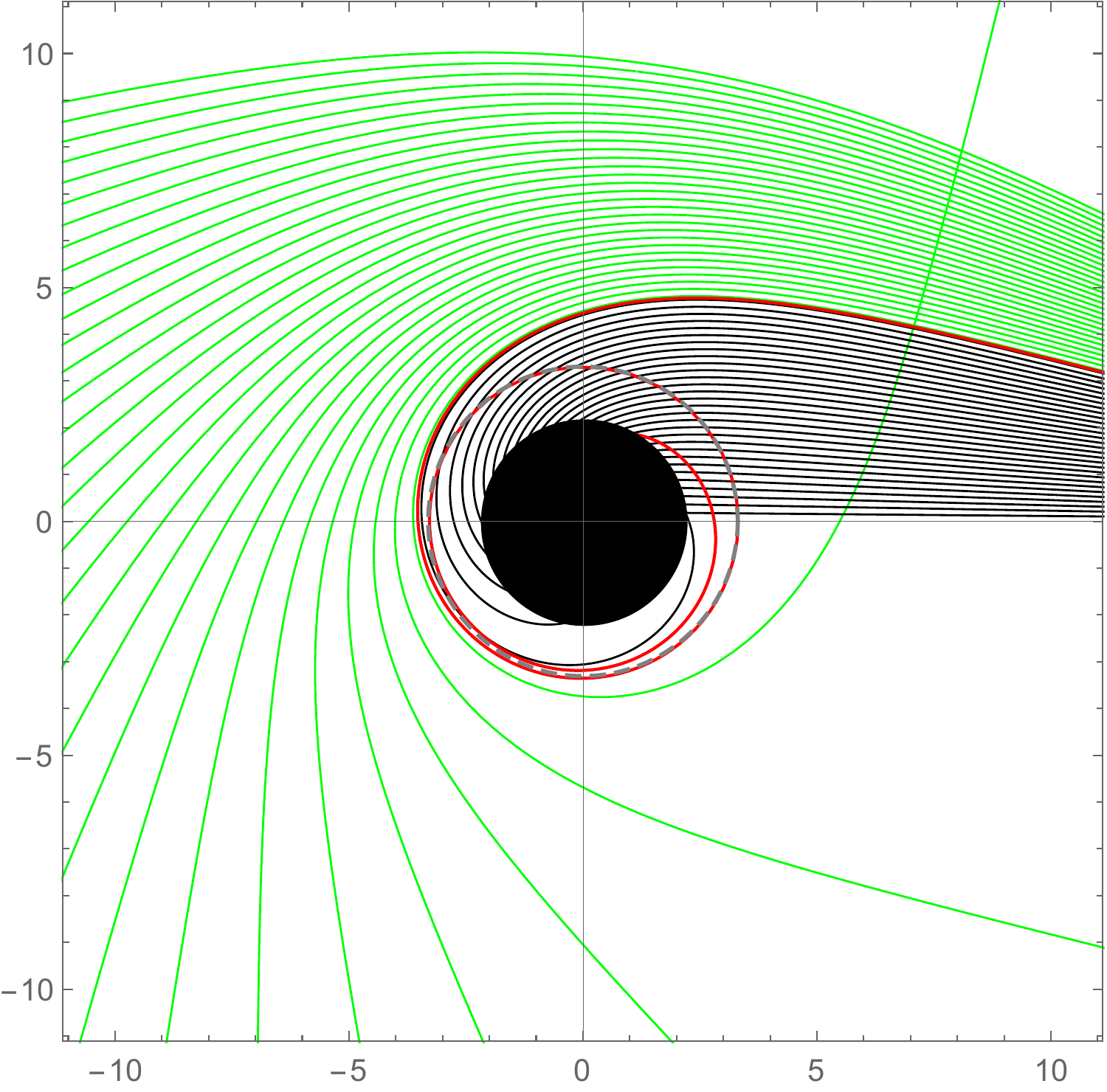}}
\subfigure[$\alpha = 0.003, \beta =0.1$]{
\includegraphics[scale=0.355]{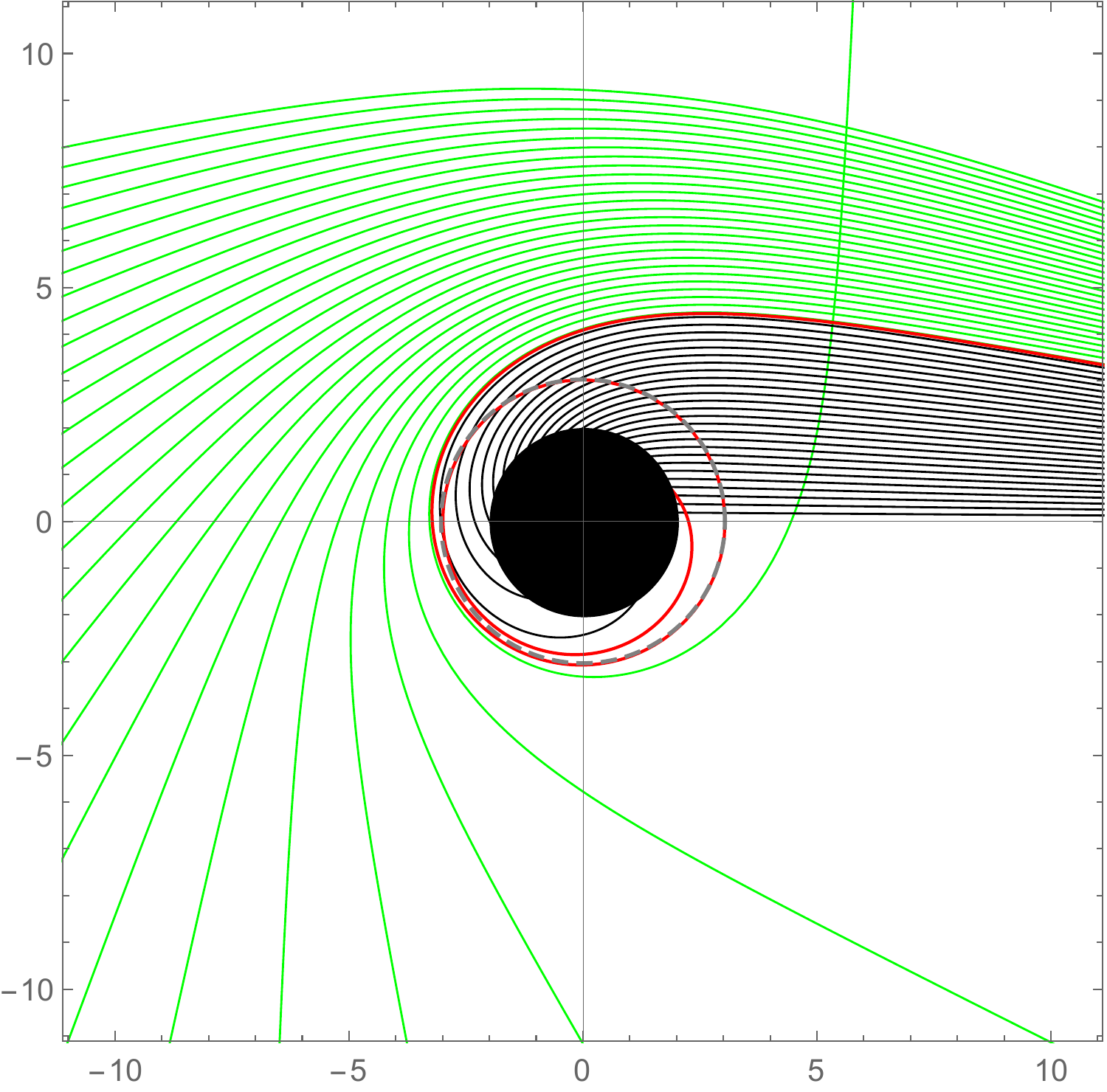}}
\caption{Trajectories of photons in the polar coordinates $(r, \phi)$ for different values of $\alpha$ and $\beta$. The spacings of impact parameter $b_c$ are all fixed to $\frac{1}{5}$. The solid disk and dashed gray circles correspond to the black hole and photon orbit, respectively.  }\label{fig2}
\end{figure}

Based on Eq.(\ref{eq16}), Figure \ref{fig2} presents the trajectories of a light ray.

In the black line in Figure \ref{fig2}, its impact parameter $b_c < b_p$ shows that the trajectories of those photons drop into a black hole. In the green line, its impact parameter $b_c > b_p$ shows that the trajectories of those photons are reflected and move to infinity. While $b_c = b_p$, photons, in this case, make a circular motion around the black hole infinitely many times and the trajectories of those photons approximately correspond to the red line in Figure \ref{fig2}. Interestingly, we can see in Figure \ref{fig2} that the incident light rays slope to the abscissa axis in subfigures (a), (b), and (c). Because of the cosmological horizon, the light rays are no longer an ingoing line from infinity. In Figure \ref{fig2}, as the observer is located at $\frac{r_c}{5}$, the ingoing rays is fixed to $\frac{r_c}{5}$.
By selecting the observer¡¯s location in this manner, nonparallel entry trajectories can be clearly represented. In addition, the parameter $\alpha$ and the dark matter parameter $\beta$ significantly affect trajectories of light rays.

\section{ Shadows and rings of thin disk emission }\label{sec3}
\label{disk}
In this section, we consider an ideal model. An optical and geometric thin disk located at the equatorial plane outside of the black hole is used to investigate the appearance of a Brane-World black hole.
Without losing generality, we assume that the static observer is located at the North Pole, and the light emitted from thin disk accretion is isotropic in the rest of the frame.
In this case, the relationship between the observed specific intensity $I_{obs}$ and the emission specific intensity $I_{{emi}}$ in the static system is:
\begin{equation}
I_{{obs}}=\frac{\nu^{'3}}{\nu^3} I_{{emi}},\label{EQ3.1}
\end{equation}
where $\nu^{'}$ and $\nu$ are the observed and emitted frequencies, respectively, and they satisfy $\nu^{'}={A(r)}^{1/2} \nu $. By integrating Eq.(\ref{EQ3.1}), we can get the total observed intensity, which is:
\begin{equation}
I(r)=\int I_{{obs}}d\nu^{'}=\int {A(r)}^{2} I_{{emi}} d \nu={A(r)}^{2} I_{{em}}(r).\label{EQ3.2}
\end{equation}
Here, $I_{em}(r)=\int I_{{emi}}d \nu$, corresponding to the total emitted intensity. According to the ray-tracing method, when the ray trajectory is emitted from the observer, it can pass through the thin disk and is accompanied by energy extracted from the thin disk. The intersection between the ray trajectory and the thin disk provides an additional light intensity; thereby, the observed appearance of the black hole is naturally affected.
Given this, the number of intersections and their locations are critical factors for the observer. Next, we investigated the trajectory and deflection angle of a light ray traveling near a black hole.

\subsection{Direct emission, lensing ring, and photon ring}\label{sec31}
\label{rings}
\begin{figure}[h]
\centering 
\includegraphics[width=0.55\textwidth]{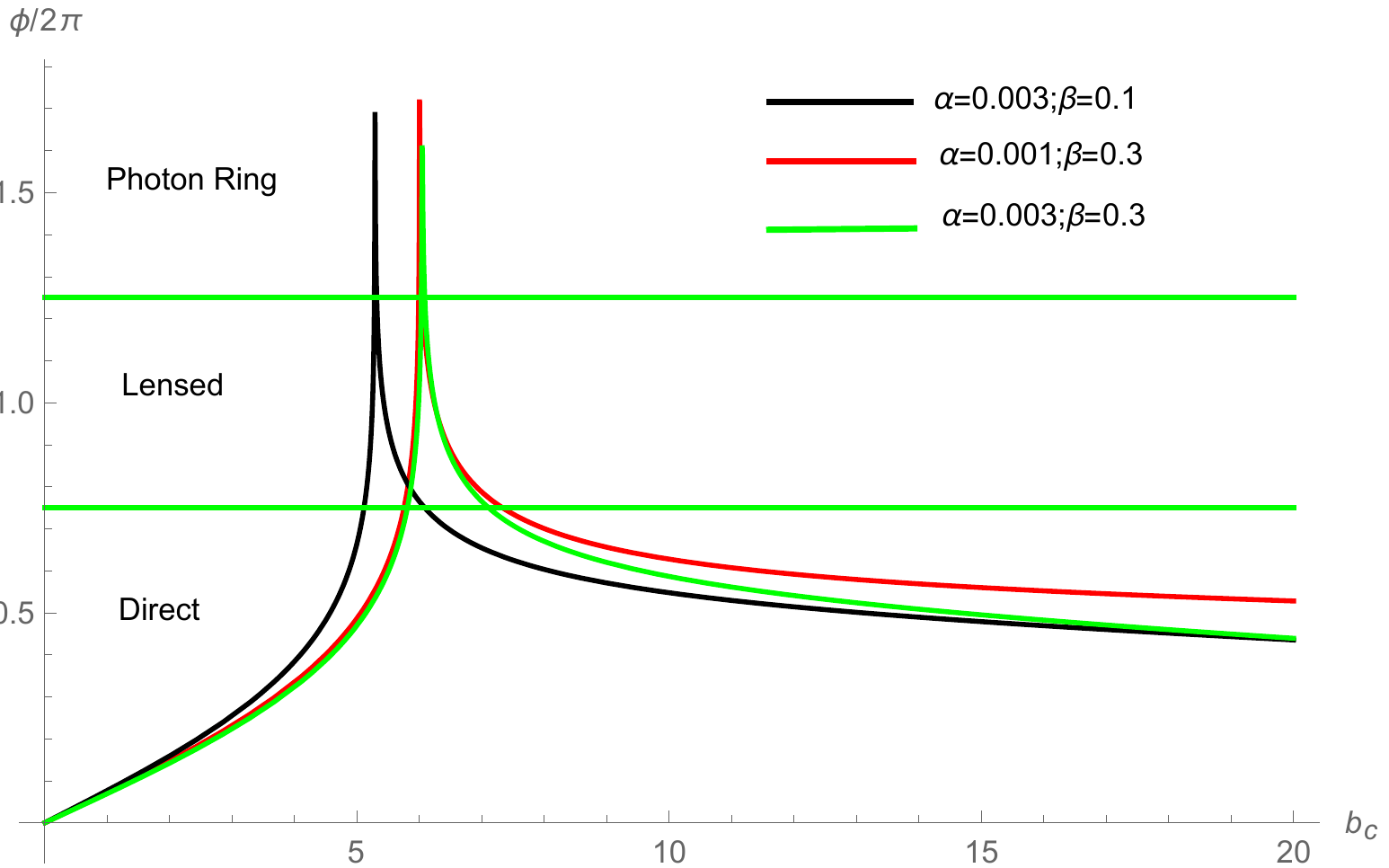}
\caption{\label{2fig1}Total number of orbits for different values of $\alpha$ and $\beta$. }
\end{figure}

In Ref.\cite{Wald}, after analyzing the behavior of light trajectories near black holes, Wald et al. found photon rings and lensing rings near black holes. Moreover, these rings can be distinguished by the total number of orbits $n=\phi/2 \pi$, a function of the impact parameter $b_c$. In particular, when $n < 3/4$, the trajectories of light rays intersect the equatorial plane only once, corresponding to the direct emission region. When $3/4 < n < 5/4$, the trajectories of light rays will intersect the other side of the thin disk again, i.e., the courses of light rays have two intersections with a thin disk, corresponding to the lensing rings. When $n > 5/4$, the deflected ray trajectories have at least three intersection points with a thin disk corresponding to the photon rings. In the context of a Brane-World black hole, we present these three regions (direct emission, lensing ring, and photon ring) vs. the impact parameter $b_c$ when the relevant parameters take different values, shown in Figure \ref{2fig1}.

\begin{figure}[h]
\centering
\subfigure[$\alpha=0.001$, $\beta=0.3$]{
\includegraphics[scale=0.35]{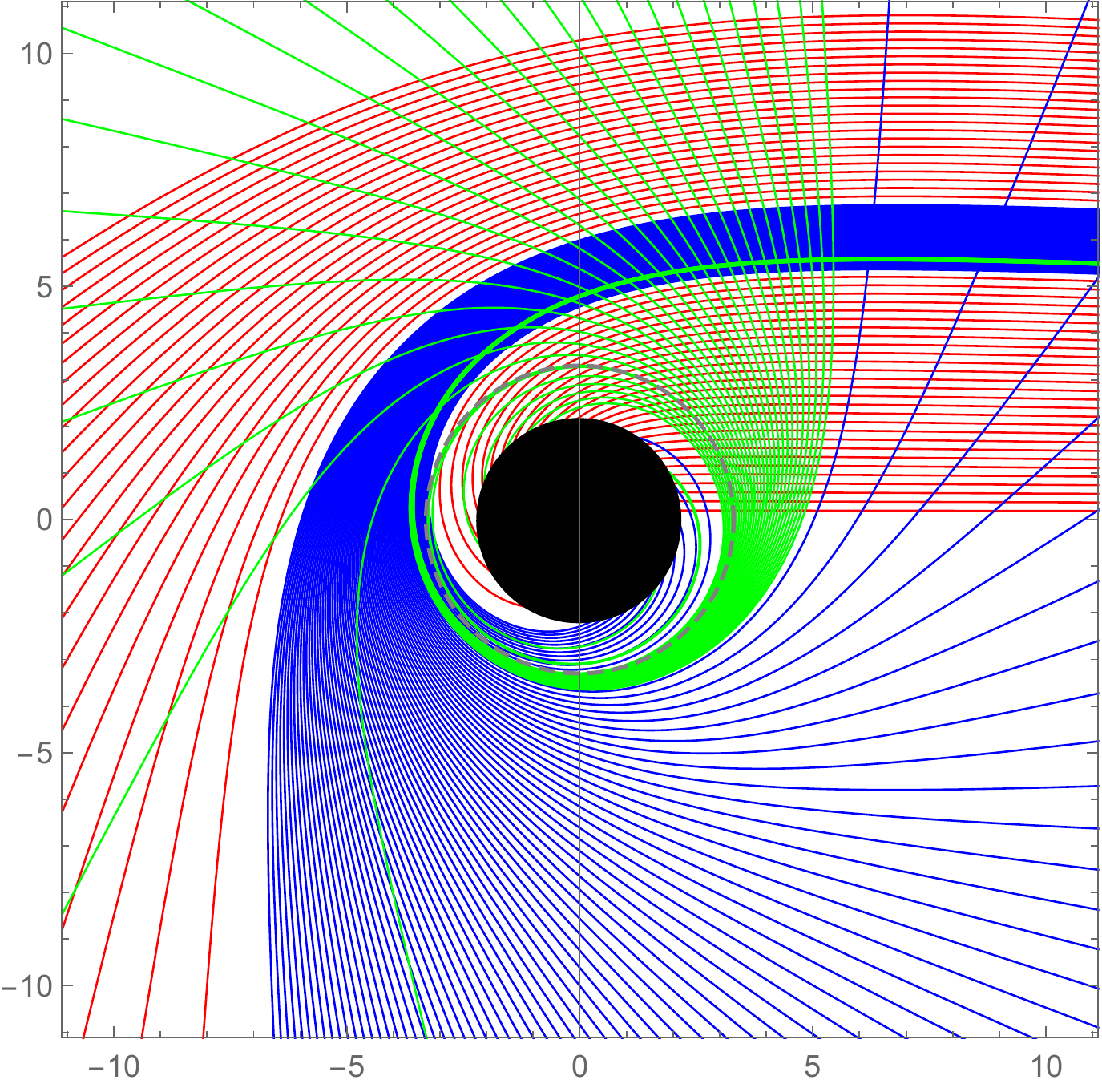}}
\subfigure[$\alpha=0.003$, $\beta=0.3$]{
\includegraphics[scale=0.35]{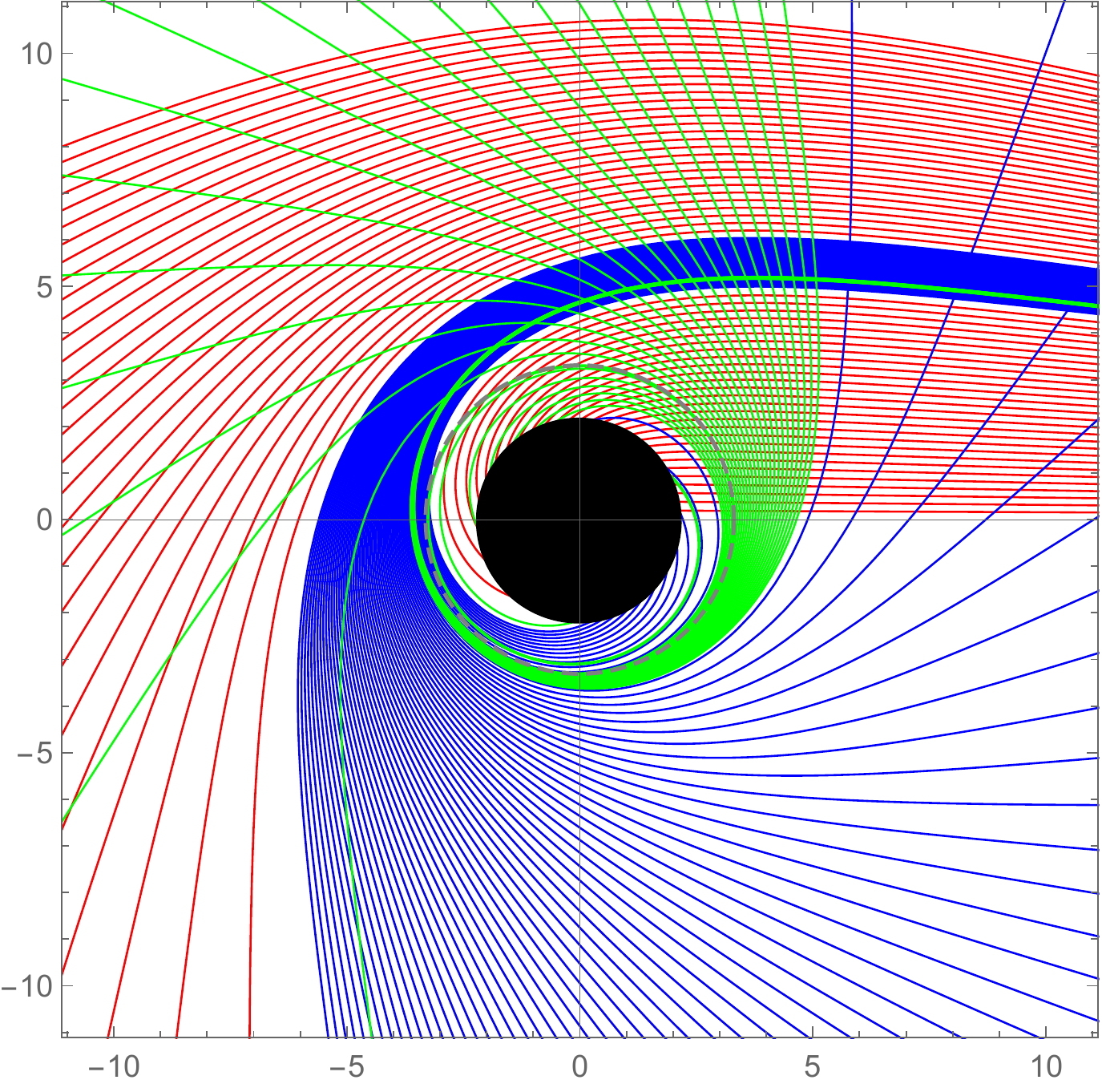}}
\subfigure[$\alpha=0.003$, $\beta=0.1$]{
\includegraphics[scale=0.35]{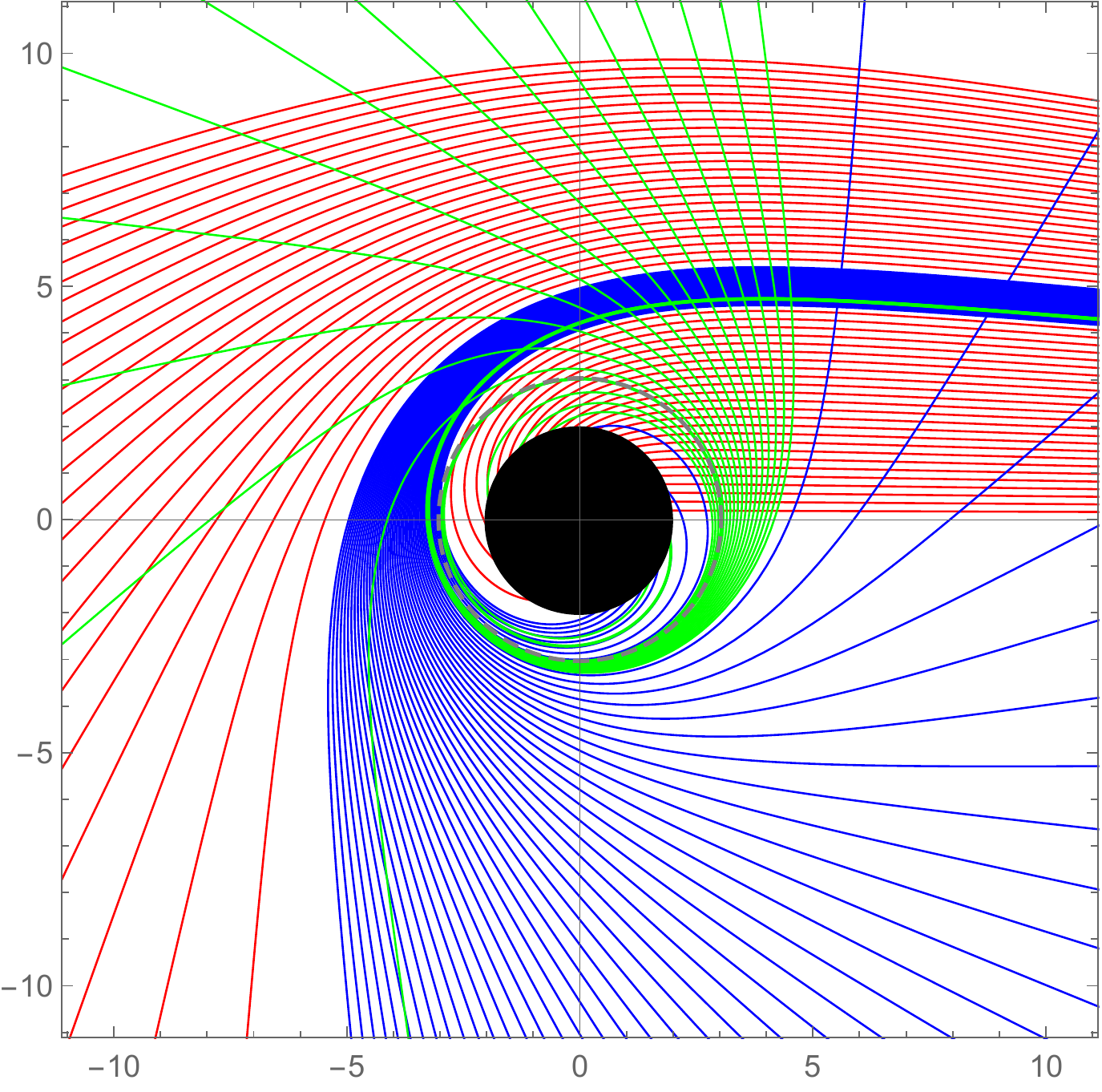}}
 \caption{Trajectory of the light ray for different parameters ($\alpha$ and $\beta$ ) in the polar coordinates $(r, \varphi)$. The black hole is shown as a black disk. Among them, the green, blue, and red lines represent photon ring, lensing ring, and direct emission. The direct (red line), lensing (blue line), and photon (green line) ring trajectories correspond to the spacing of $b_c$ are $1/5$, $1/50$, and $1/500$, respectively.}\label{2fig2}
\end{figure}

As shown in Figure \ref{2fig1}, the ranges of photon and lensing rings are squeezed in a tiny area, but the range of the lensing ring is more extended than that of the photon ring.
With the increase of dark matter parameter $\beta$ in the system, the ranges of $b_c$ corresponding to photon and lensing rings are different. In contrast, the influence of parameter $\alpha$ is minimal\footnote{The reason is that the value of $\alpha$ is minimal. This choice of $\alpha$ is just for obtaining a suitable value when plotting the figures.}.
When the value of $\beta$ increases, the value of $b_c$ corresponding to the lensing ring increases, and its range is amplified.
On the contrary, the peaked value of the lensing ring decreases, and its ranges dwindle with the increase of cosmological parameter $\alpha$.
Those facts imply that two parameters $\beta$ have some important effects on space-time geometry. In the polar coordinates ($b_c$, $\phi$), the corresponding photon trajectories in the Brane-World black holes are shown in Figure \ref{2fig2}.

\begin{figure}[h]
\centering
\subfigure[$\alpha=0.001$, $\beta=0.3$]{\includegraphics[scale=0.35]{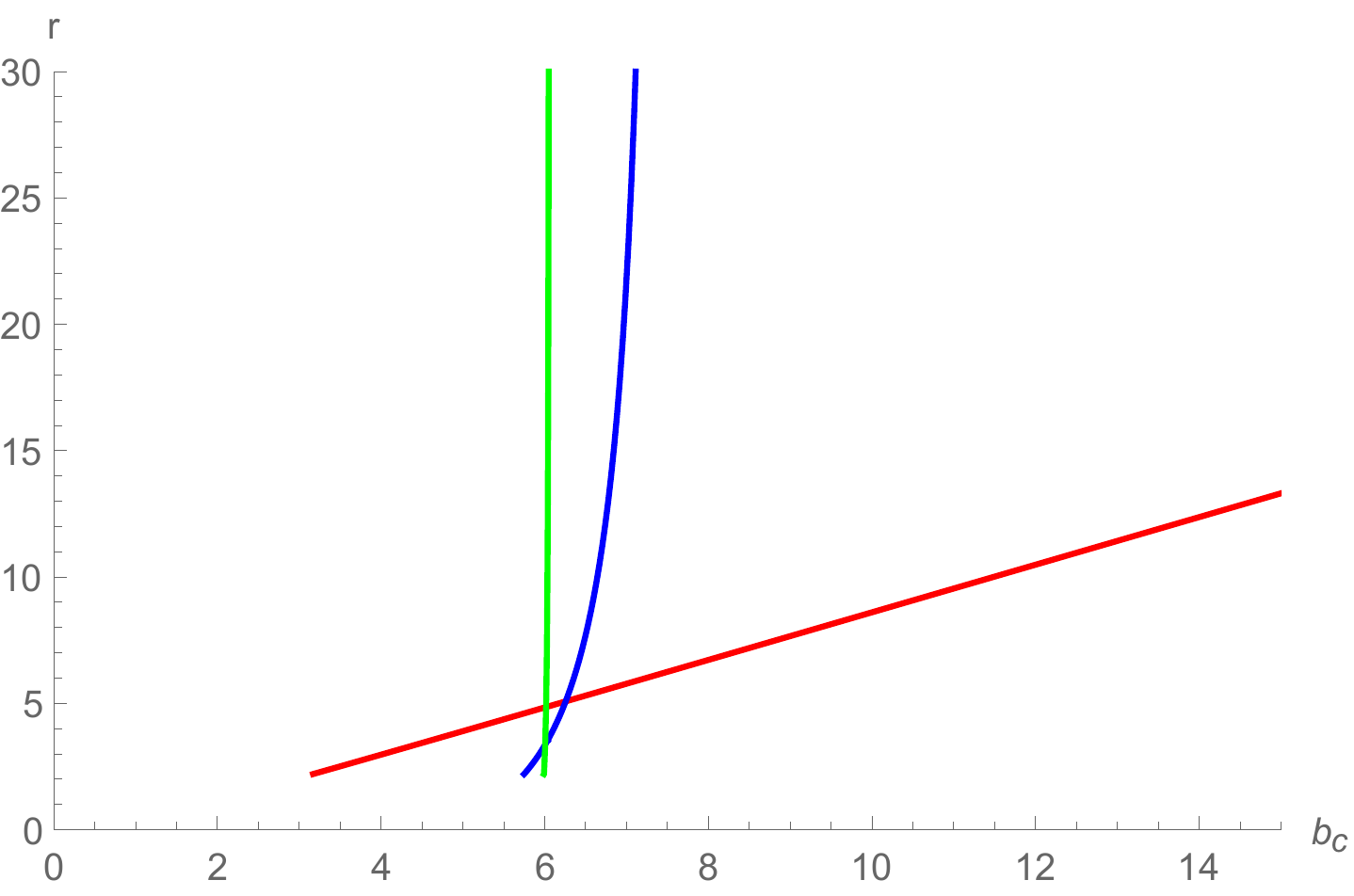}}
\subfigure[$\alpha=0.003$, $\beta=0.3$]{\includegraphics[scale=0.35]{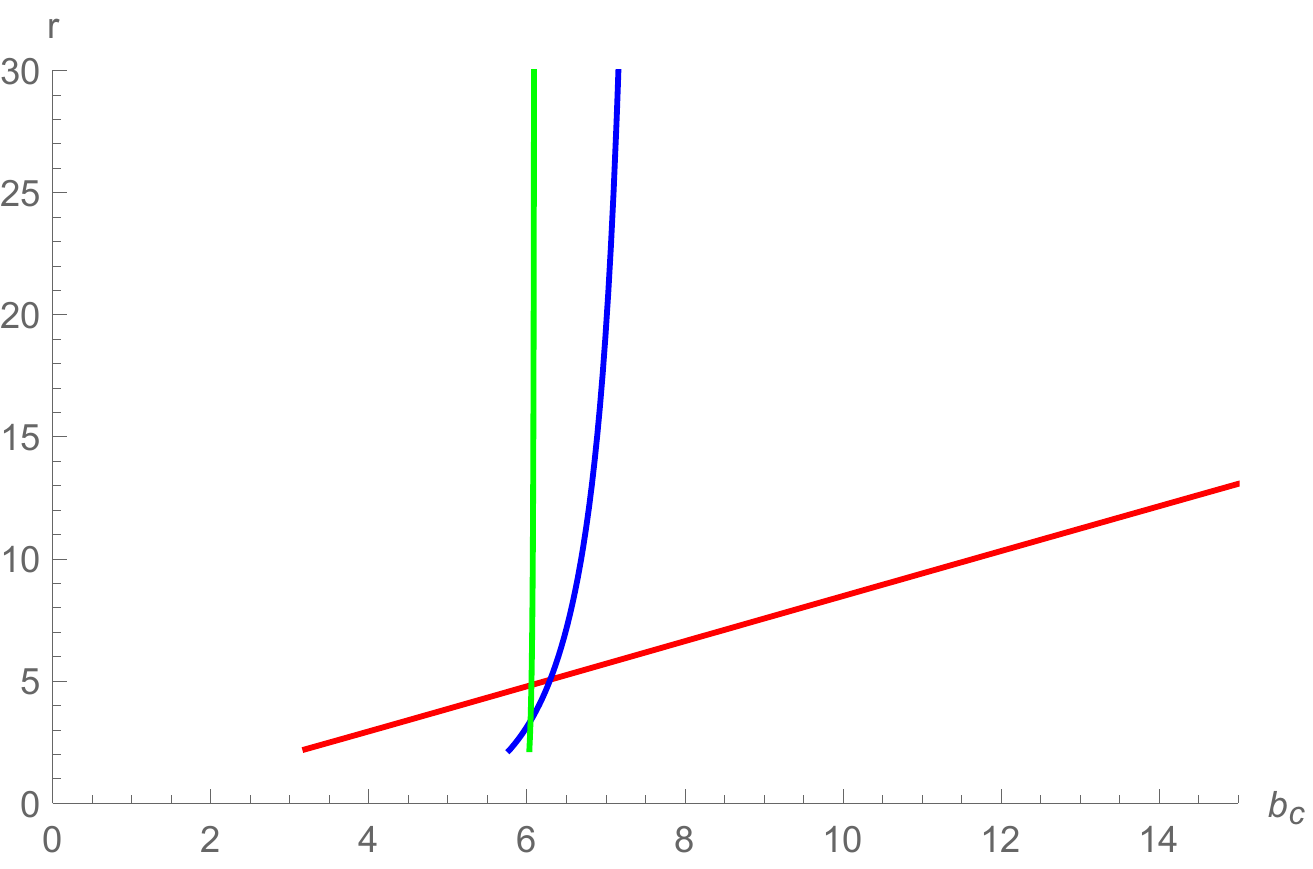}}
\subfigure[$\alpha=0.003$, $\beta=0.1$]{\includegraphics[scale=0.325]{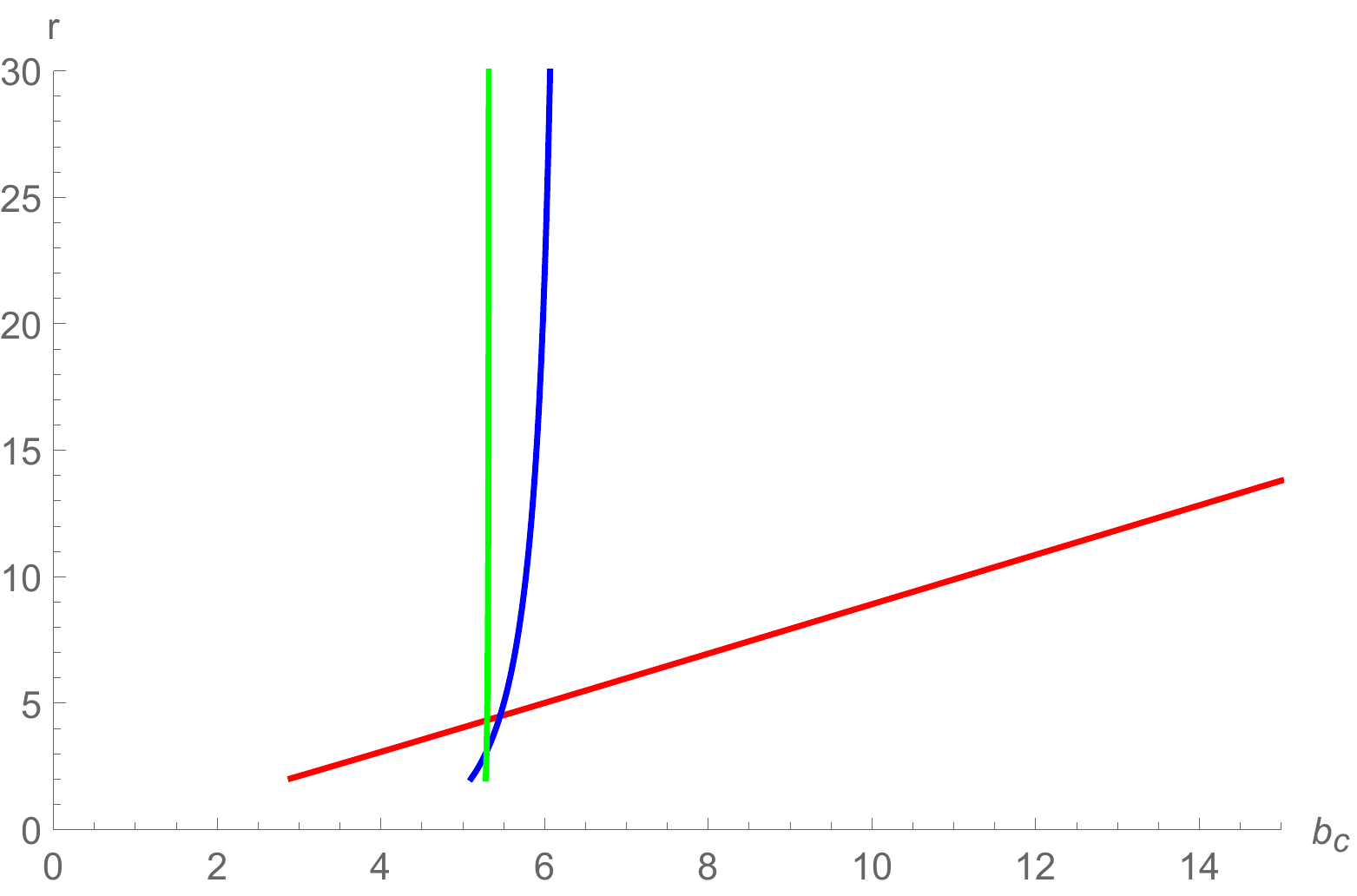}}
 \caption{First three transfer functions for the Brane-World black hole. The red, blue, and green lines represent the first, second, and third transfer functions, respectively. }\label{3fig2}
\end{figure}

In Figure \ref{2fig2}, when the value of parameter $\beta$ is $0.3$, the range of the lensing ring is broader than that when $\beta = 0.1$. Namely, the increase of parameter $\beta$ improves the contribution of the lensing ring to the observed luminosity. The change in parameter $\alpha$ significantly affects the photon and lensing rings near the black hole. Besides, the increase of $\alpha$ makes the light rays emitted from the observer no longer parallel.
Since the cosmological horizon exists, the observer is assumed to be located inside the cosmological horizon rather than infinity. So, the observer¡¯s position directly depends on the parameter $\alpha$.
Using Ref.\cite{Wald}, we can get the total observed intensity at different positions outside the black hole, which is expressed as:
\begin{equation}
I(r)=\sum _n A(r)^2 I_{em}(r)|_{r=r_n(b)}.\label{EQ3.3}
\end{equation}
where $r_n(b)$ is the transfer function. It describes the radial position of the $n^{th}$ intersection point between the ray trajectory and the plane of the disk. The slope of the transfer functions $dr/db$ represents the demagnetization factor, reflecting the demagnified scale of the transfer function. When the parameters in the system changed, we obtained the relationship between the transfer function $r_n(b)$ and the impact parameter $b_c$, which are shown in Figure \ref{3fig2}.

In Figure \ref{3fig2}, the red line corresponds to the first transfer function, and its slope is almost equal to 1. It reflects the direct image of a thin disk, which is essentially the redshifted source profile. Then, the blue line corresponds to the second transfer function, the lensing ring. It represents a highly demagnified image of the backside of the disk. The green line is the third transfer function, which is the photon ring. We can see from Figure \ref{3fig2} that the slope of the green line tends to be infinite, i.e., there is a highly demagnified image from the front side of the disk.
In addition, we can find that the change in parameter $\beta$ will affect the range of transfer function; that is, the intersection position between the light rays and the thin disk will be different when the parameter $\beta$ changes.

\subsection{Observed appearances of direct emissions and rings}\label{sec33}

Combined with the above discussion, we now explore three ideal emission models to further study the observed specific intensity.
Model I: the emission profile occurs at the innermost stable circular orbit $r_{{isco}}$, and it is a decay function suppressed by the second power, which is:
\begin{align}
    I^1_{{em}}(r) =\begin{cases}\left(\frac{1}{r-(r_{isco}-1)}\right)^2, &  r>r_{{isco}}  \\
    0, &r \leq r_{{isco}} \label{Eq.16}
    \end{cases}
\end{align}
{Model II}: the emission profile is still a decay function, but the position of the emission located at the photon sphere $r_p$, which is:
\begin{align}
    I^2_{{em}}(r) =\begin{cases} \frac{3-\tanh(r-r_p)}{3\cdot e^{(r-r_p)}}, &  r>r_{p}  \\
    0, &r \leq r_{p} \label{Eq.16}
    \end{cases}
\end{align}

\begin{figure}[h]
\centering
\subfigure[$$]{
\includegraphics[scale=0.35]{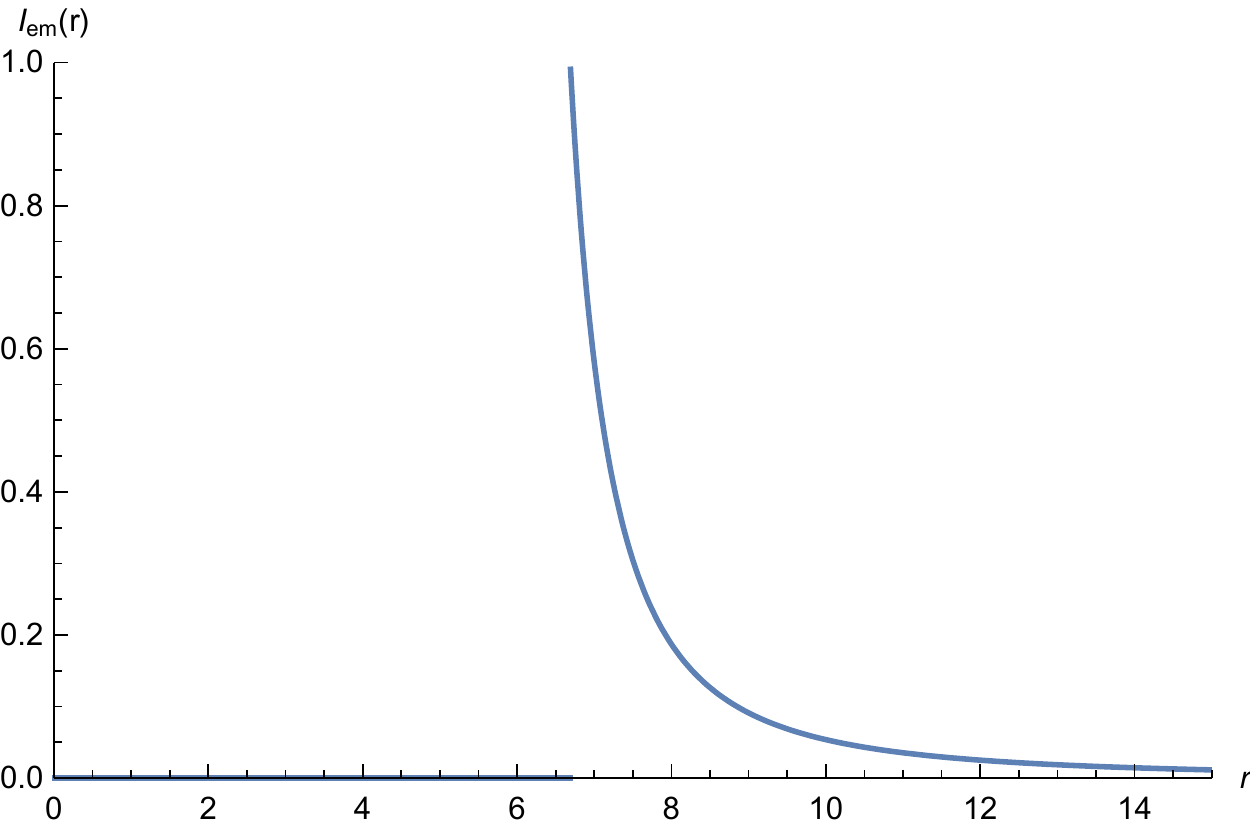}}
\subfigure[$$]{
\includegraphics[scale=0.35]{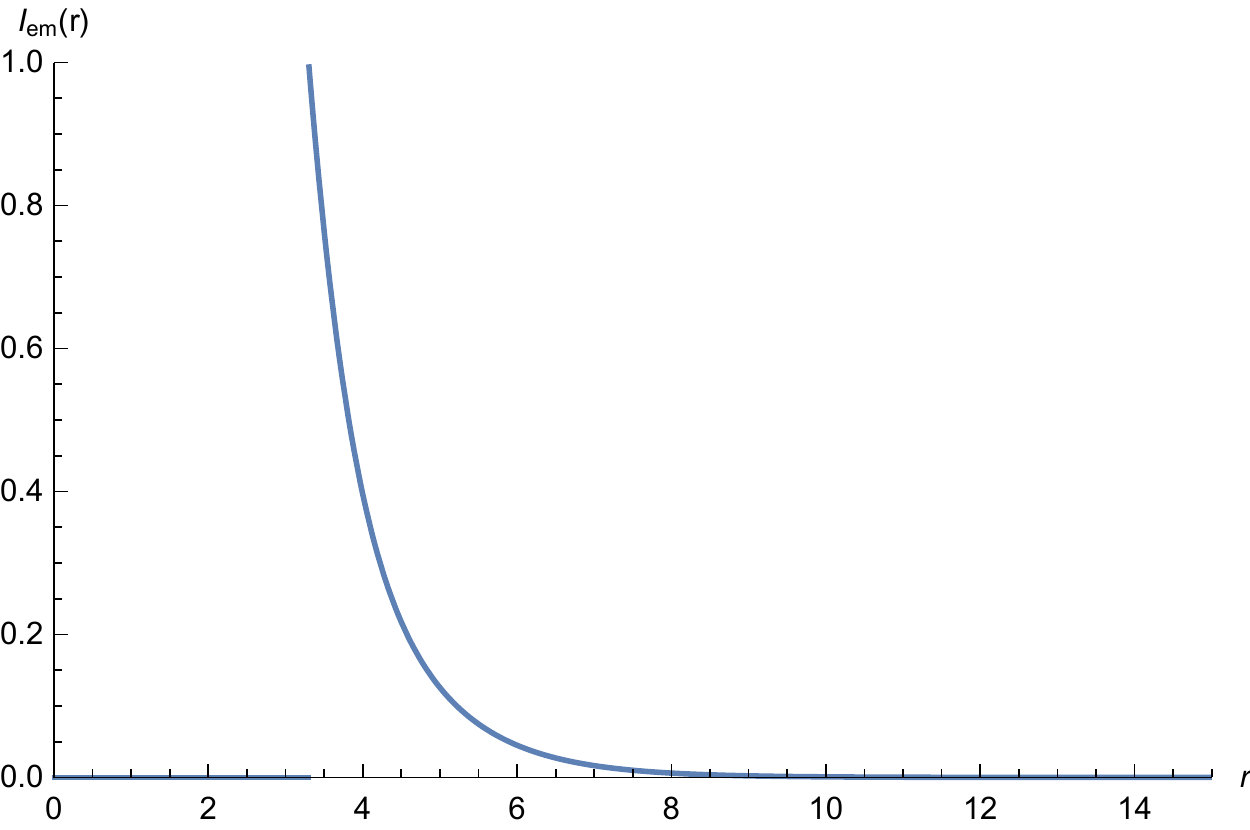}}
\subfigure[$$]{
\includegraphics[scale=0.35]{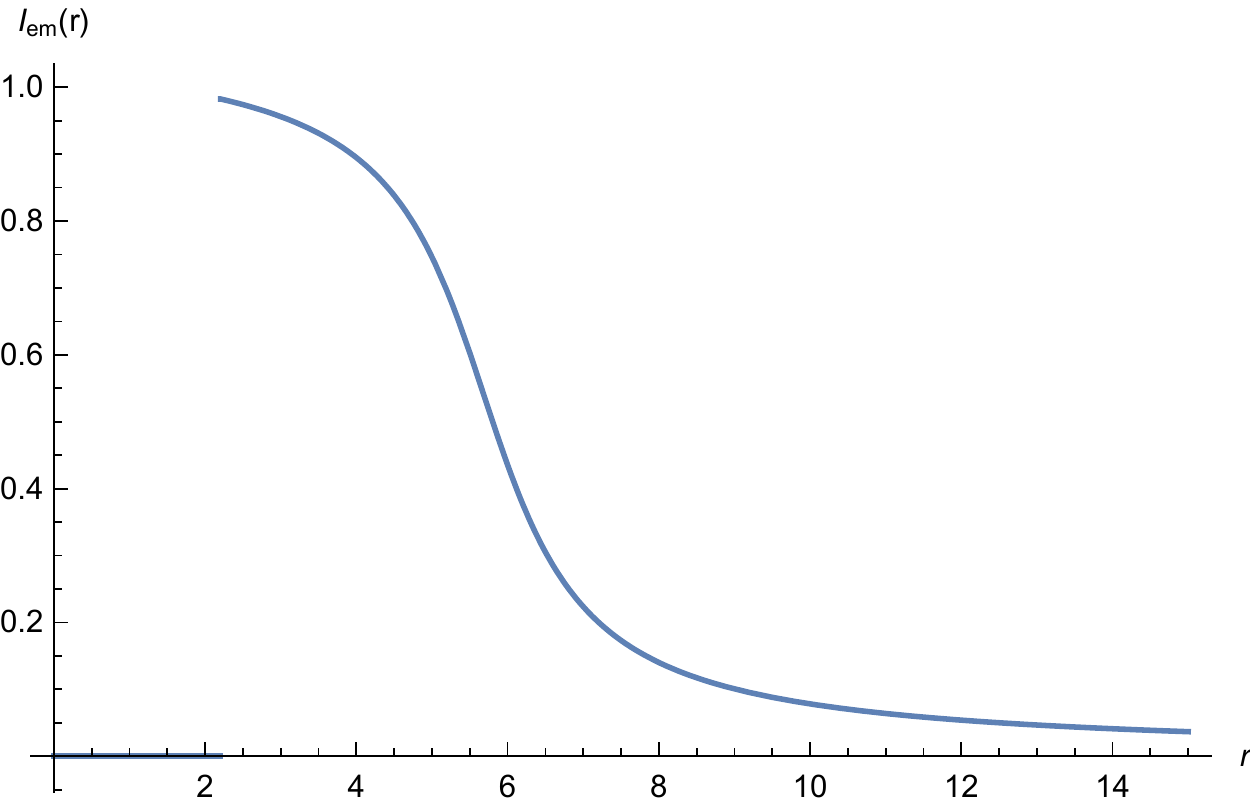}}
 \caption{Emission profiles of the three models when the values of relevant parameters are $\alpha=0.001$ and $\beta=0.3$. From left to right, it corresponds to the emissions of Models I, II, and III.}\label{4fig2}
\end{figure}

Model III: The emission is located at the event horizon, and it is a moderate decay function, which is:

\begin{align}
    I^3_{em}(r) =\begin{cases} \frac{\frac{\pi }{2}-\arctan(r- r_{isco})}{\frac{\pi }{2}+\arctan (r_{p}+1)}
    , &  r>r_h  \\
    0,  &r \leq r_h  \label{Eq.18}
    \end{cases}
\end{align}
Taking $\alpha=0.001$ and $\beta=0.3$ as an example, we give the relevant emission profiles of these three models in \ref{4fig2}. We also give the corresponding plots and density plots of observed intensity, which are shown in Figure \ref{5fig2}\footnote{The arrangement of the following figures are the same as Figure \ref{5fig2}.}.

\begin{figure}[h]
\centering 
\includegraphics[width=.35\textwidth]{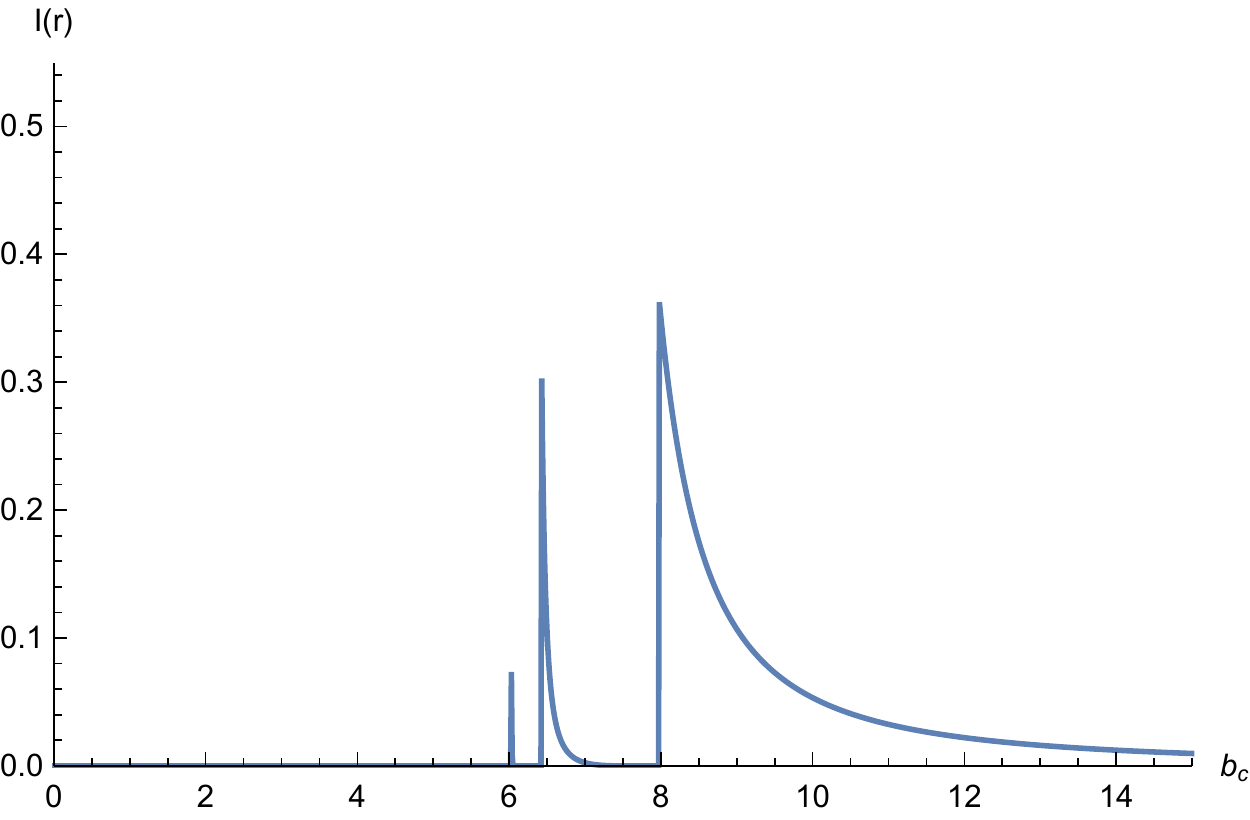}
\includegraphics[width=.235\textwidth]{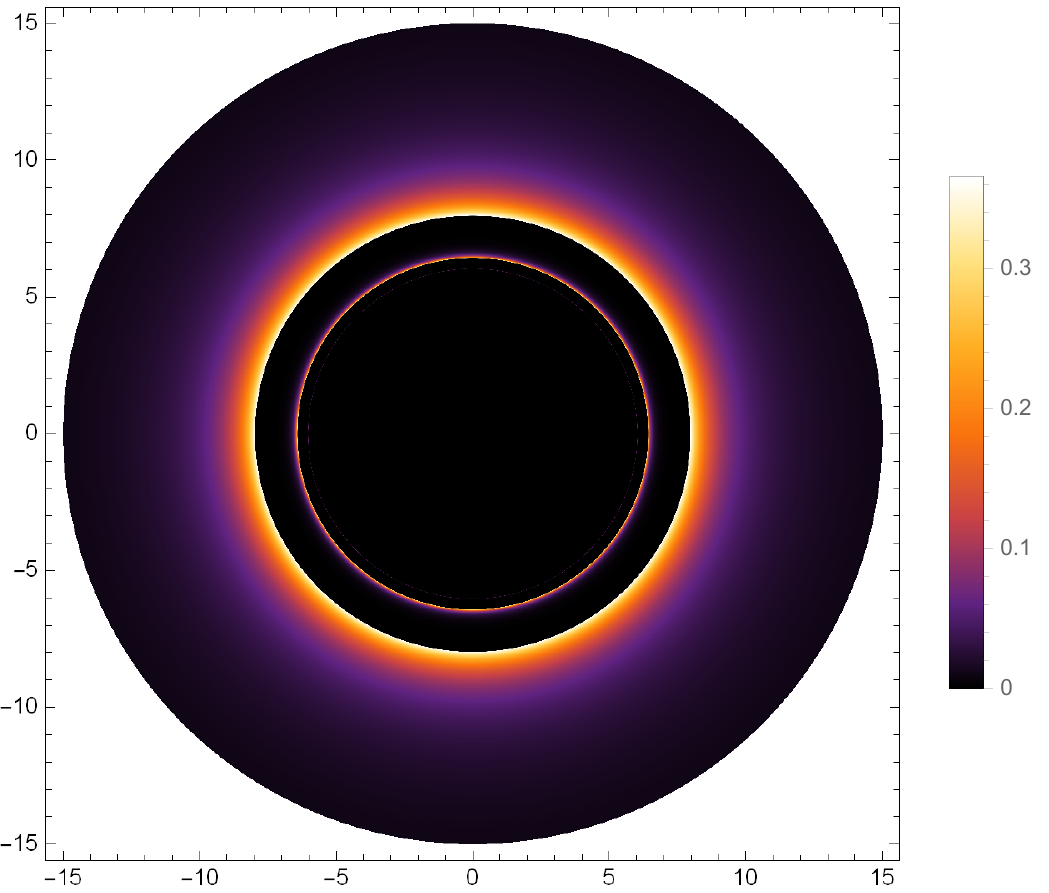}
\includegraphics[width=.2\textwidth]{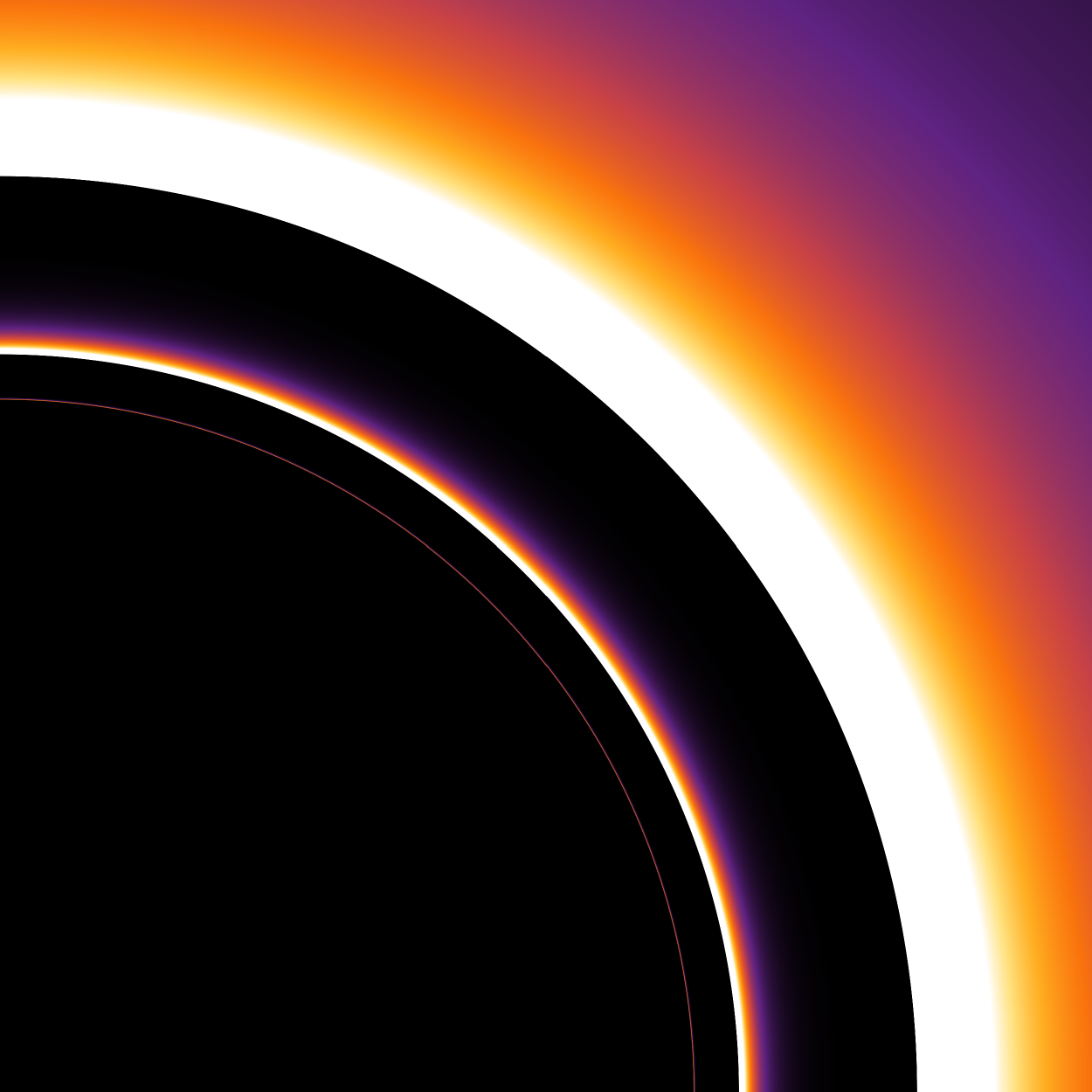}
\includegraphics[width=.35\textwidth]{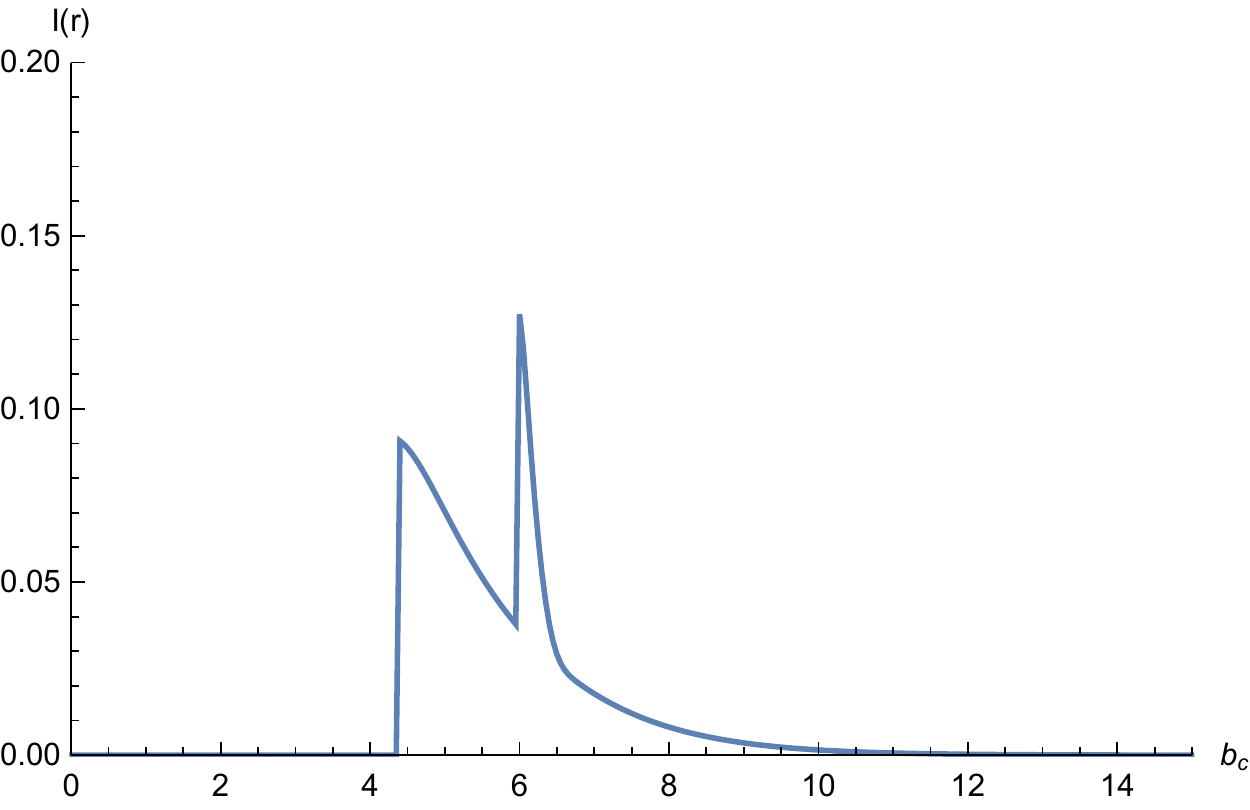}
\includegraphics[width=.235\textwidth]{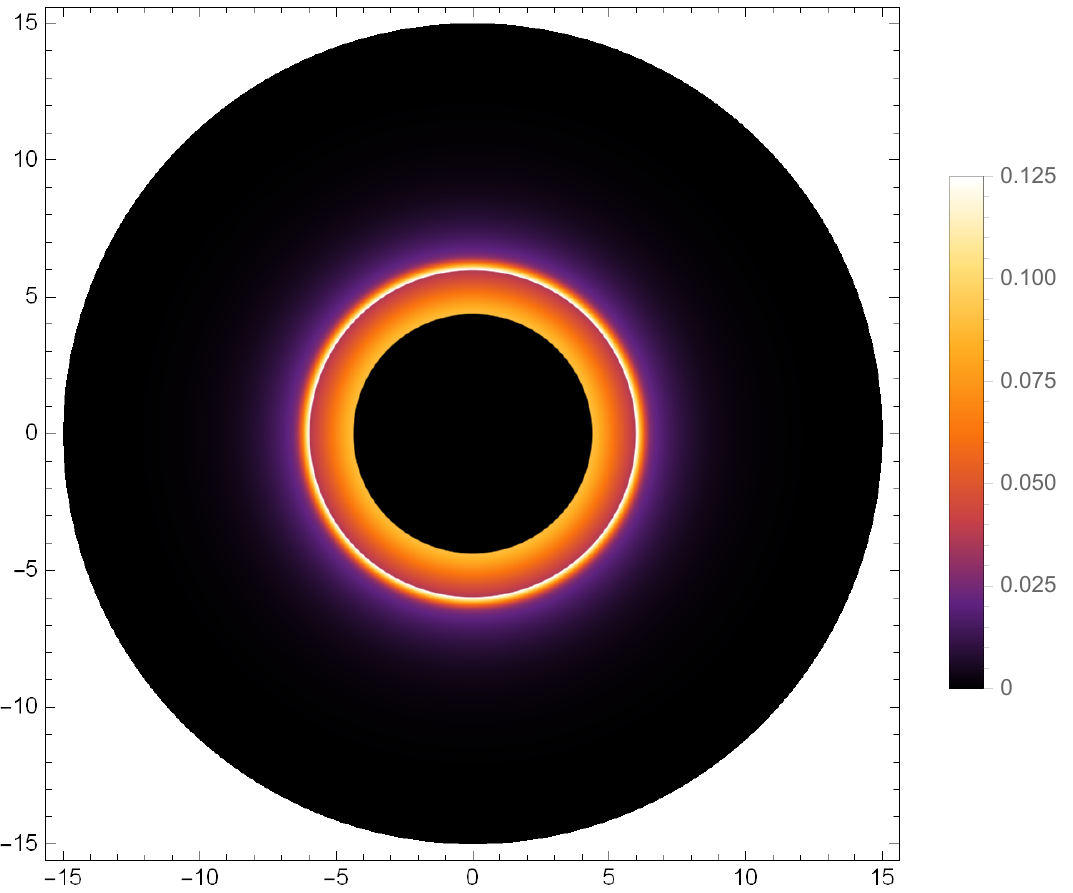}
\includegraphics[width=.2\textwidth]{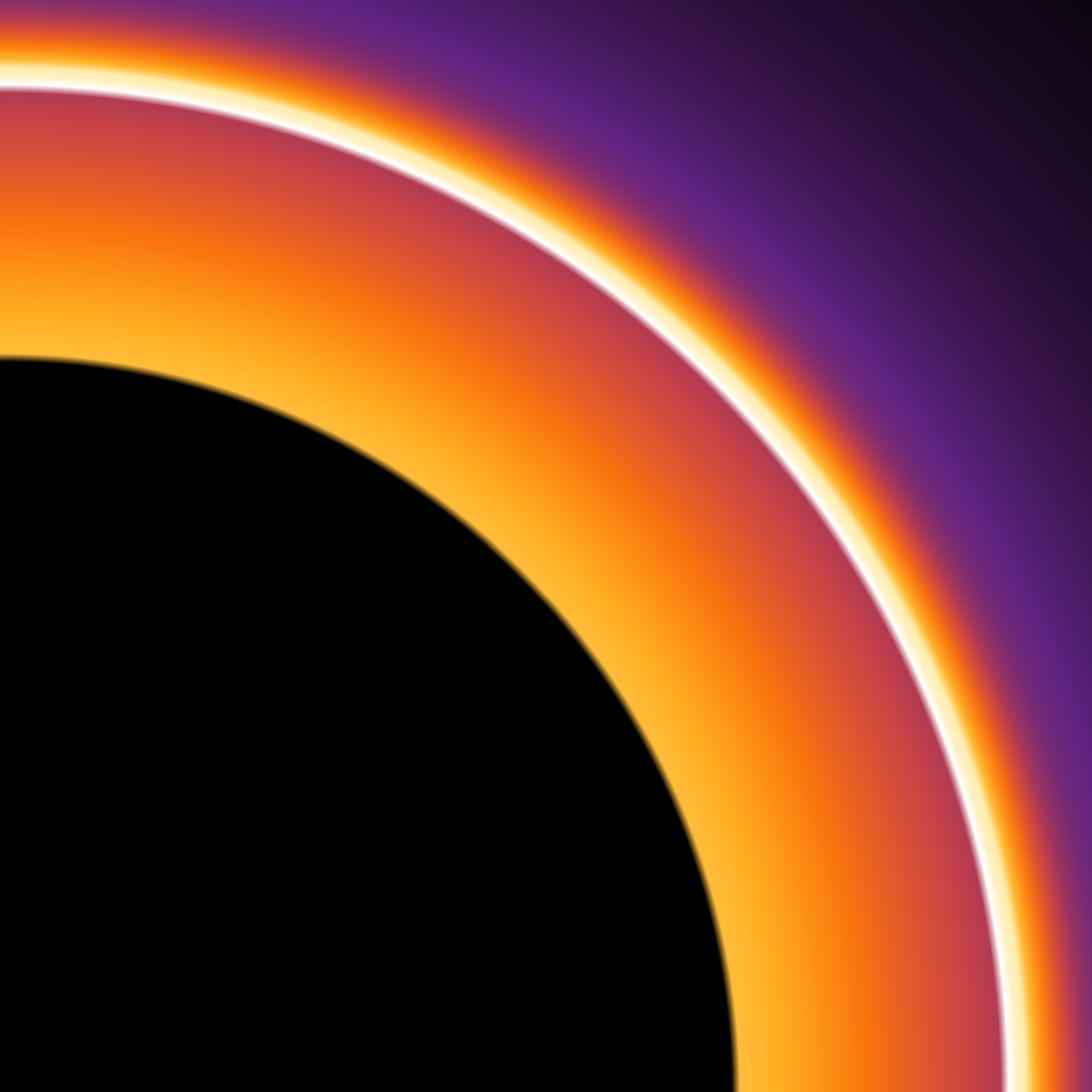}
\includegraphics[width=.35\textwidth]{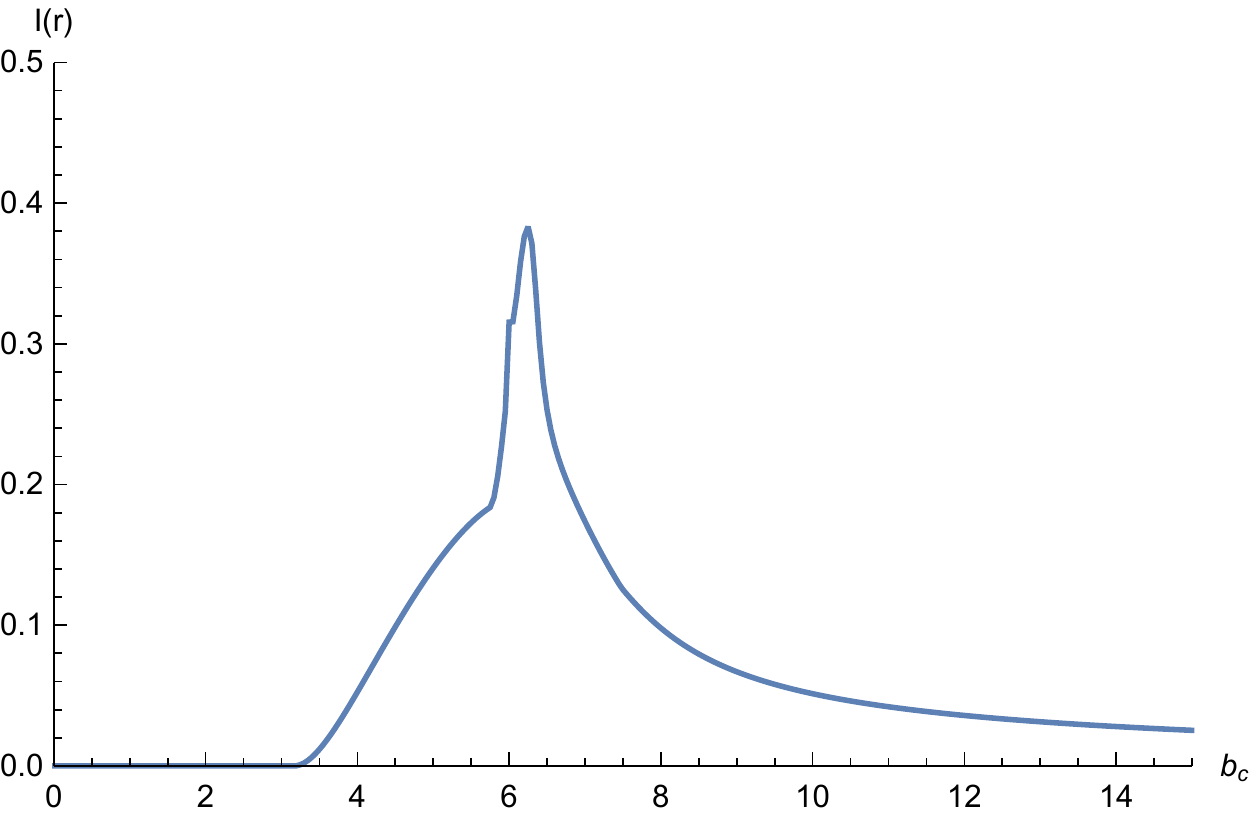}
\includegraphics[width=.235\textwidth]{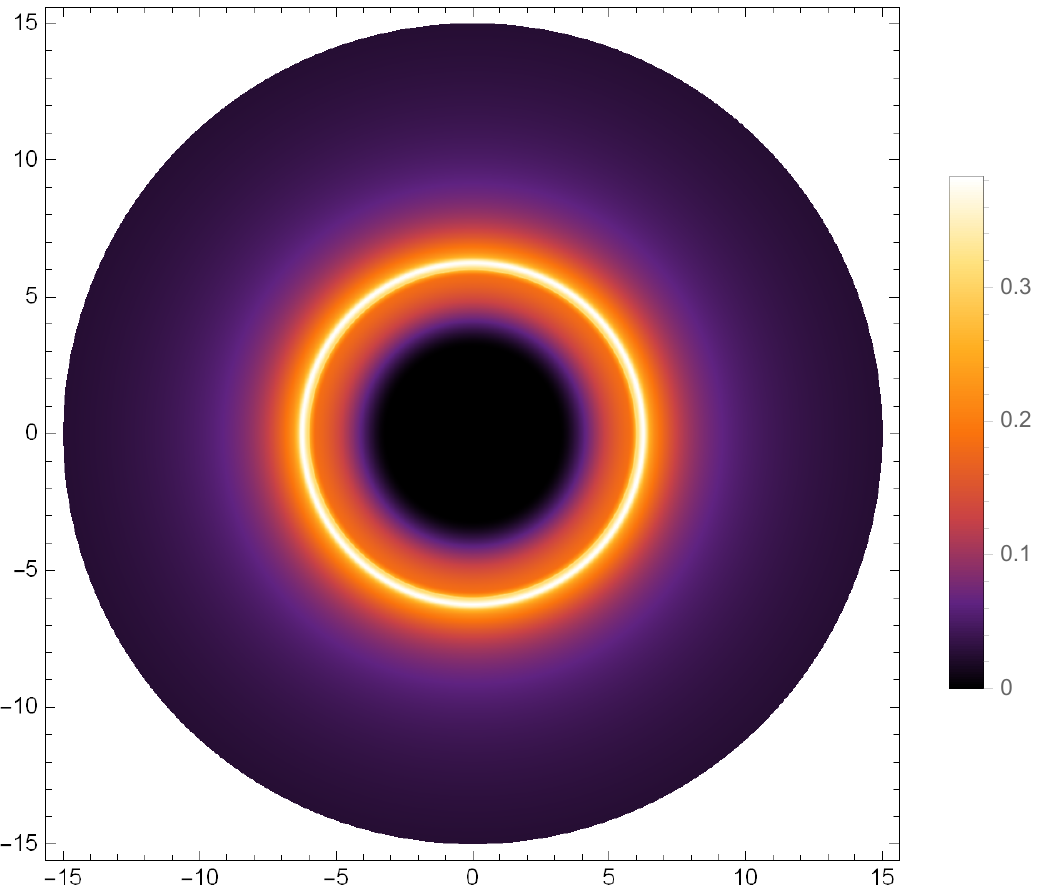}
\includegraphics[width=.2\textwidth]{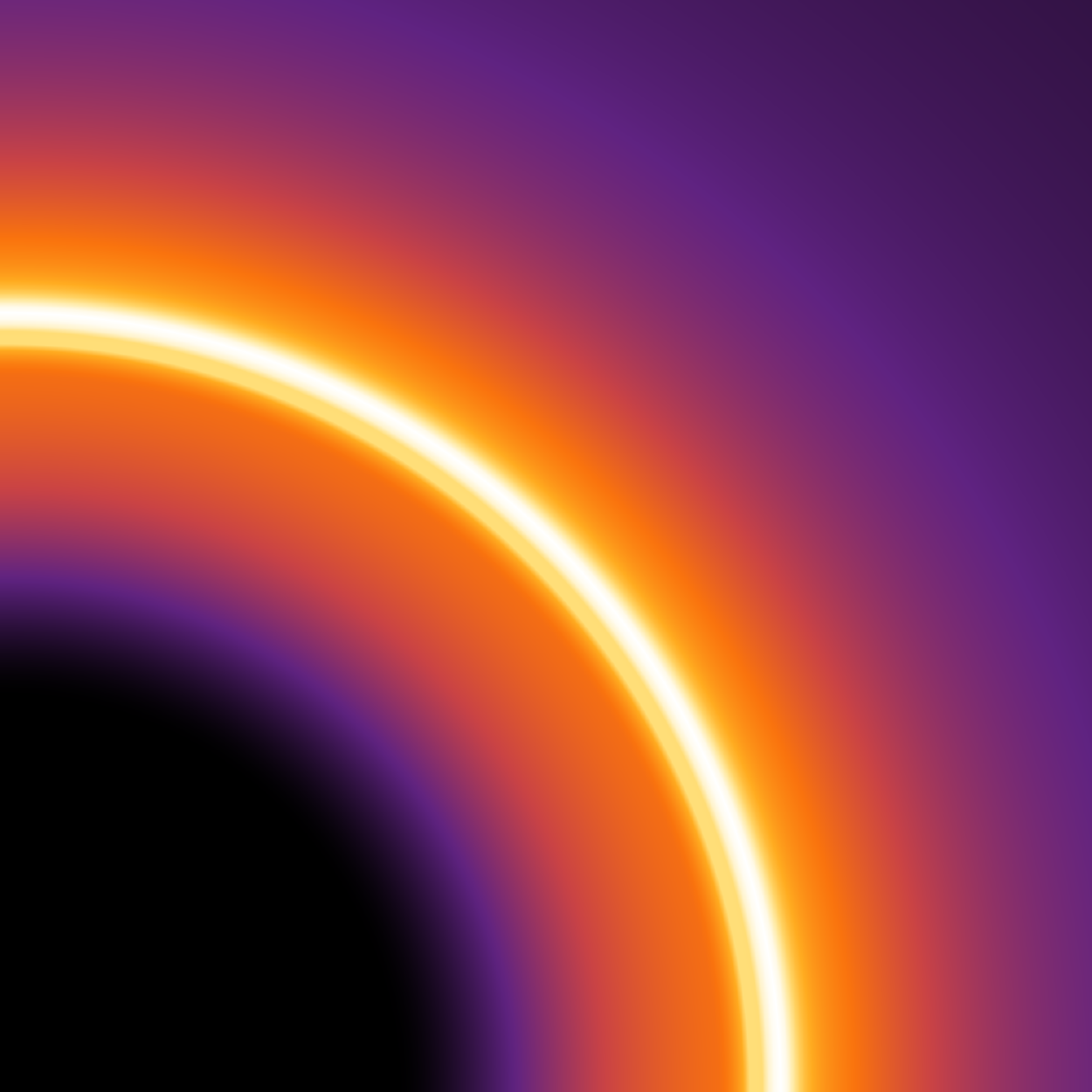}
\caption{\label{5fig2}Distribution plots (left) and density plots (middle) of the observed intensity and local density plots (right) when the relevant parameters take $\alpha=0.001$ and $\beta=0.3$.}
\end{figure}

In Figure \ref{5fig2}, the first row is the corresponding observed intensity for Model I. Although the regions of the photon and lensing ring are located on the inner side of the emission disk, there are still obse The observed appearances of black holes rved intensities of the lensing and photon rings are due to the gravitational lens effect. However, the observed intensity regions of the photon and lensing rings are very narrow, leading to the fact that the contributions of the two regions to the total observed flux are minimal. In the density plots, there is only one thin line in the photon ring region. In the area of the lensing ring, there is a more or less bright ring. However, its observed intensity is much lower than the direct emission area, particularly in the density plots. In other words, the contribution of the lensing ring to the total observed intensity is also minimal so that it can be ignored. Hence, the observed intensity that the observer can obtain mainly originates from direct emission.
In the second row of Figure \ref{5fig2}, the emitted function is fixed to Model II. The observed intensities of the photon ring and lensing ring overlapped, making them indistinguishable from the photon and lensing rings. In this case, there will be a brighter and broader bright band around the outside of the black hole. Although the total observed intensity mainly depends on direct emission, the photon and lensing rings also contribute to each part to the observer. For the emission equation in Model III, the observed intensity increases rapidly from the event horizon, peaks at $b_c\approx6.25M$, and gradually decreases to zero.
The observed intensities of photon ring, lensing ring, and direct emission overlap in a more extensive range.
In this case, the observed intensity increases sharply in the overlapped region, very different from that of the first two models. Therefore, the exterior of the black hole shadow is only accompanied by a distinct bright ring, which can also be seen in the density plots.

\begin{figure}[h]
\centering 
\includegraphics[width=.35\textwidth]{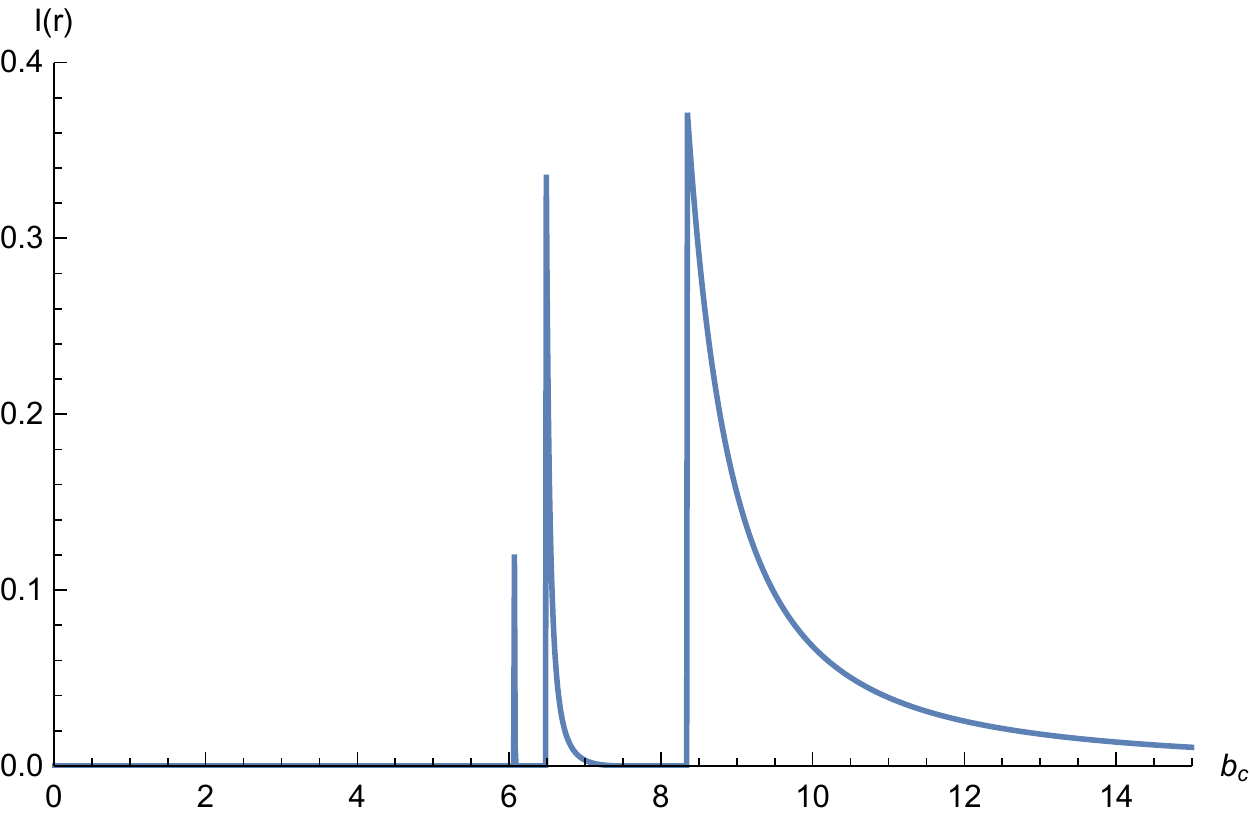}
\includegraphics[width=.235\textwidth]{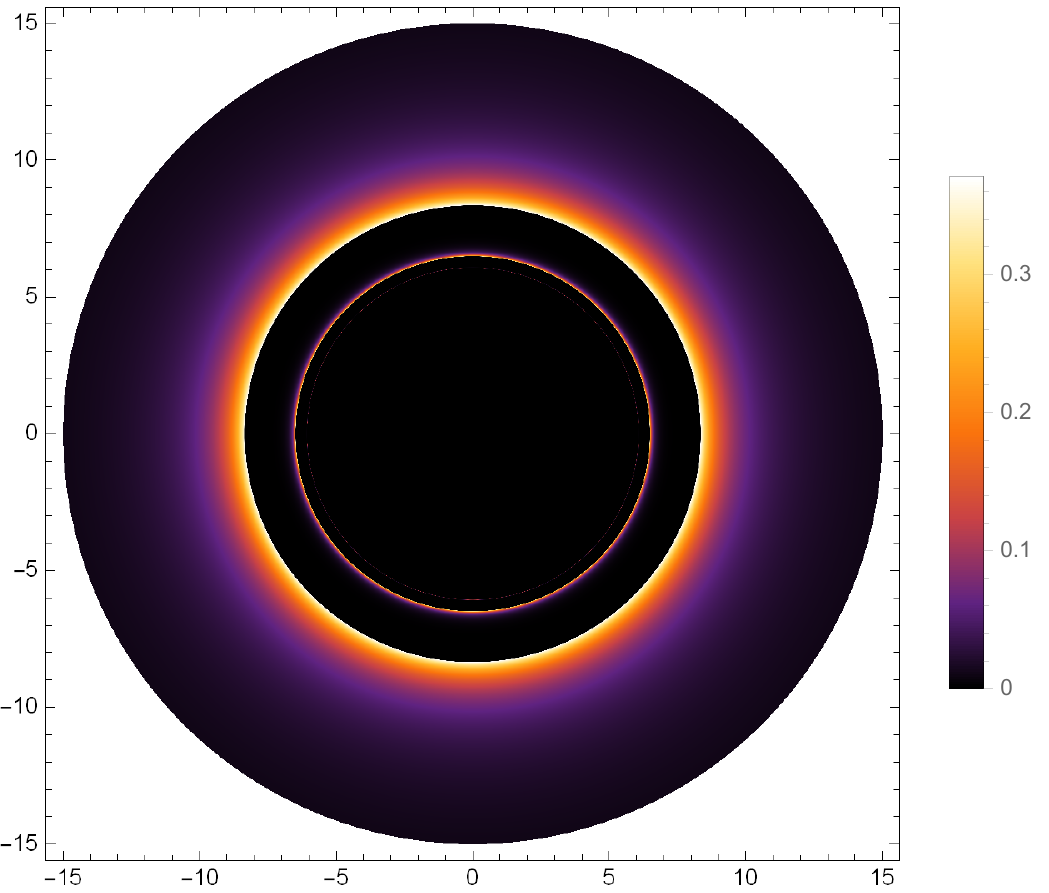}
\includegraphics[width=.2\textwidth]{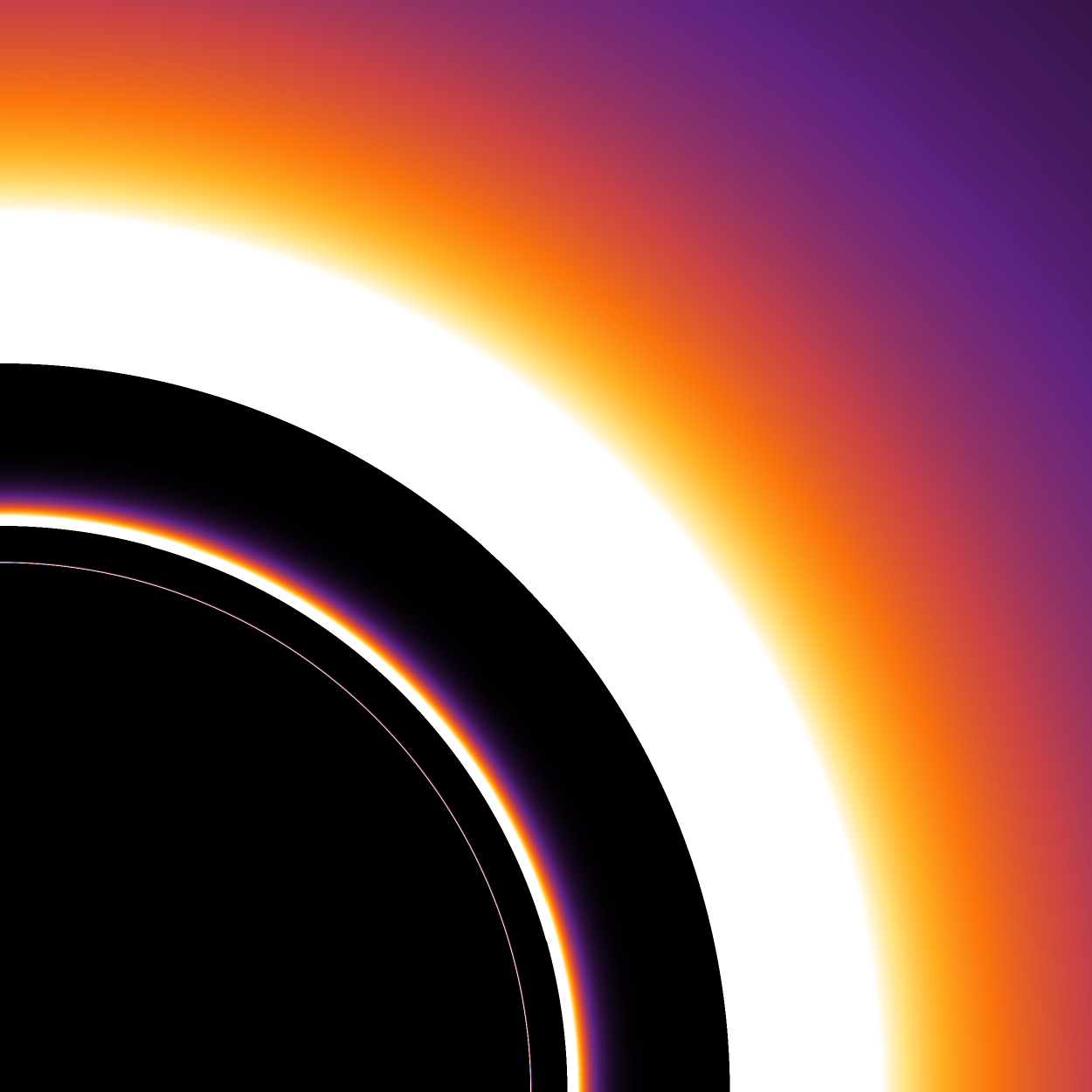}
\includegraphics[width=.35\textwidth]{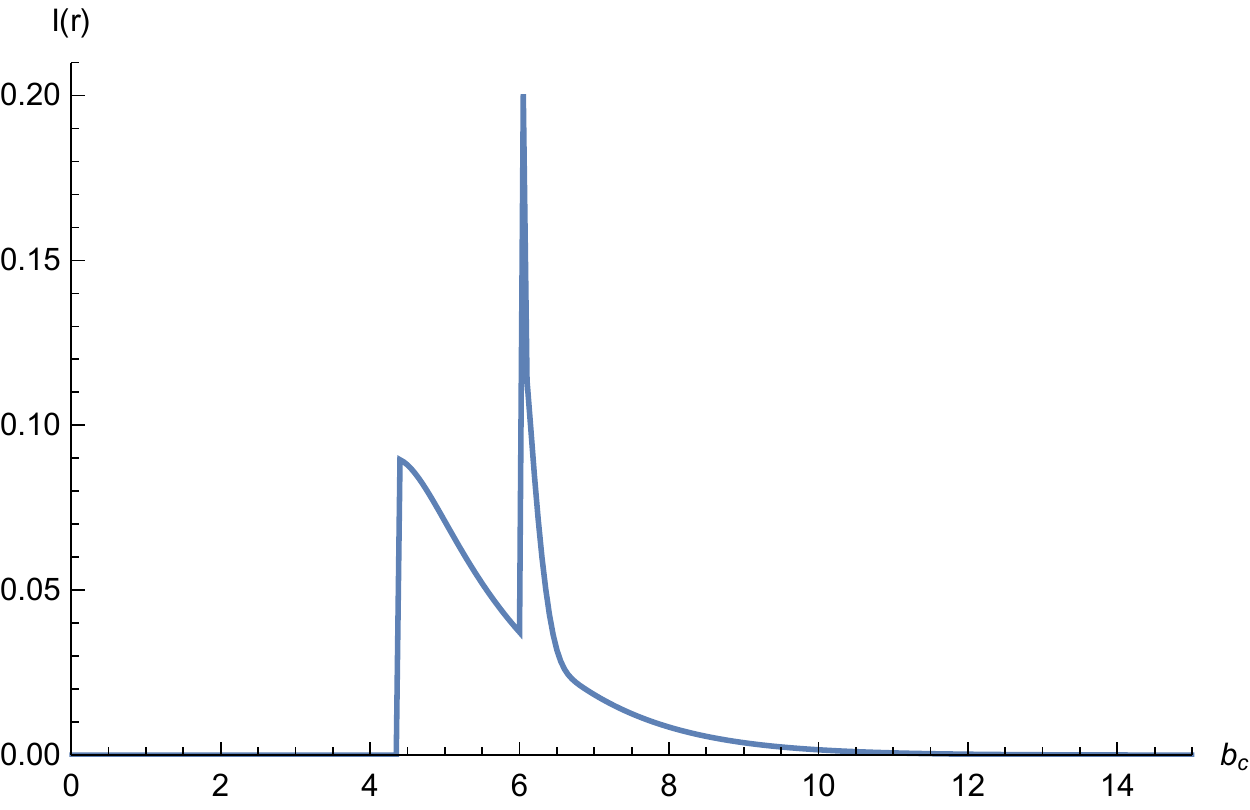}
\includegraphics[width=.235\textwidth]{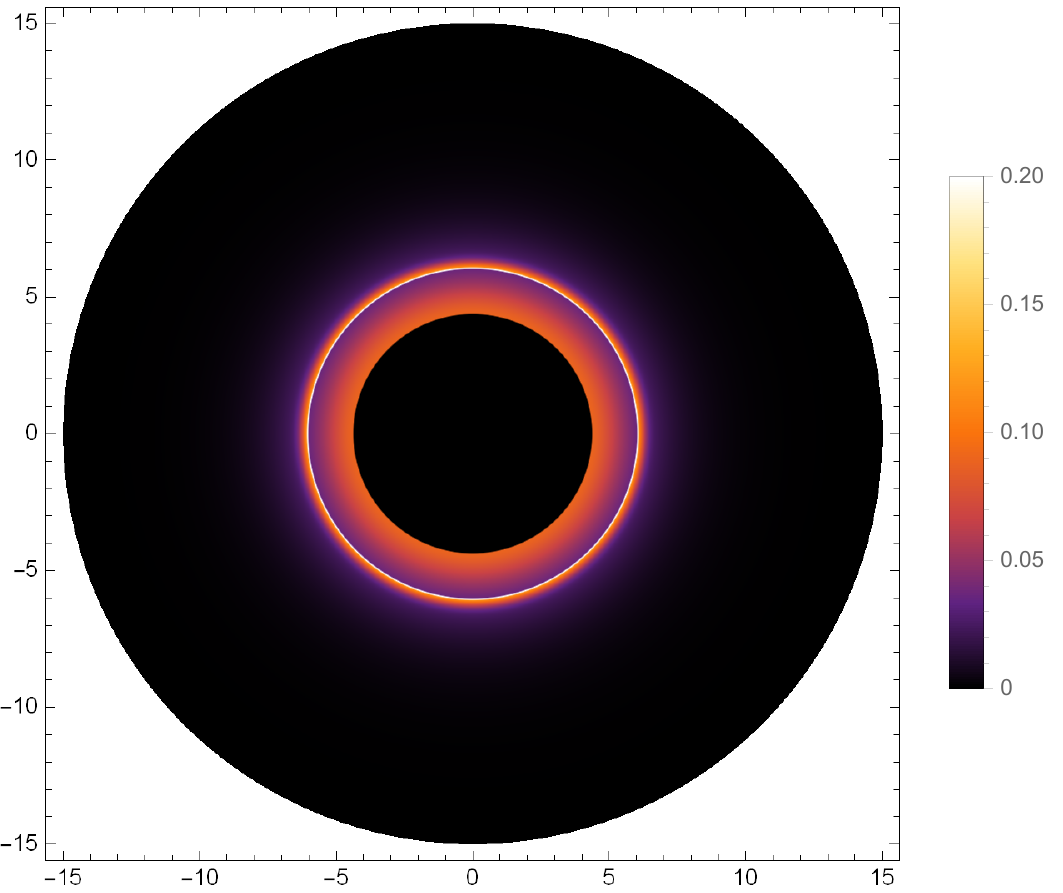}
\includegraphics[width=.2\textwidth]{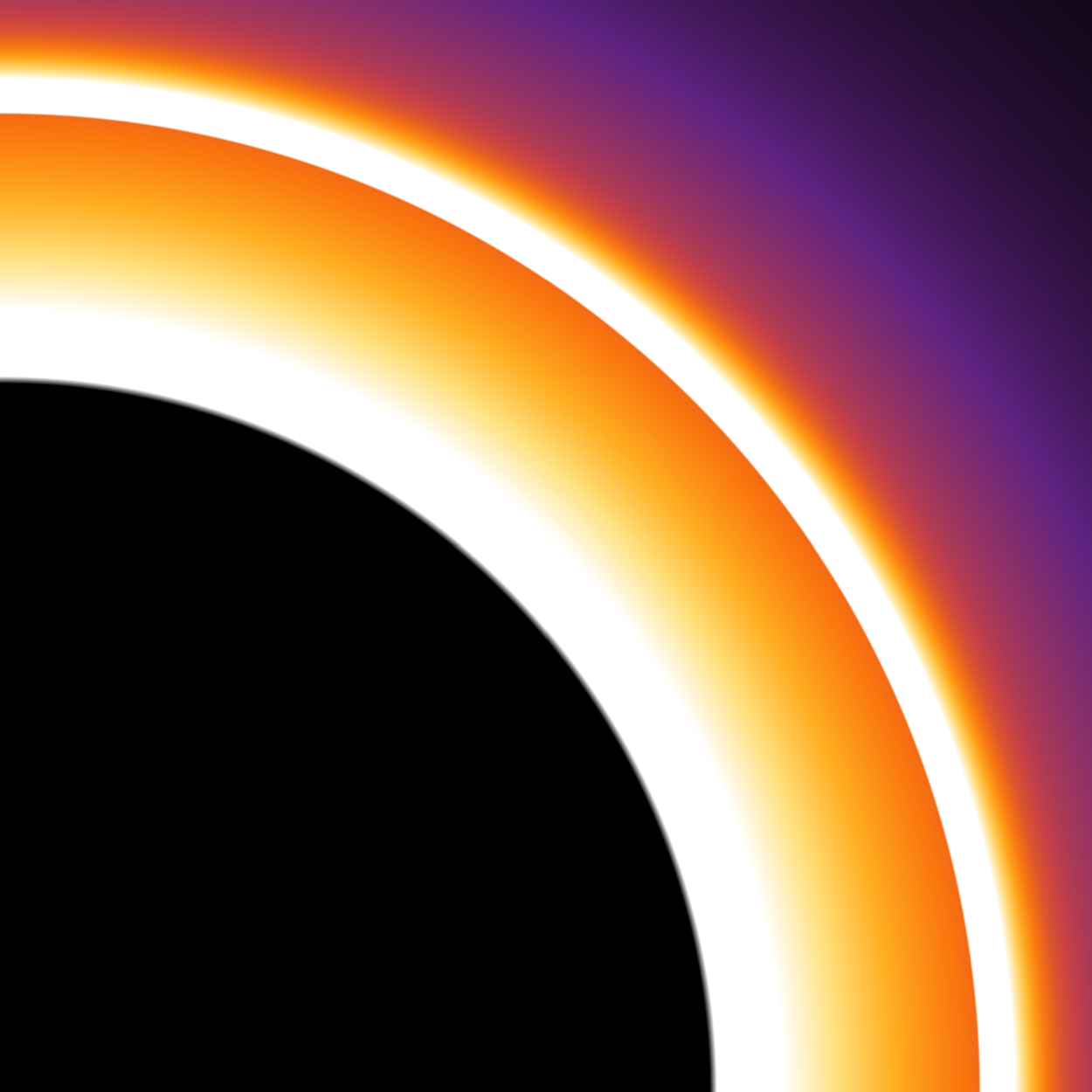}
\includegraphics[width=.35\textwidth]{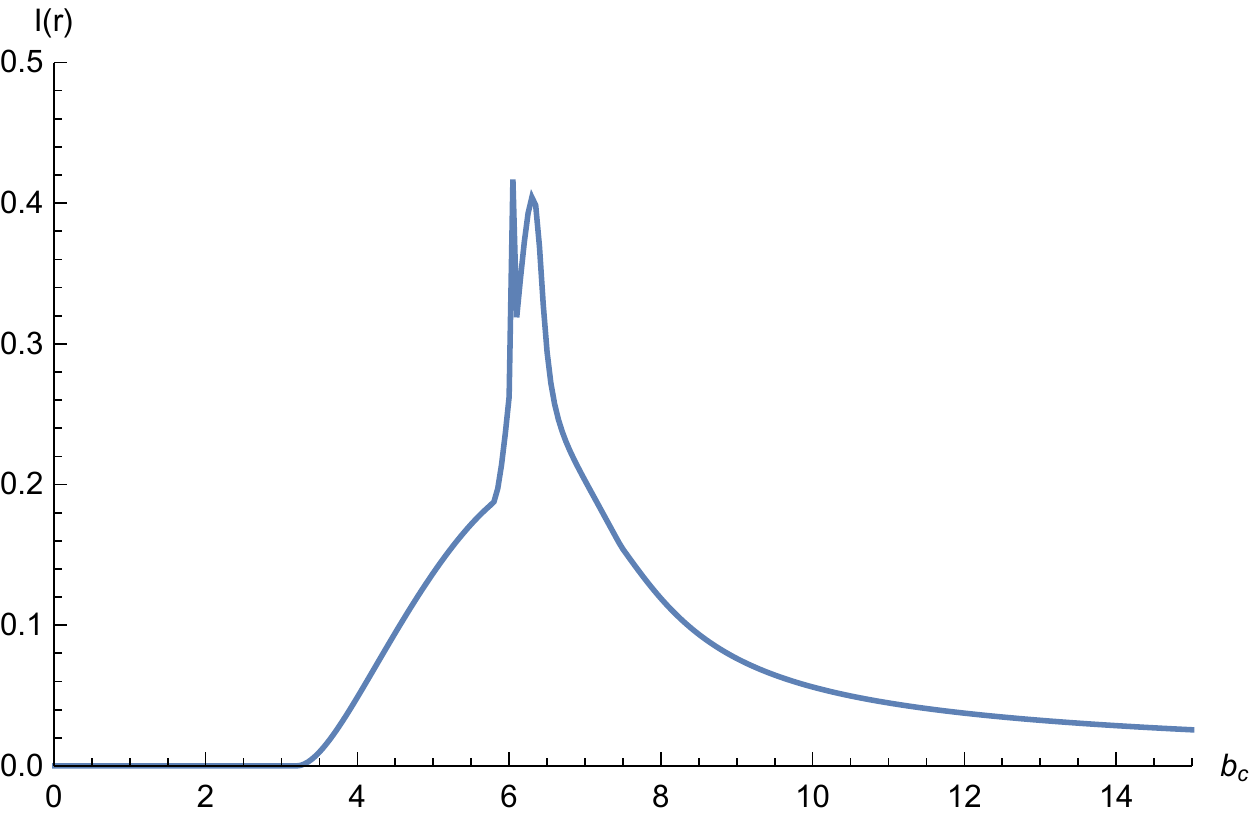}
\includegraphics[width=.235\textwidth]{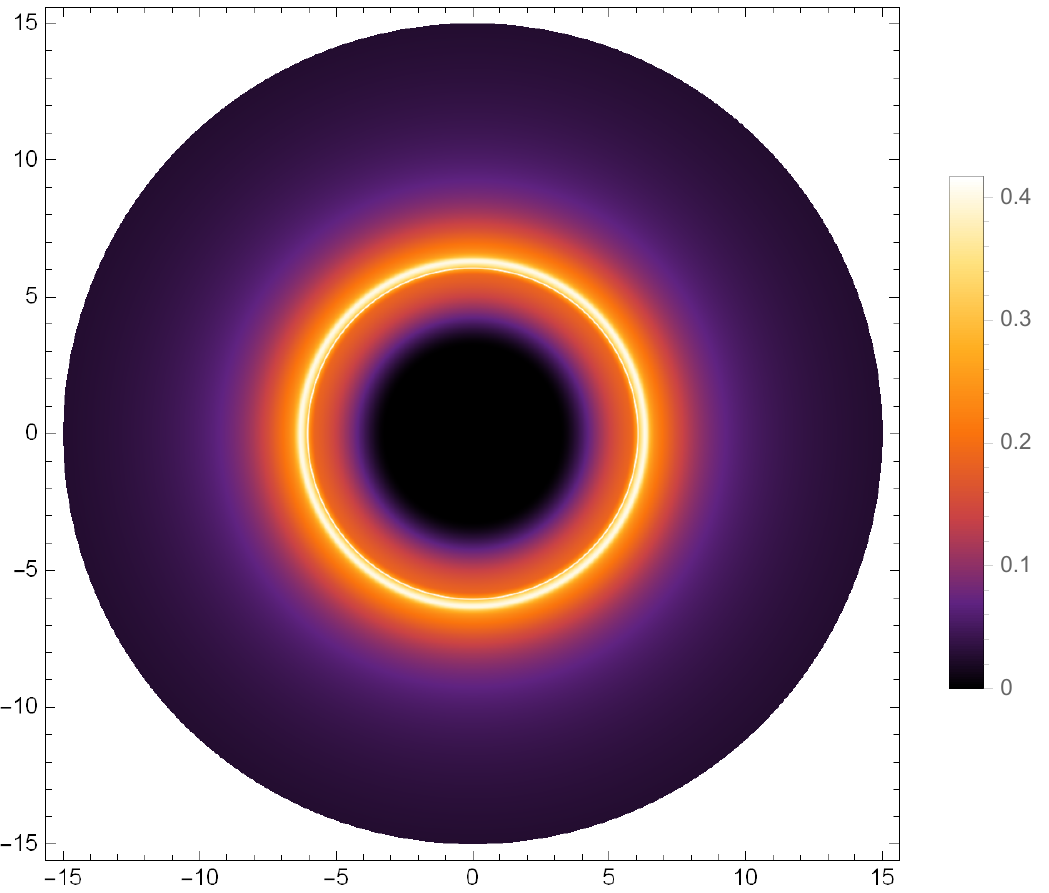}
\includegraphics[width=.2\textwidth]{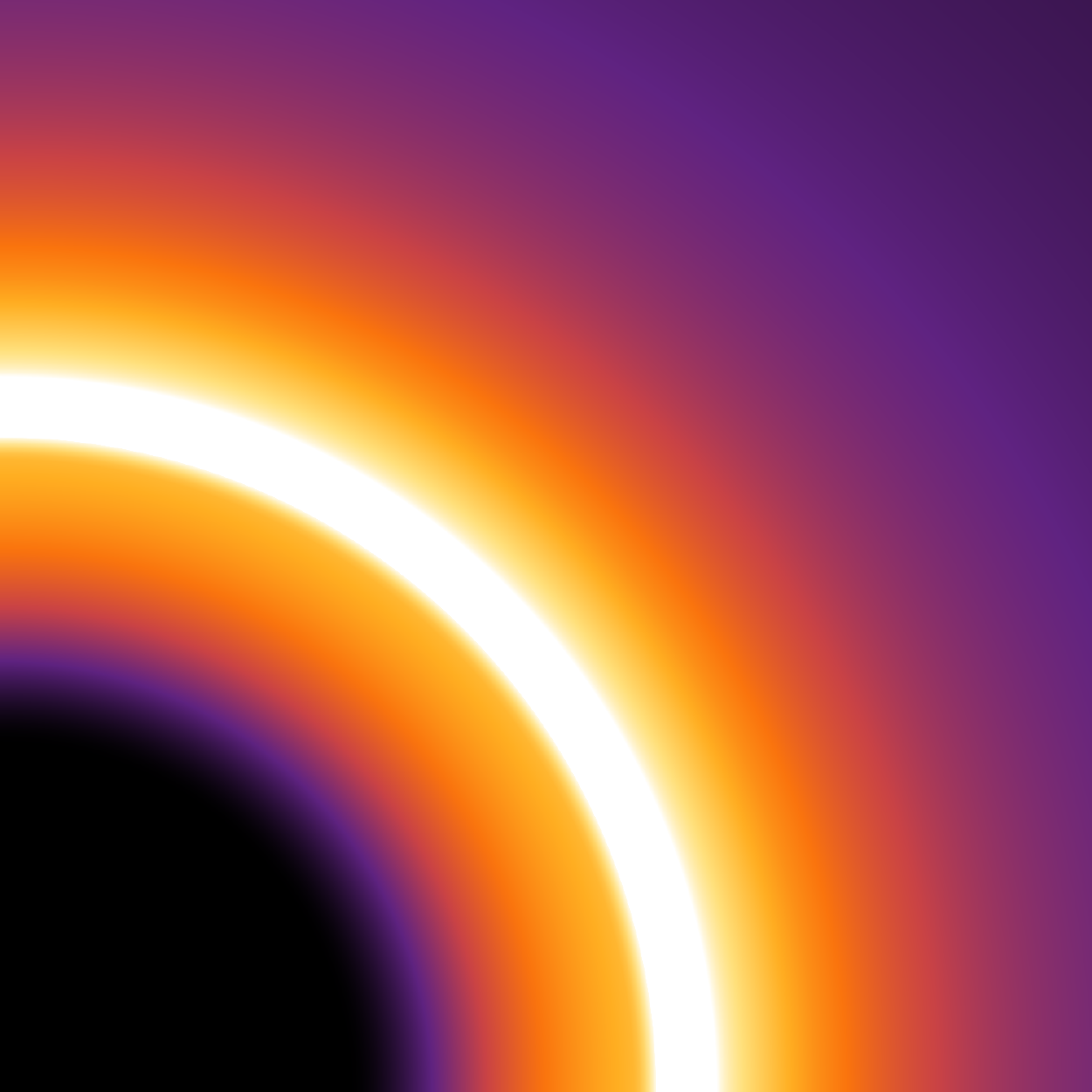}
\caption{\label{6fig6}Observed appearances of black holes when the relevant parameters take $\alpha=0.003$ and $\beta=0.3$.}
\end{figure}

\begin{figure}[h]
\centering 
\includegraphics[width=.35\textwidth]{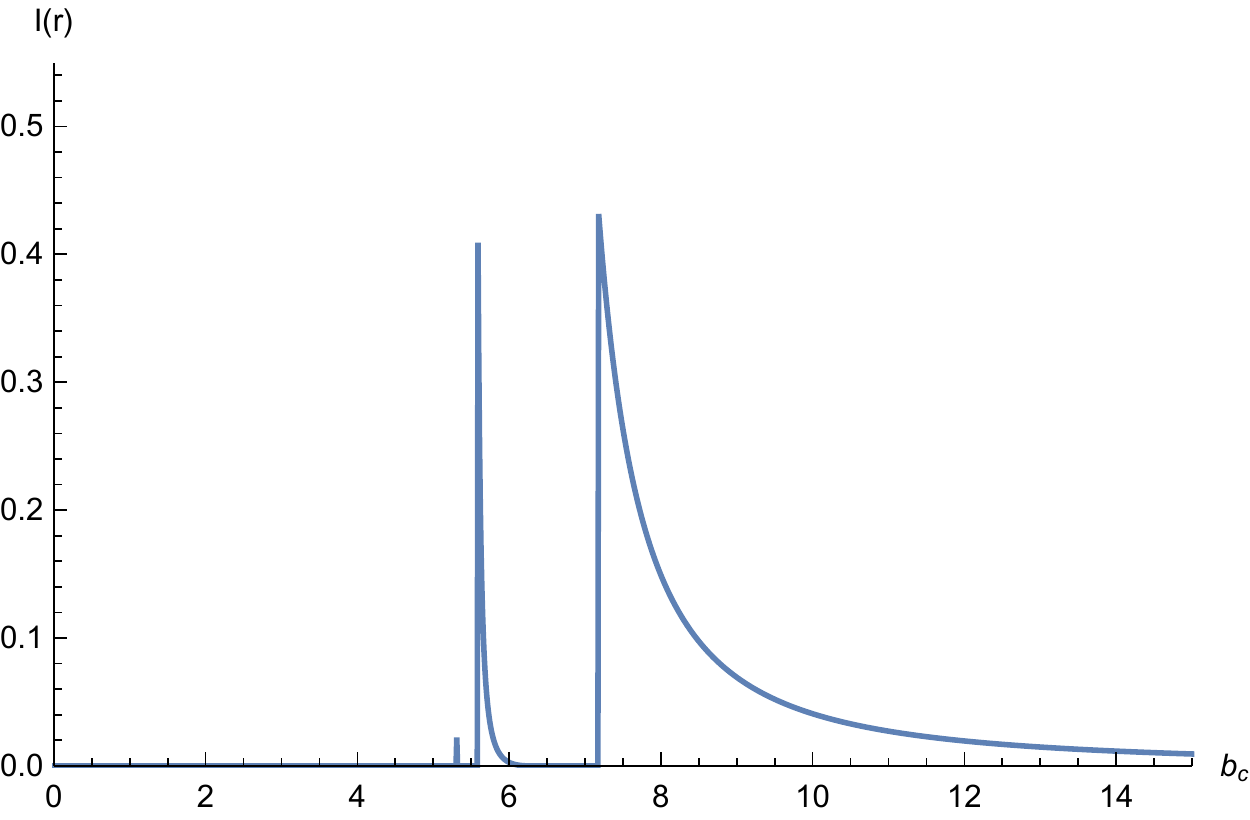}
\includegraphics[width=.235\textwidth]{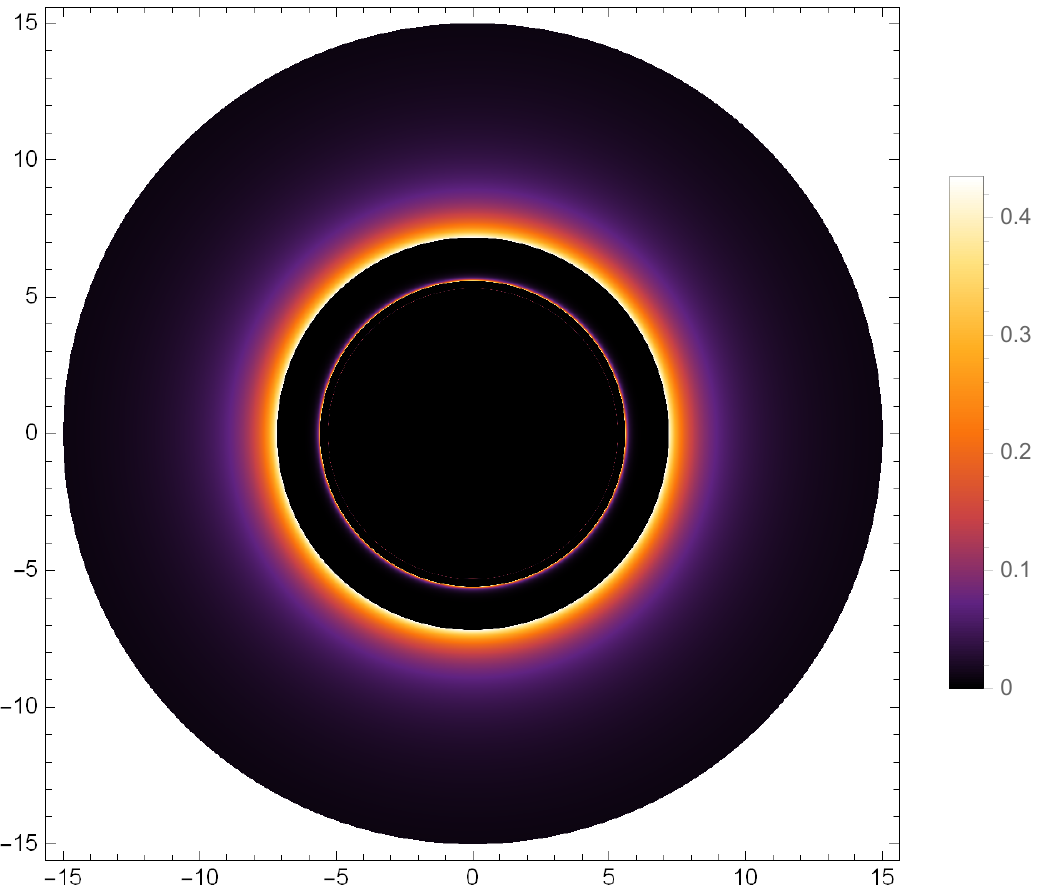}
\includegraphics[width=.2\textwidth]{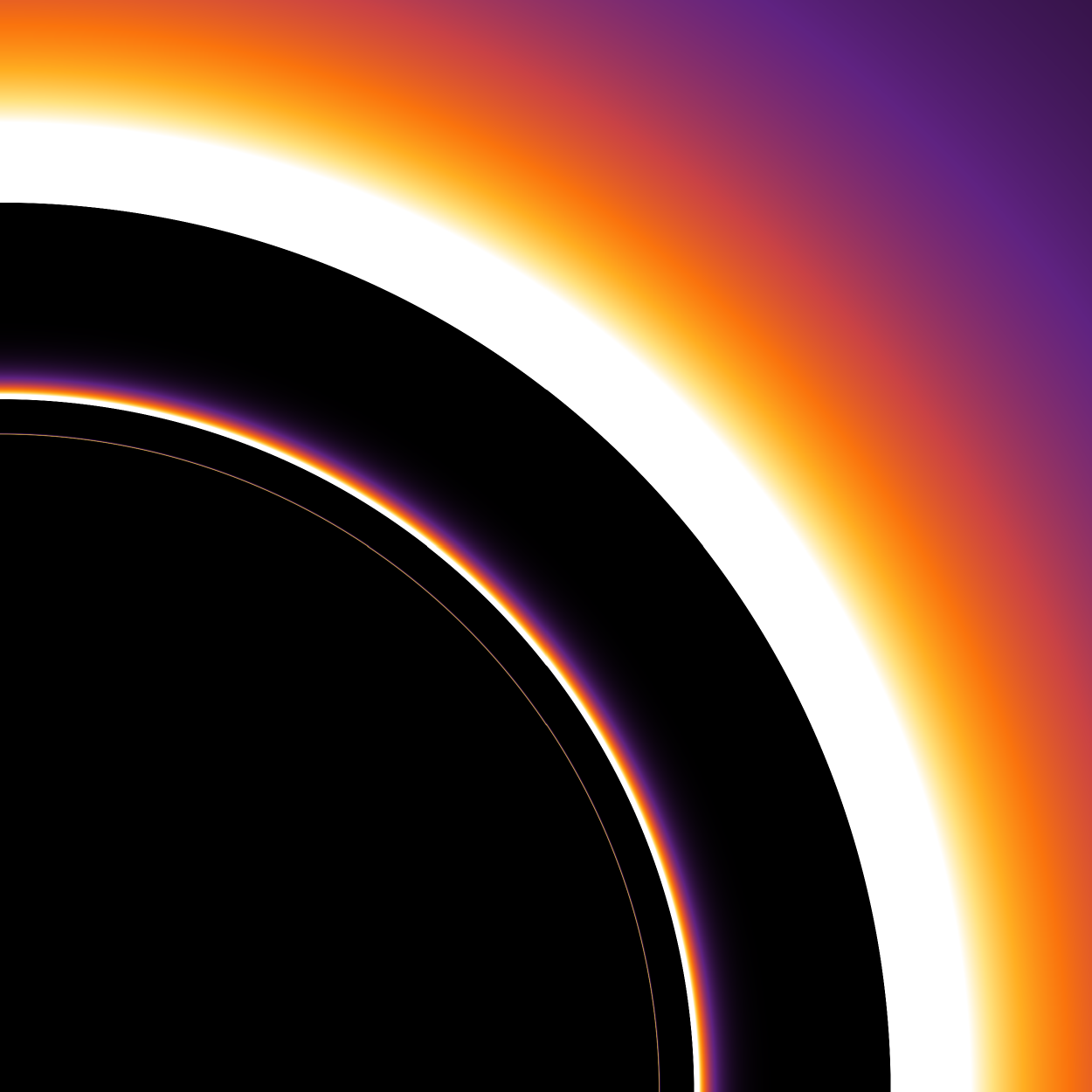}
\includegraphics[width=.35\textwidth]{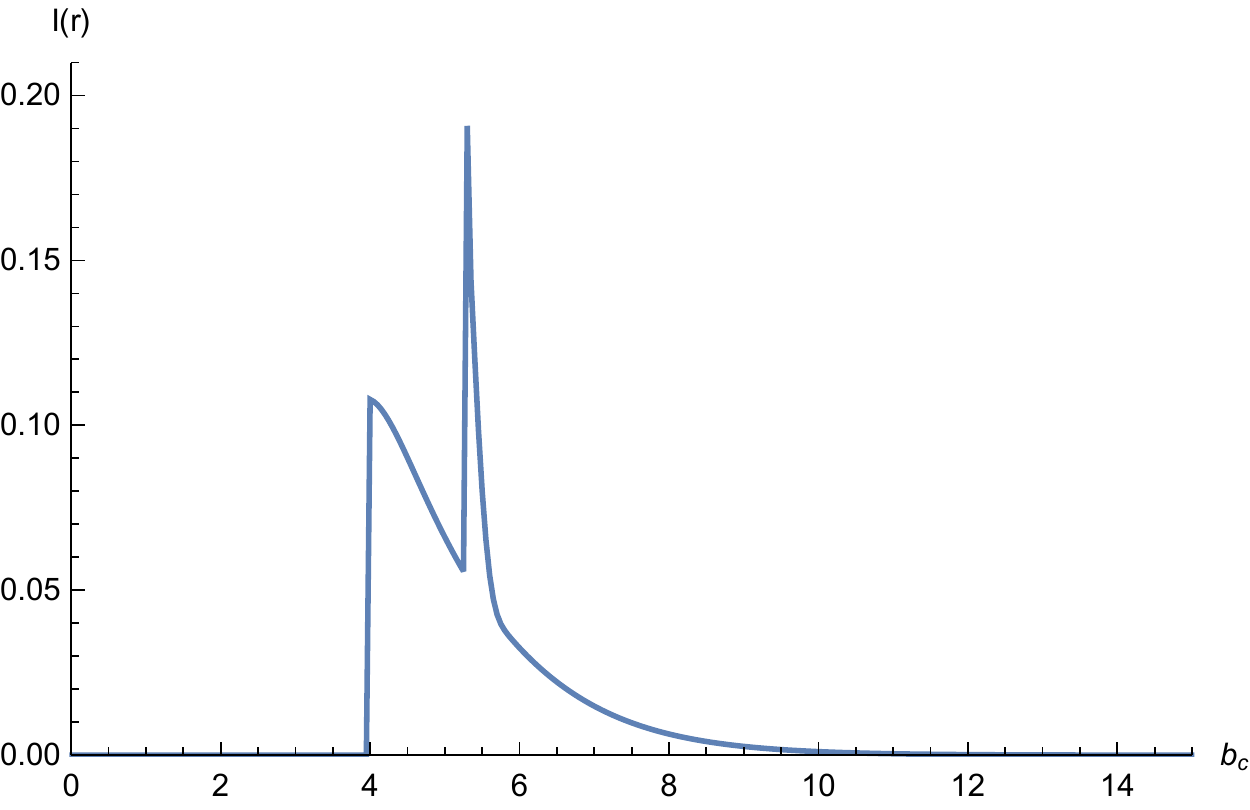}
\includegraphics[width=.235\textwidth]{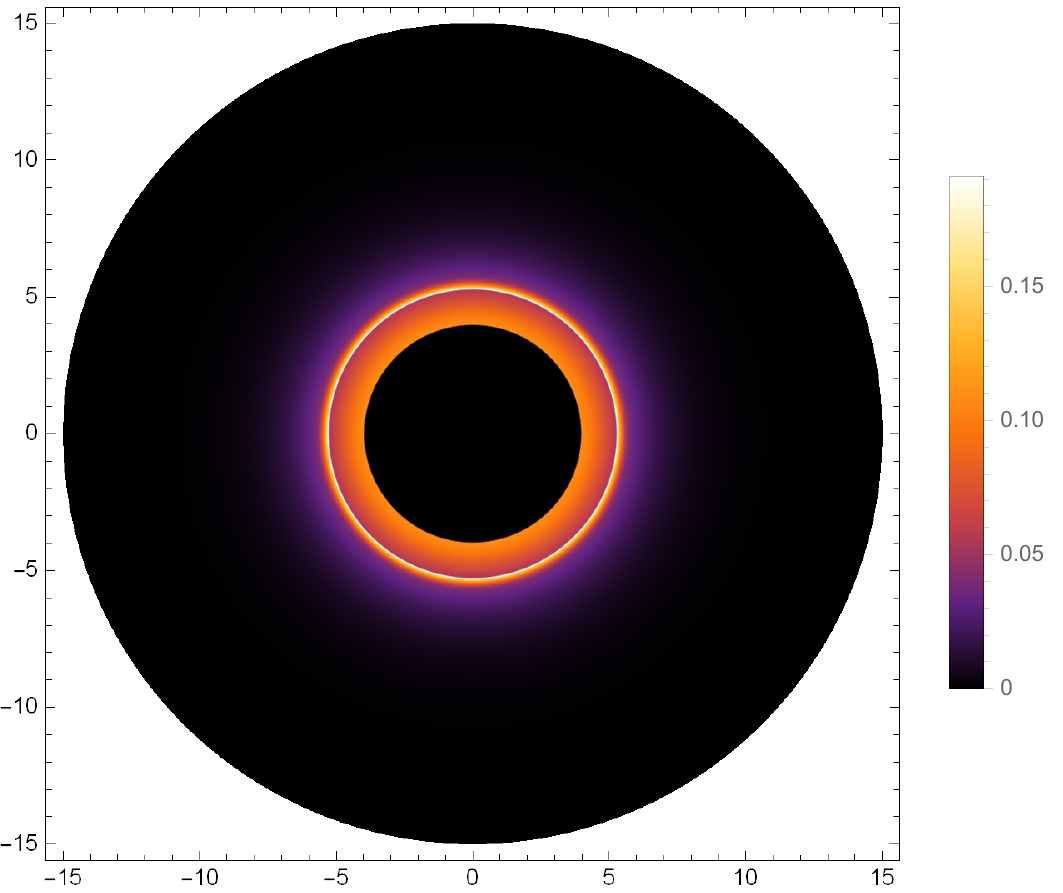}
\includegraphics[width=.2\textwidth]{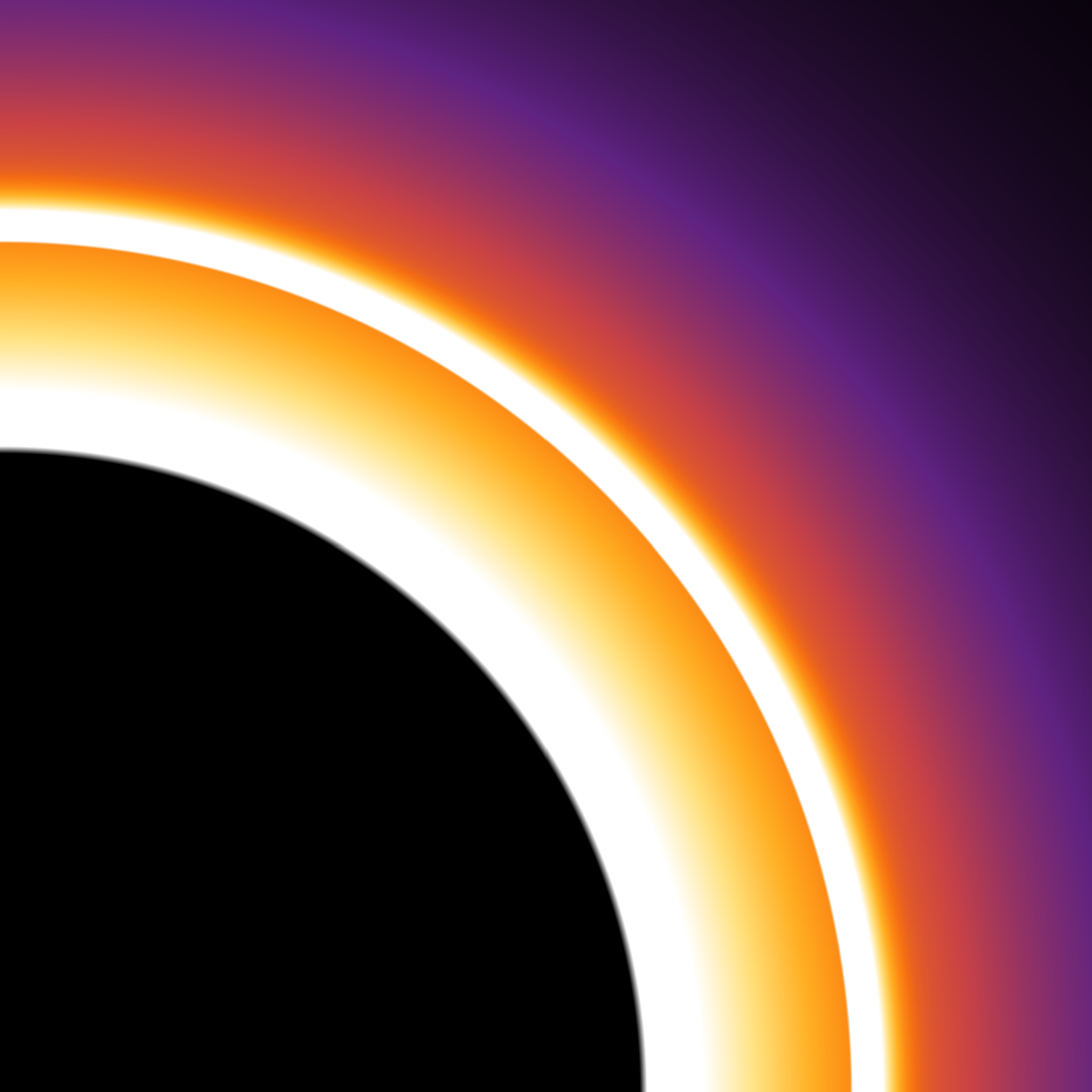}
\includegraphics[width=.35\textwidth]{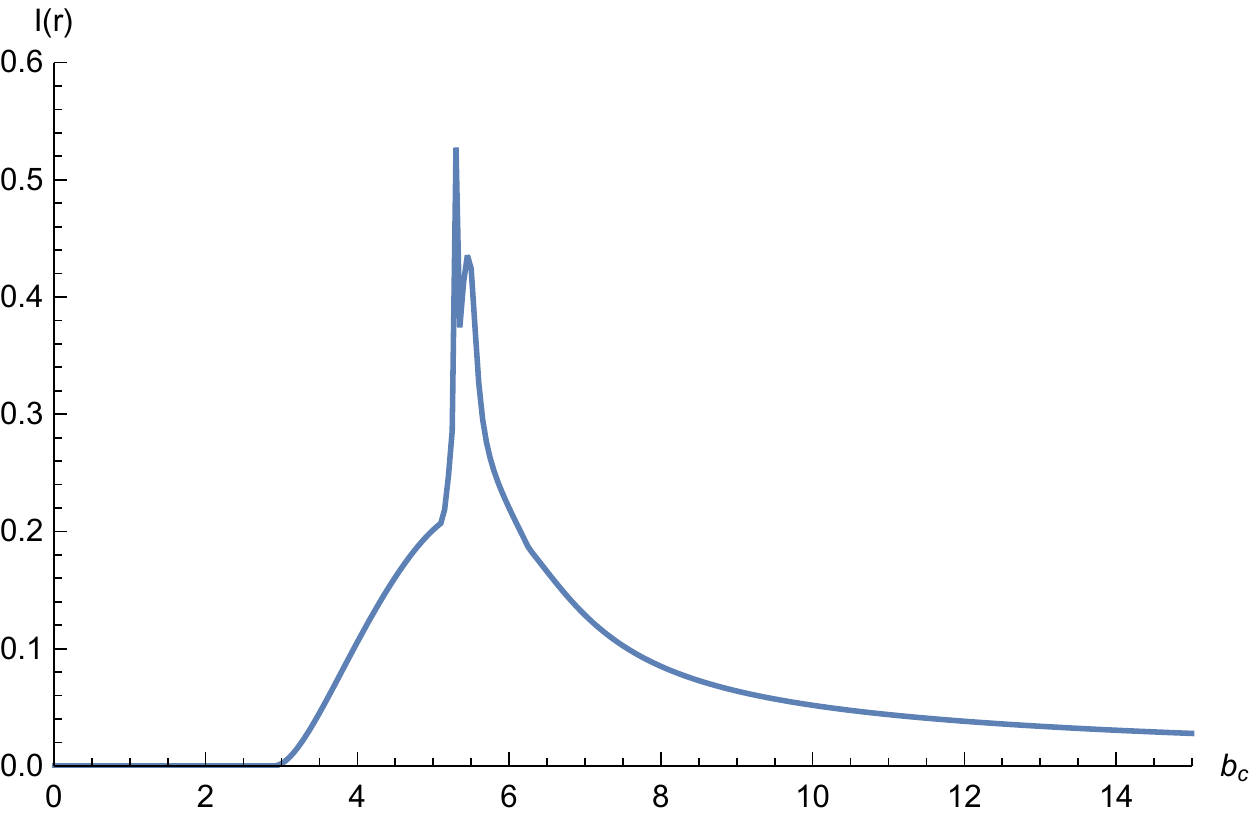}
\includegraphics[width=.235\textwidth]{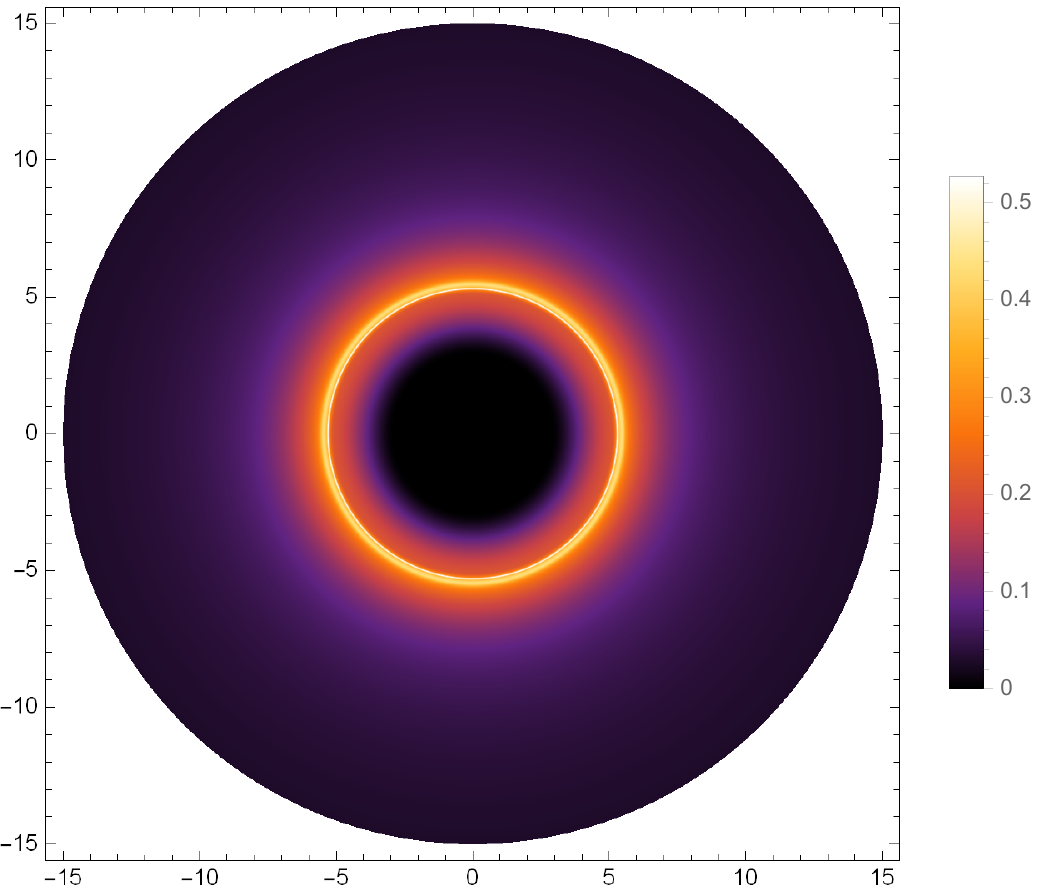}
\includegraphics[width=.2\textwidth]{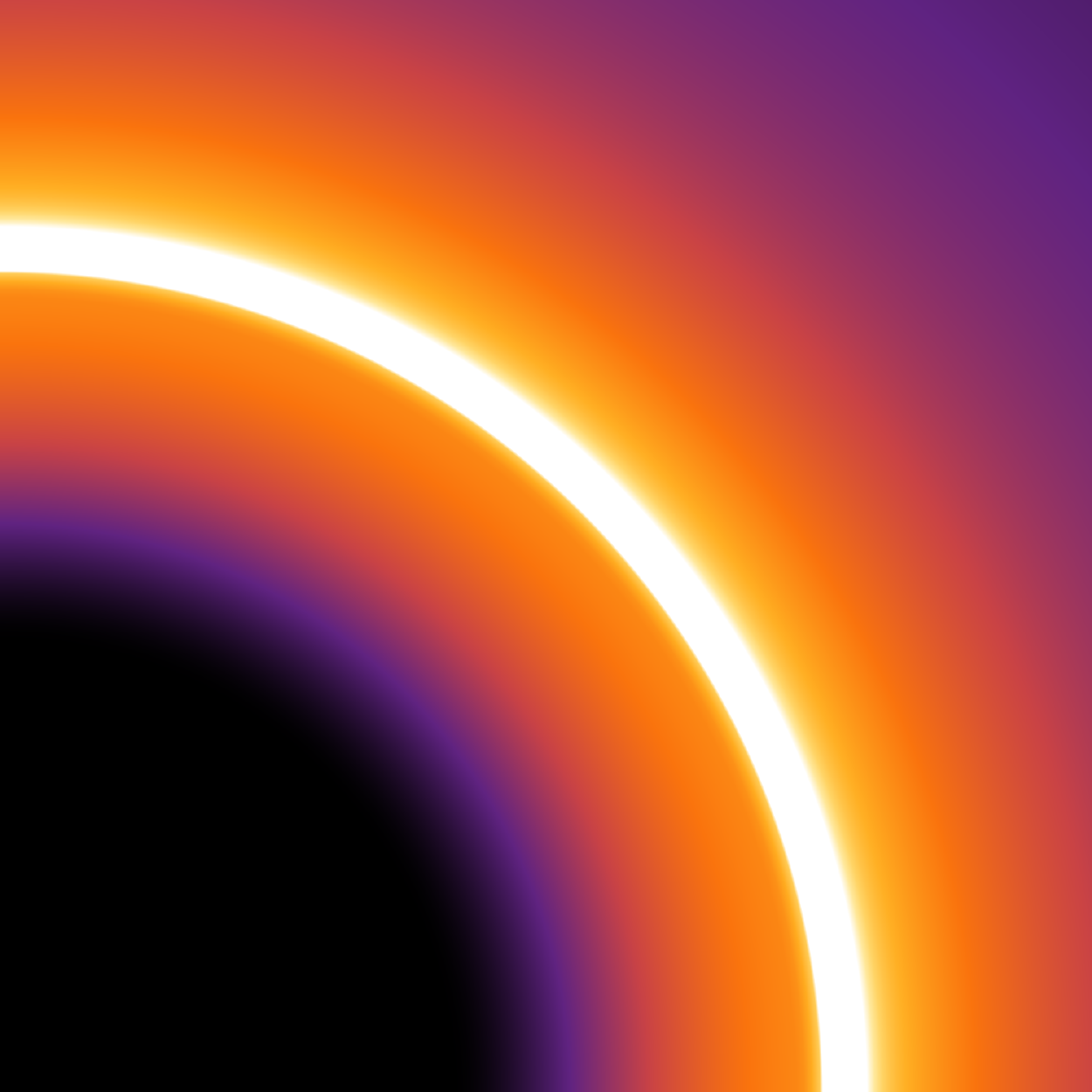}
\caption{\label{7fig7} Observed appearances of black holes when the relevant parameters take $\alpha=0.003$ and $\beta=0.1$. }
\end{figure}

\begin{figure}[h]
\centering
\subfigure[]{\includegraphics[scale=0.25]{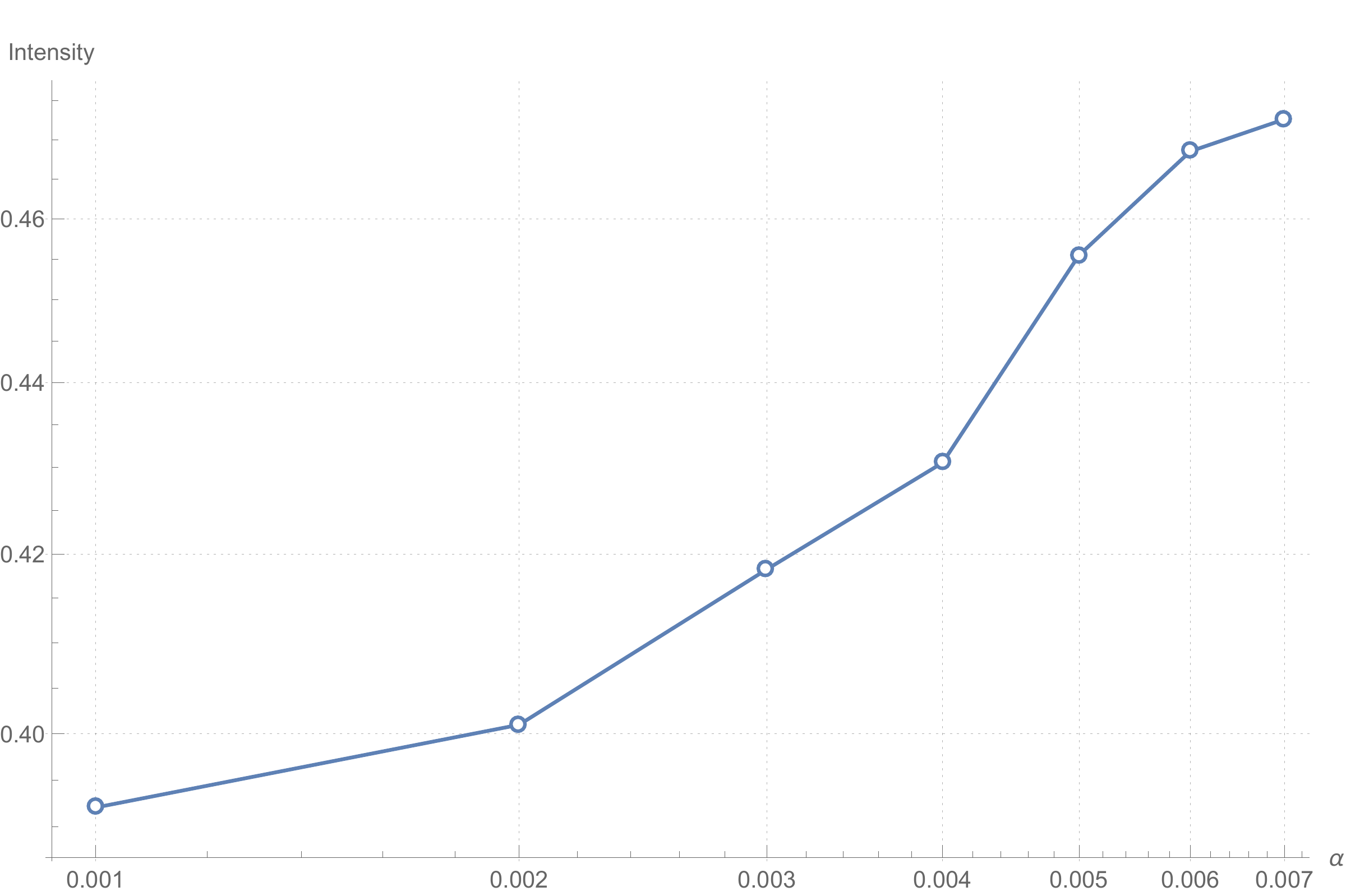}}\quad
\subfigure[]{\includegraphics[scale=0.25]{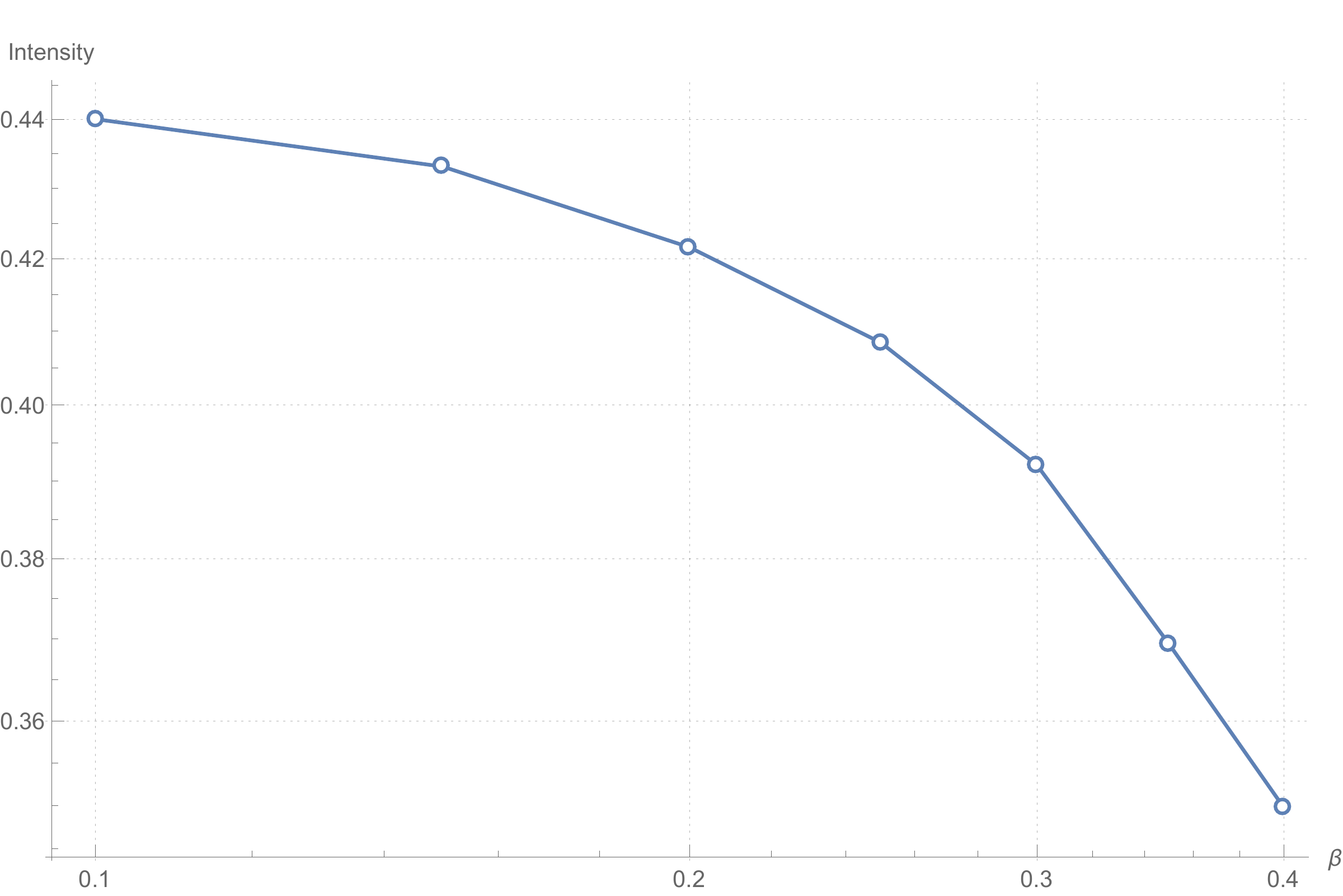}}
\caption{Maximal intensities for the Brane-World black hole for Model III $I^3_{{em}}(r)$, where the (a) represent the relationship between the maximal intensities and $\alpha$, and (b) is the relationship between the maximal intensities and $\beta$.}\label{add4fig23456}
\end{figure}

Using these three emission modes, we change the values of relevant state parameters and show the appropriately observed intensities in Figures \ref{6fig6} and \ref{7fig7}. The change in state parameters $\alpha$ and $\beta$ affects black holes¡¯ observed appearance and intensity. At fixed $\beta$ and increased value of $\alpha$, although the distribution range of observed intensity remains unchanged, the observed intensity is strengthened. In particular, the peak value of the observed intensity has a noticeable increase. In other words, the light band around the black hole becomes brighter with the rise of $\alpha$, but the position and range of the light band do not have a noticeable change.
However, when we fixed the value of $\alpha$ and changed parameter $\beta$, the radius of photon and lensing rings does not increase, and their range becomes wide with the increase of $\beta$. Moreover, the observed intensity and peak value of the photon ring increase, but the lensing ring and direct emission decrease. Hence, the optically observed appearance of the black hole will appear darker, and the observation range will become wider when the parameter $\beta$ in the system has a larger value. {In addition, by taking Model III $I^3_{{em}}(r)$ as an example, we show maximal intensities for the Brane-World black hole as the parameters $\alpha$ or $\beta$ are changed. In Figure \ref{add4fig23456}, we can see that the maximal intensities increase with the parameter $\alpha$ but decrease with $\beta$.
}

\subsection{Inclined thin disk accretion}\label{sec3333}
\begin{figure}[h]
\centering 
\includegraphics[width=0.55\textwidth]{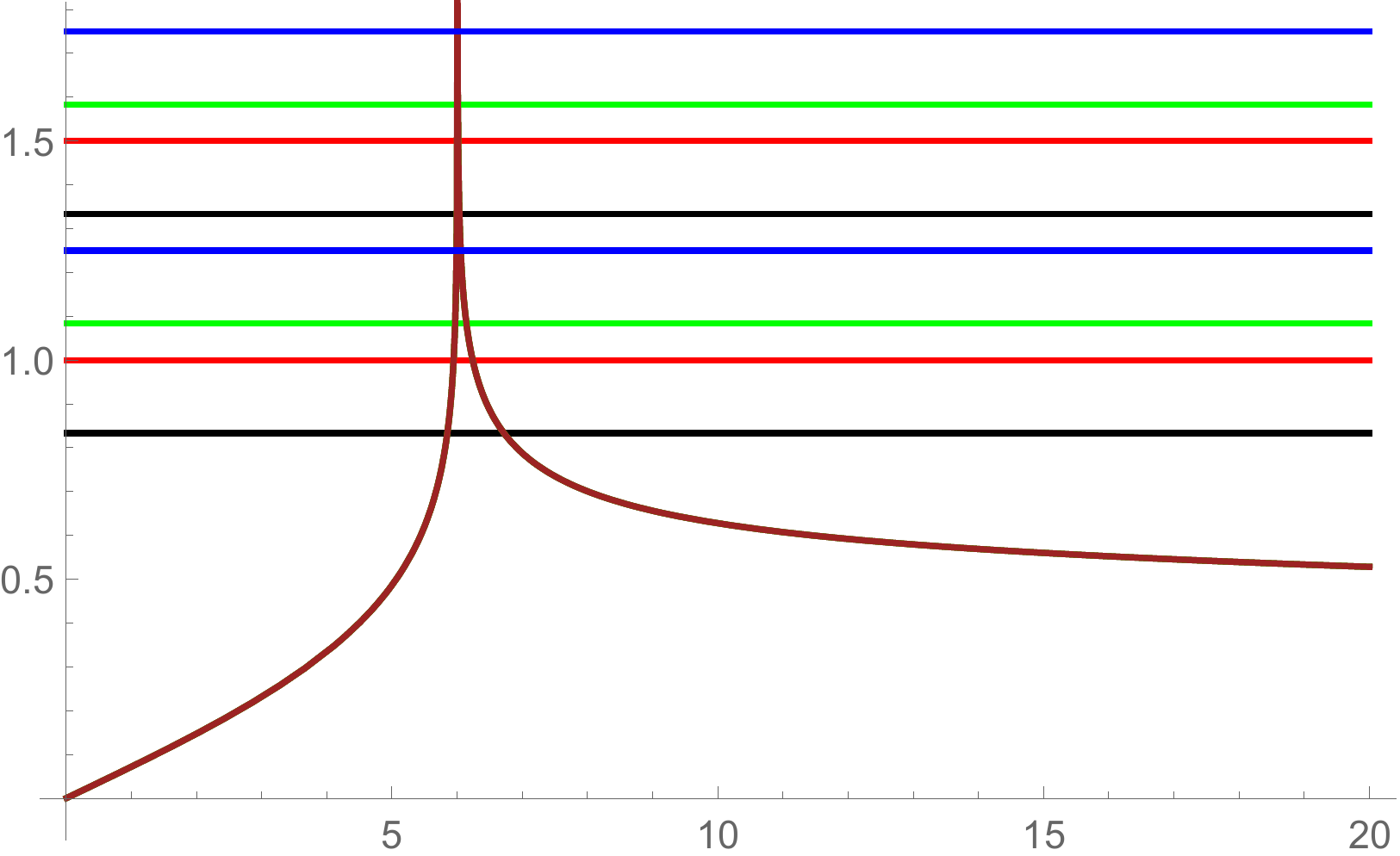}
\caption{\label{2fig222222}Total number of orbits for different values of $\theta$.}
\end{figure}

{For simplicity, we only studied the thin disk accretion located at the equatorial plane in the above section.
We can also study the case of the disk at other places.
The study of black hole shadow when the disk is located at different inclination angles is also interesting. In \cite{aa1}, the black hole images for inclination angle $\theta=30^\circ$ of static spherically symmetric space-time are captured. The polarization profile of a lensed accretion disk for inclination angle $\theta=45^\circ,75^\circ$ is also shown in \cite{aa2}. However, the detailed study of shadow with the inclusion of the lensing and photon rings remains unknown.
Given this, we will discuss the lensing and photon rings when the disk is located at different places in this subsection.
Since the regions of lensing and photon rings obtain additional brightness from the disk, the definition of those regions is also defined as the trajectories that intersect the disk $2$ and $3+$ times. Given this, one can see that the total number of orbits $n = \frac{\varphi}{2 \pi}$ for direct emission, and lensing and photon rings correspond to the regions $n < \frac{3}{4}$, $\frac{3}{4} > n > \frac{5}{4}$ and $n > \frac{5}{4}$ when the disk lies in the equatorial plane.
Since the disk is tipped only toward the South Pole, the total number of orbits $n$ and the trajectory of direct emission, lensing and photon rings for different inclination angles should be changed with different inclination angles.}
{To clearly show the change in the regions of direct emission, lensing and photon rings with different inclination angles, here we only consider a simple case where the photon is located at the plane perpendicular to both the disk and the equatorial plane.
By taking the case $\alpha=0.001,\beta=0.3$ as an example, we present the regions of direct emission, and lensing and photon rings of a Brane-World black hole when the disk is located at the position with different inclination angles.
\begin{center}
{\footnotesize{\bf Table 2.} The total number of orbits of the direct emission, and lensing and photon rings of a Brane-World black hole}.\\
\vspace{1mm}\label{t111l}
\begin{tabular}{ccccc}
\hline \hline
\footnotesize{Angle} &\footnotesize{$\theta=30^\circ$} &\footnotesize{$\theta=45^\circ$}  &\footnotesize{$\theta=60^\circ$} &\footnotesize{$\theta=90^\circ$}\\
 \hline
\footnotesize{Direct emmision } &\footnotesize{$n<\frac{11}{12}$, or $n>\frac{17}{12}$} &\footnotesize{$n<1$, or $n>\frac{3}{2}$} &\footnotesize{$n<\frac{13}{12}$, or $n>\frac{19}{12}$} &\footnotesize{$n<\frac{5}{4}$, or $n>\frac{7}{4}$}\\
\hline
\footnotesize{Lensing ring}   &\footnotesize{$\frac{11}{12}<n<\frac{17}{12}$}          &\footnotesize{$1<n<\frac{3}{2}$}         &\footnotesize{$\frac{13}{12}<n<\frac{19}{12}$}  &\footnotesize{$\frac{5}{4}<n<\frac{7}{4}$}\\
\hline
\footnotesize{photon ring}    &\footnotesize{$n>\frac{17}{12}$}                        &\footnotesize{$n>1$}                      &\footnotesize{$n>\frac{13}{12}$}                &\footnotesize{$n>\frac{5}{4}$}\\
\hline \hline
\end{tabular}
\end{center}

\begin{center}
{\footnotesize{\bf Table 3.} The numerical results of the regions of direct emission, and lensing and photon rings of a Brane-World black hole.}\\
\vspace{1mm}\label{t111ll}
\begin{tabular}{ccccc}
\hline \hline
 \footnotesize{Angle}   &\footnotesize{Direct emission}        &\footnotesize{Lensing ring}       &\footnotesize{Photon ring}                            \\
 \hline
{\footnotesize{$\theta=30^\circ$}}    &{\footnotesize{$b_c<5.8456$, or $b_c>6.7848$}}  &{\footnotesize{$5.8456<b_c<5.9971$ and $6.0374<b_c<6.7848$}} &{\footnotesize{$5.9971<b_c<6.0374$}}\\
\hline
{\footnotesize{$\theta=45^\circ$}}    &{\footnotesize{$b_c<5.8796$, or $b_c>6.5851$}}  &{\footnotesize{$5.8796<b_c<5.9989$ and $6.0303<b_c<6.5851$}} &{\footnotesize{$5.9989<b_c<6.0303$}}\\
\hline
{\footnotesize{$\theta=60^\circ$}}    &{\footnotesize{$b_c<5.9066$, or $b_c>6.4411$}}  &{\footnotesize{$5.9066<b_c<6.0004$ and $6.0248<b_c<6.4411$}} &{\footnotesize{$5.9971<b_c<6.0374$}}\\
\hline
{\footnotesize{$\theta=90^\circ$}}    &{\footnotesize{$b_c<5.9447$, or $b_c>6.2570$}}  &{\footnotesize{$5.9447<b_c<6.0024$ and $6.0172<b_c<6.2570$}} &{\footnotesize{$6.0024<b_c<6.0172$}}\\
\hline \hline
\end{tabular}
\end{center}

From Tables 2 and 3, it can be seen that the total number of orbits $n$ always increases for direct emission and rings and that the regions become smaller as the inclination angle increases. These features can be seen in Figure \ref{2fig222222}.
}

{In Figure \ref{2fig222222}, the black, red, green and blue lines correspond to the cases of $\theta=30^\circ$, $\theta=45^\circ$, $\theta=60^\circ$, and $\theta=90^\circ$ for the number of orbits $n$, respectively. These four lines move upward gradually, leading to narrower regions of rings. In the following figure \ref{3fig2345}, we continue to show the trajectories of light ray of the regions of direct emissions and rings, where the disk is located at different places.}

\begin{figure}[h]
\centering
\subfigure[$\theta=30^\circ$]{\includegraphics[scale=0.17]{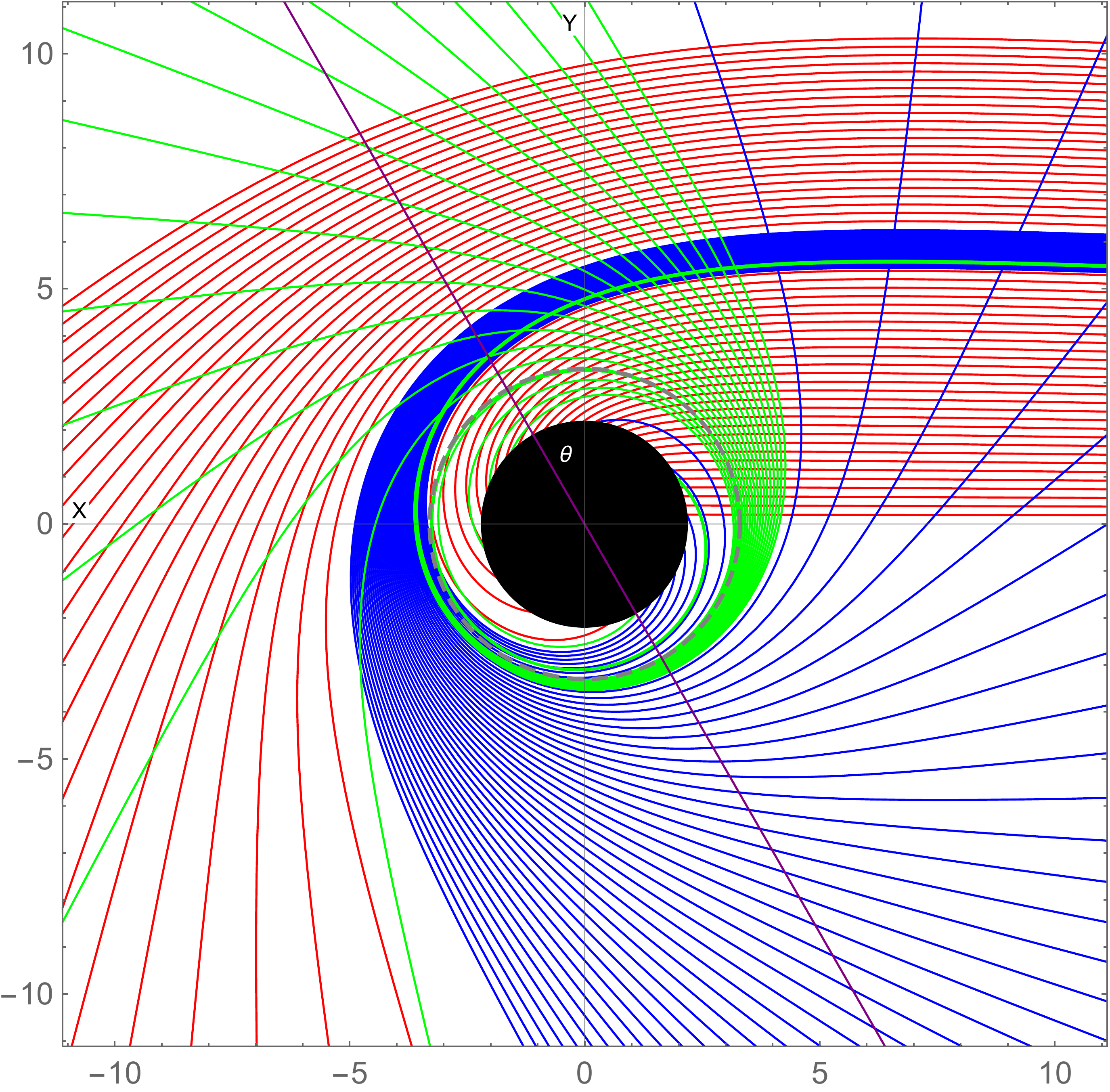}}\quad
\subfigure[$\theta=45^\circ$]{\includegraphics[scale=0.1875]{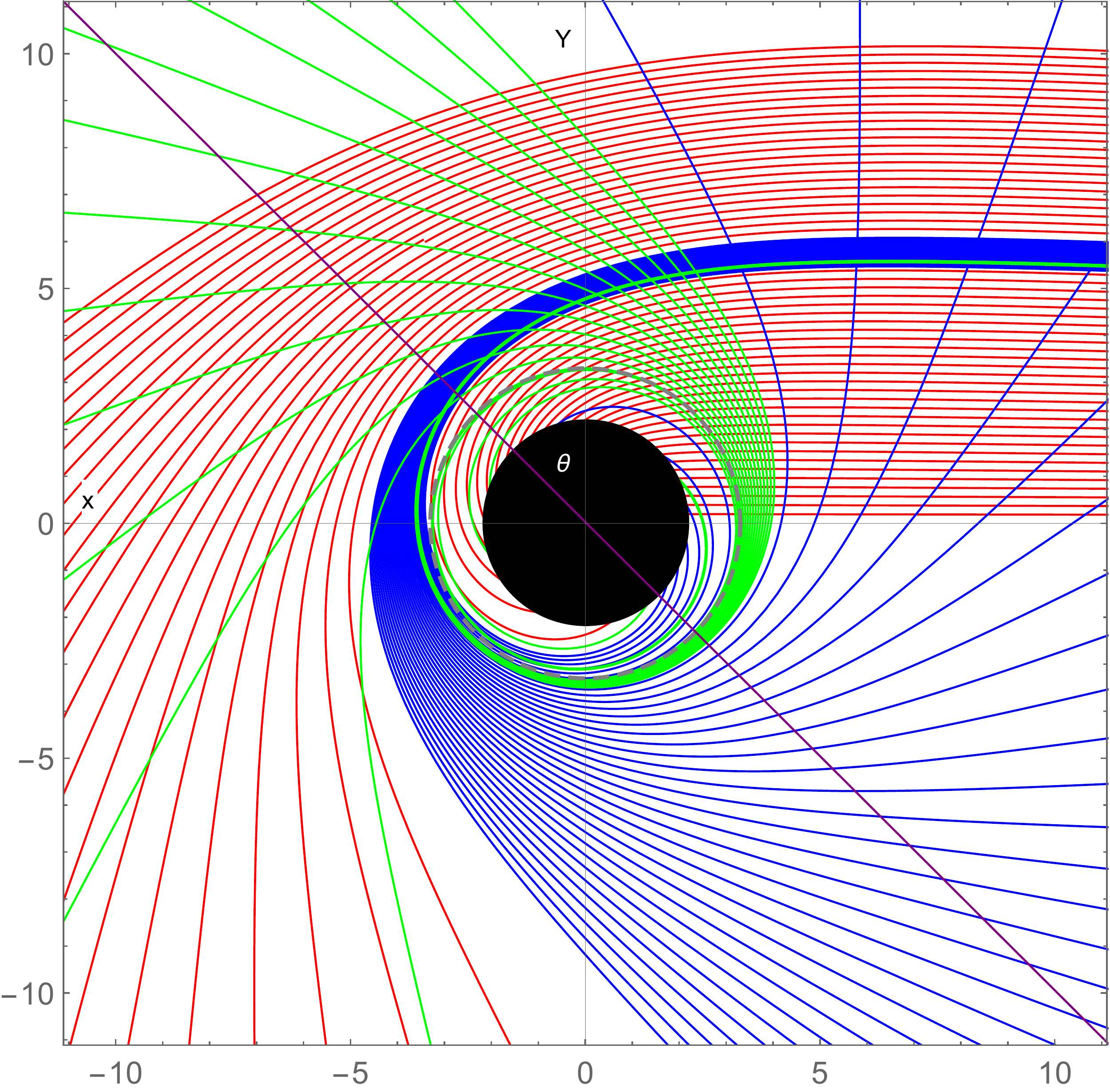}}\\
\subfigure[$\theta=60^\circ$]{\includegraphics[scale=0.17]{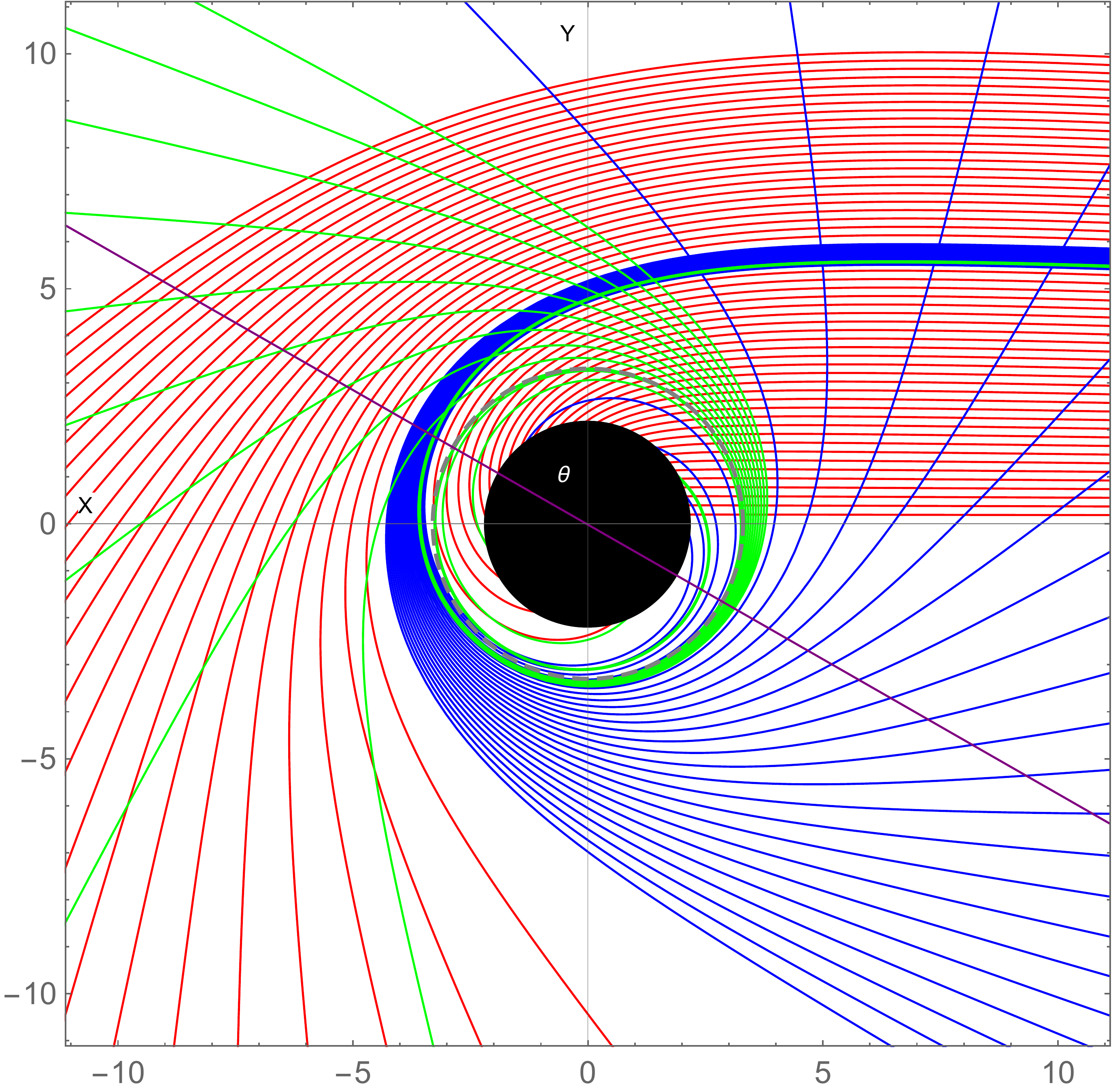}} \quad
\subfigure[$\theta=90^\circ$]{\includegraphics[scale=0.163]{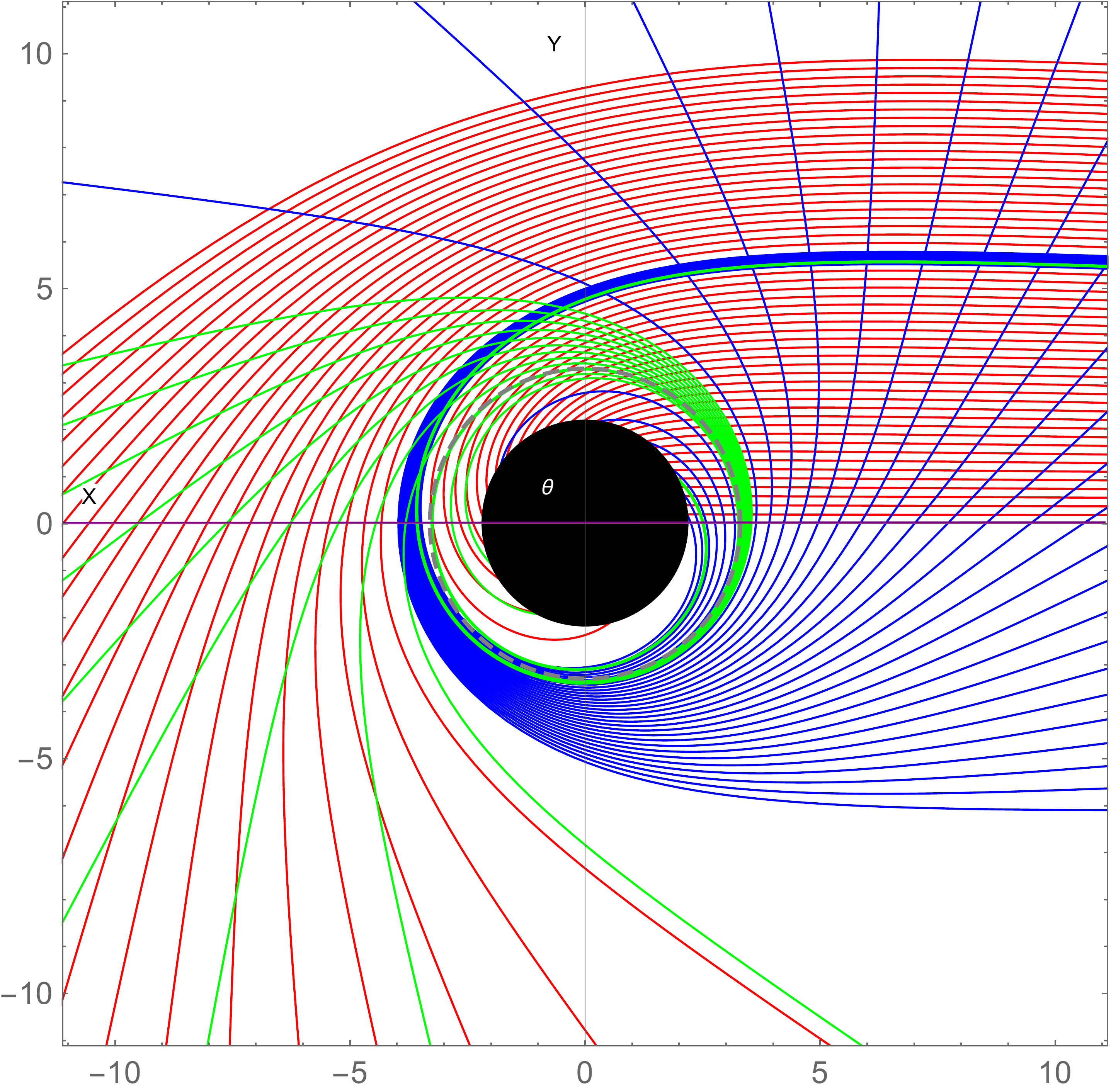}}
 \caption{Direct emission, lensing ring and photon ring for the Brane-World black hole, when the disk is located at different places, where $\alpha = 0.001$ and $\beta=0.3$ .}\label{3fig2345}
\end{figure}
{Similarly, the direct, lensing ring, and photon ring trajectories correspond to the red, blue, and green lines, for which the related spacing of $b_c$ is 1/5, 1/50, and 1/500 for the cases $\theta=30^\circ, 45^\circ, and 60^\circ$, respectively. For $\theta=90^\circ$, the related spacings are fixed to 1/5, 1/100, and 1/1000. The purple line represents the location of the disk, and the angle between the disk and the $Y$ axis is defined as the inclination angle $\theta$ deviated from the equatorial plane. This angle $\theta$ is also the inclination between the line of sight to the observer and the disk axis. From Figure \ref{3fig2345}, one can see that the regions of blue and green lines become narrower with the increase of $\theta$.

Although the regions of lensing and photon rings are small, we can see that the photons in the rings intersect the disk $2$ or $3+$ times. Hence, they also obtain additional brightness from the disk, implying that the brightness enhancement may reach a factor of $2+$. This feature is similar to the case that the disk is located at the equatorial plane. Therefore, the main conclusion that the contribution of the rings is small to the total observed intensity remains unaltered in the case of an inclined disk in this simple.
For the cases that the photons are at the other planes, we can infer that the regions of rings should also be changed, but the conclusion that the lensing ring only makes a small contribution and the photon ring makes a negligible contribution still holds true\footnote{{For the thin disk accretion, the observed shadows of a black hole for an inclined disk will be presented in our future work since we only made a simple discussion here.}}.}

\section{Shadows and photon spheres with spherical accretions}\label{Sec4}
\label{sec:5}
Next, we will continue to investigate the properties of shadows and photon spheres of a Brane-World black hole by considering spherical accretions. Primarily, when the spherical accretion is assumed to be optically thin, the static spherical accretion and infalling spherical accretion will be carefully addressed for various values of parameters $\alpha$ and $\beta$.
\subsection{The static spherical accretion }\label{sec41}
\begin{figure}[H]
\centering
\leftline {\hspace{4mm} \includegraphics[scale=0.37]{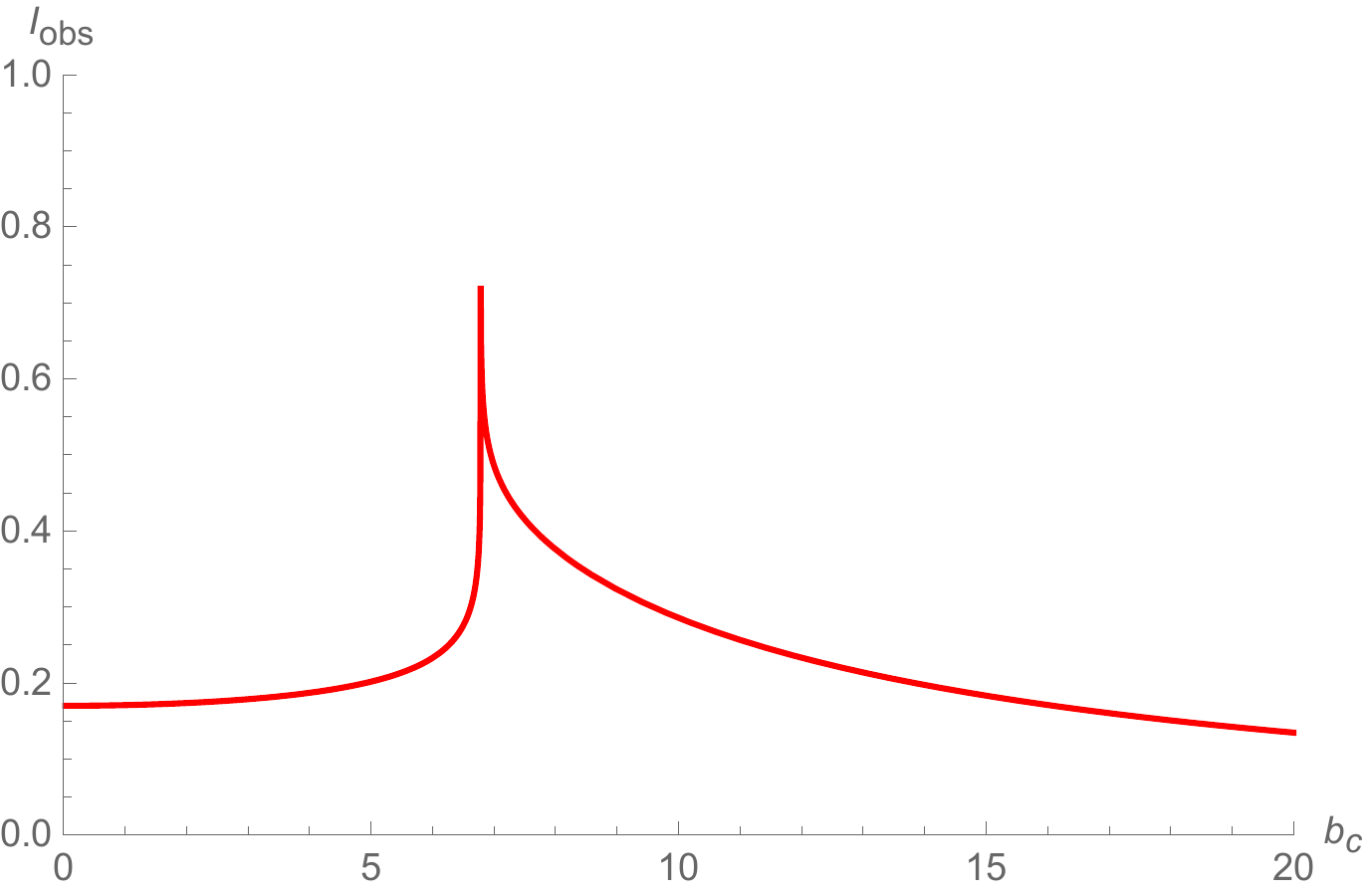} \hspace{0mm}
\includegraphics[scale=0.37]{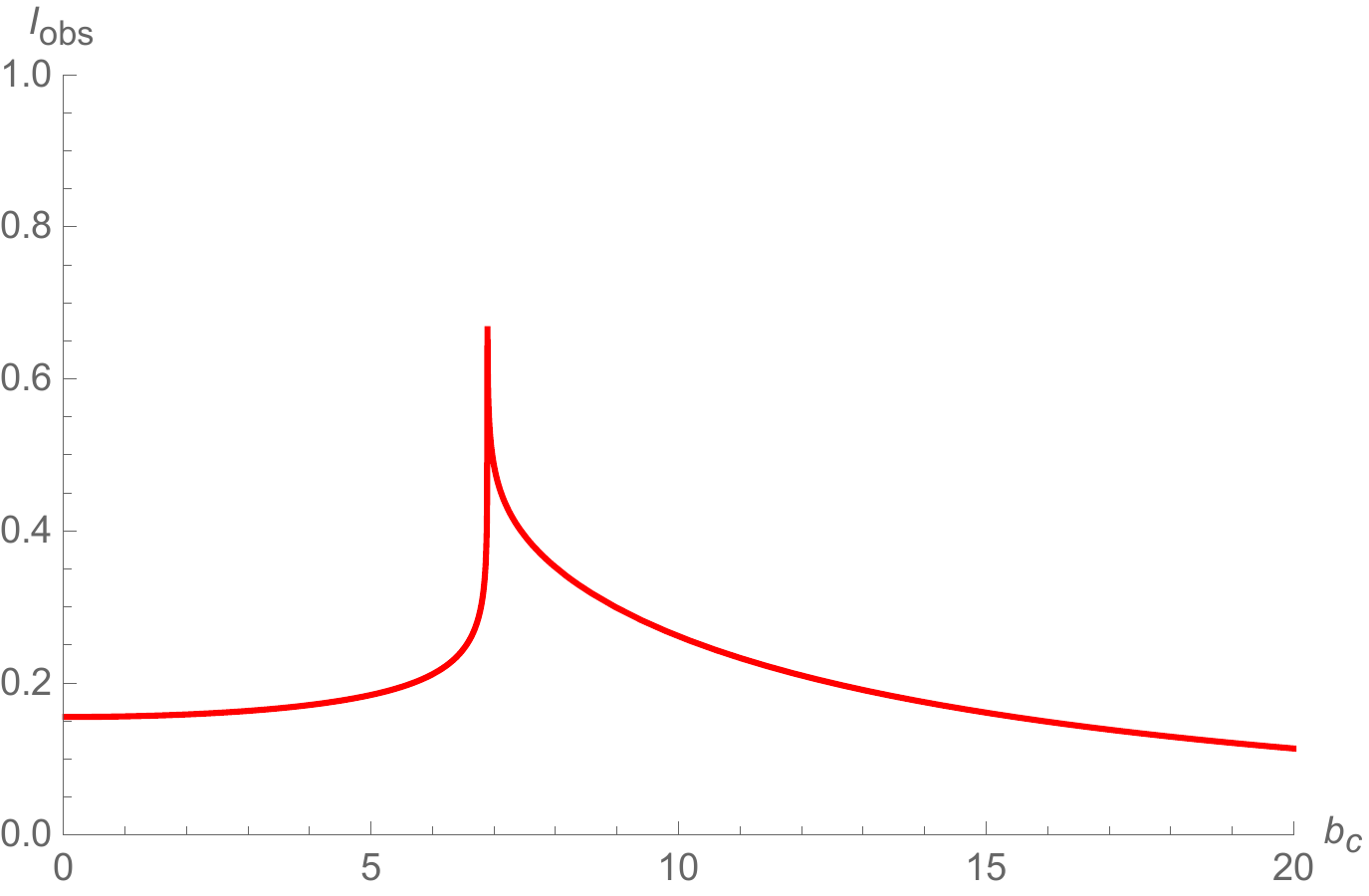} \hspace{0mm}
\includegraphics[scale=0.37]{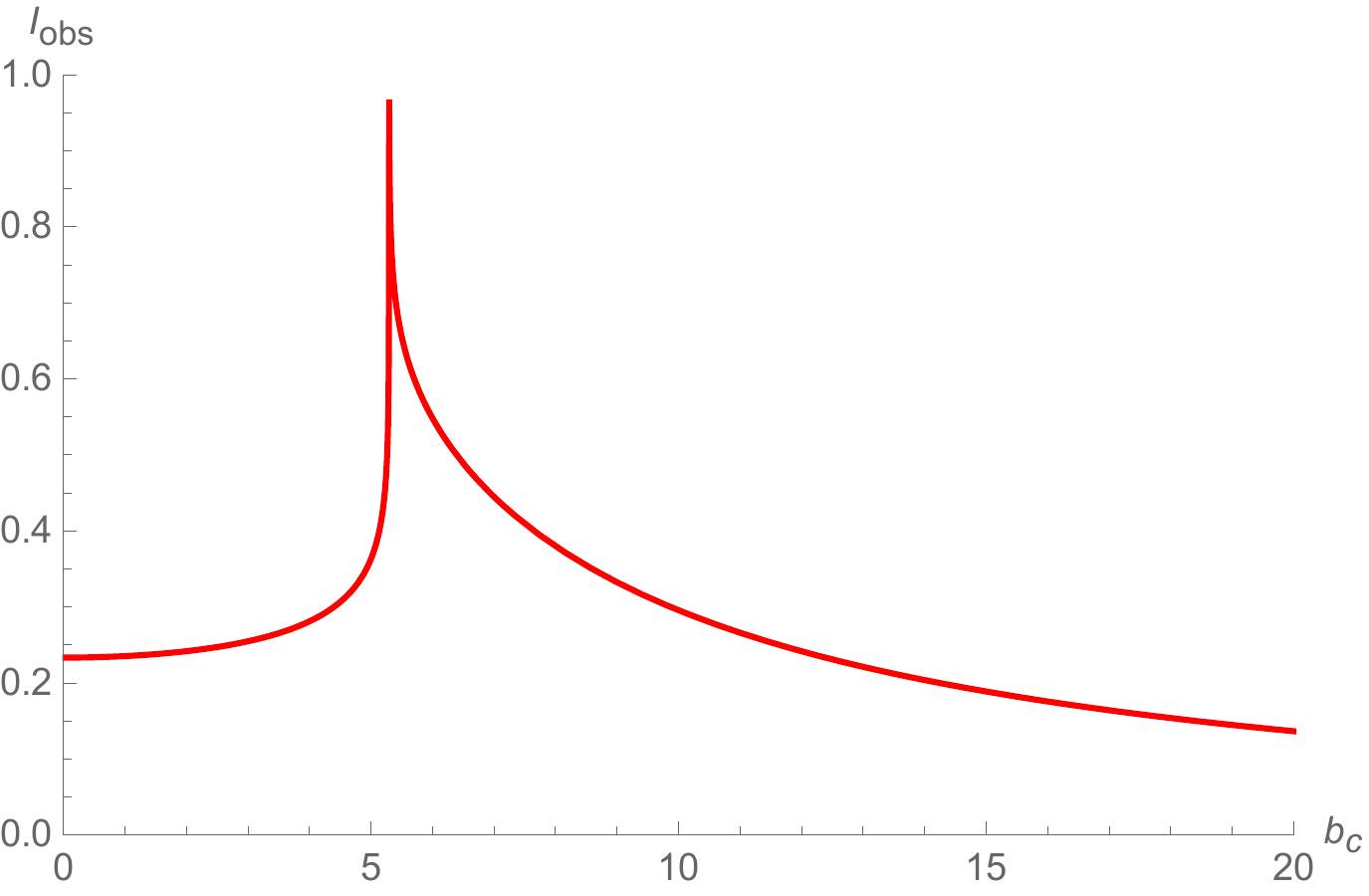}}
\subfigure[$\alpha = 0.001, \beta =0.4$]{
\includegraphics[scale=0.35]{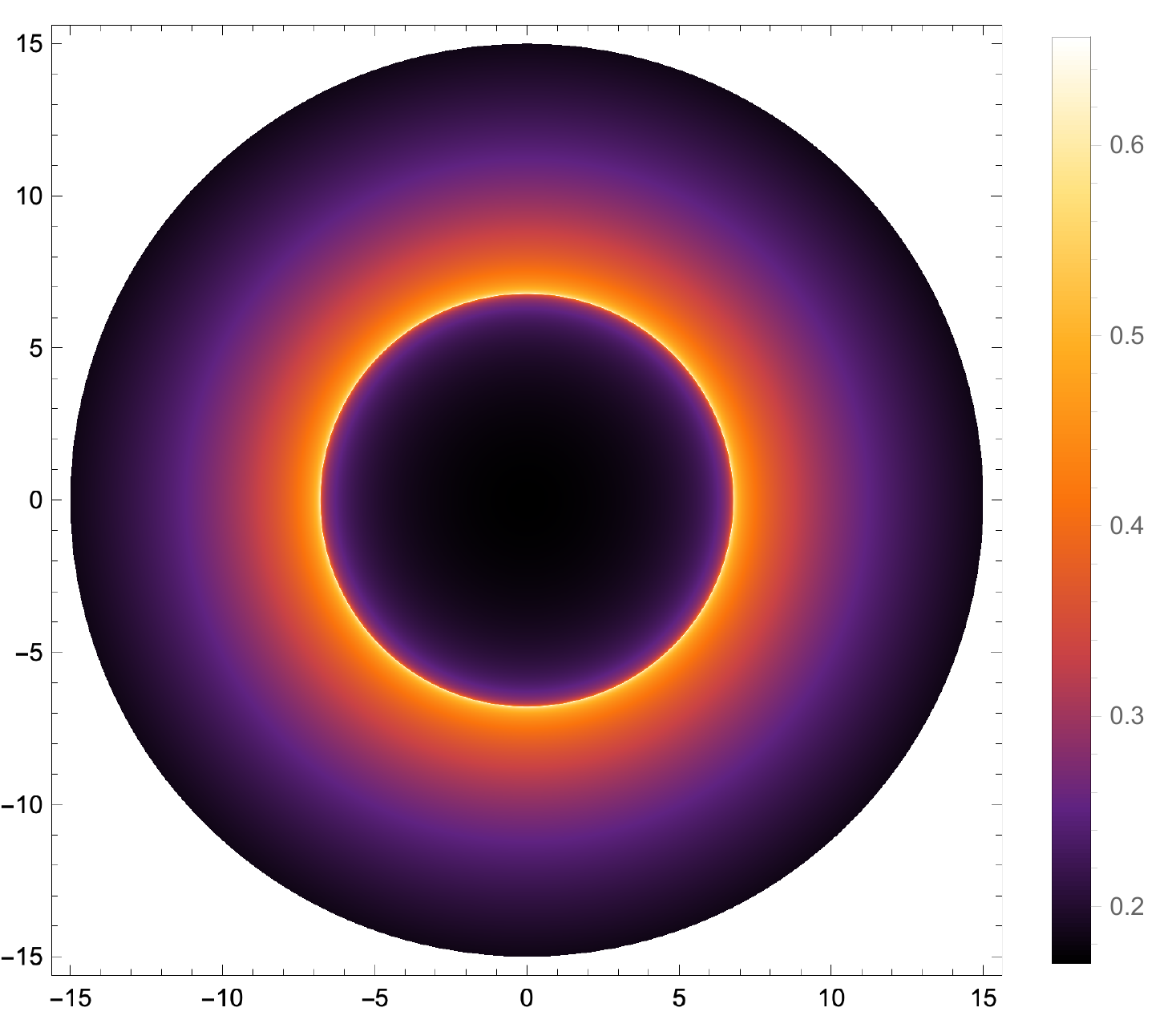}}
\subfigure[$\alpha = 0.004, \beta =0.4$]{
\includegraphics[scale=0.35]{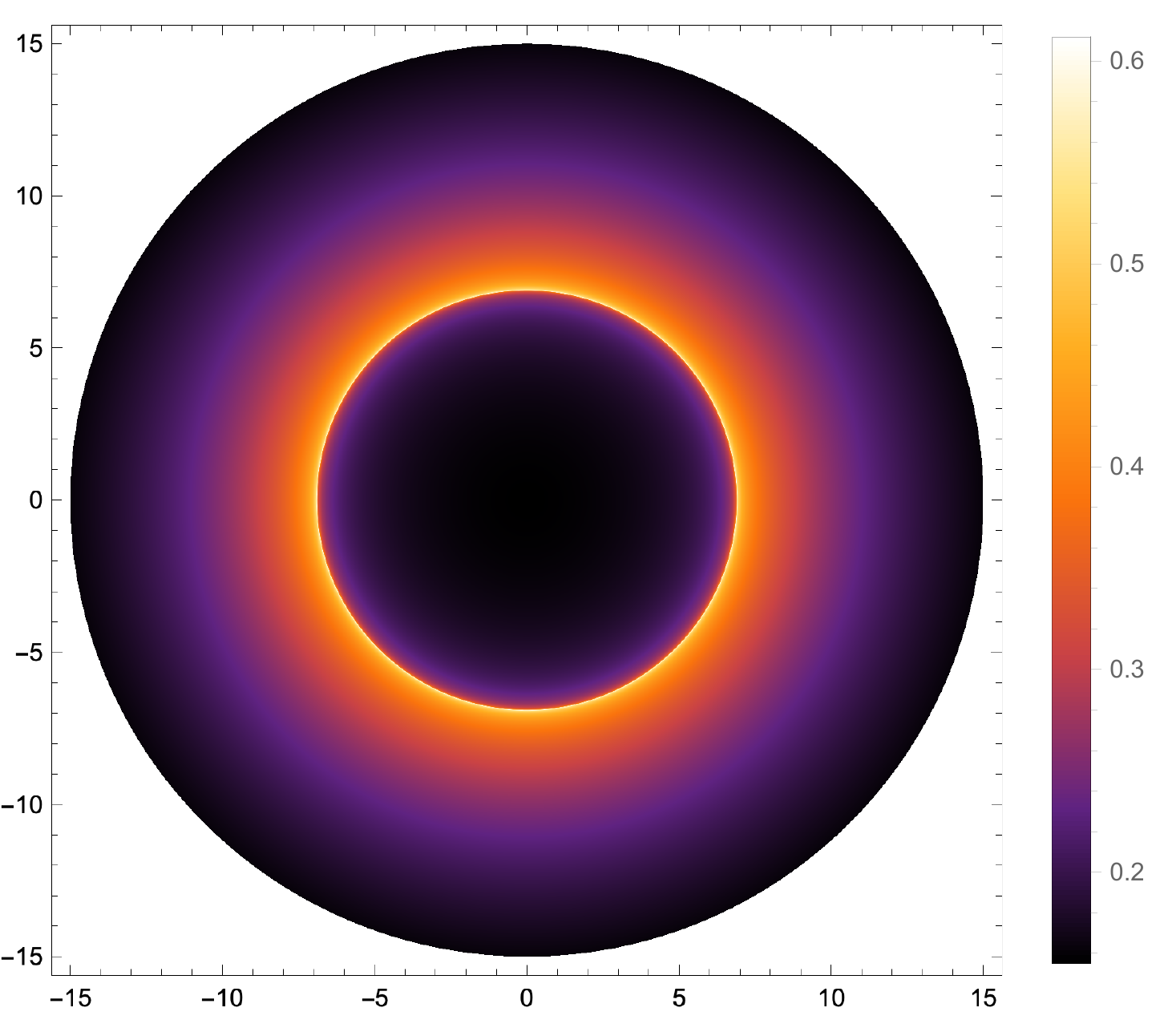}}
\subfigure[$\alpha = 0.004, \beta =0.1$]{
\includegraphics[scale = 0.35]{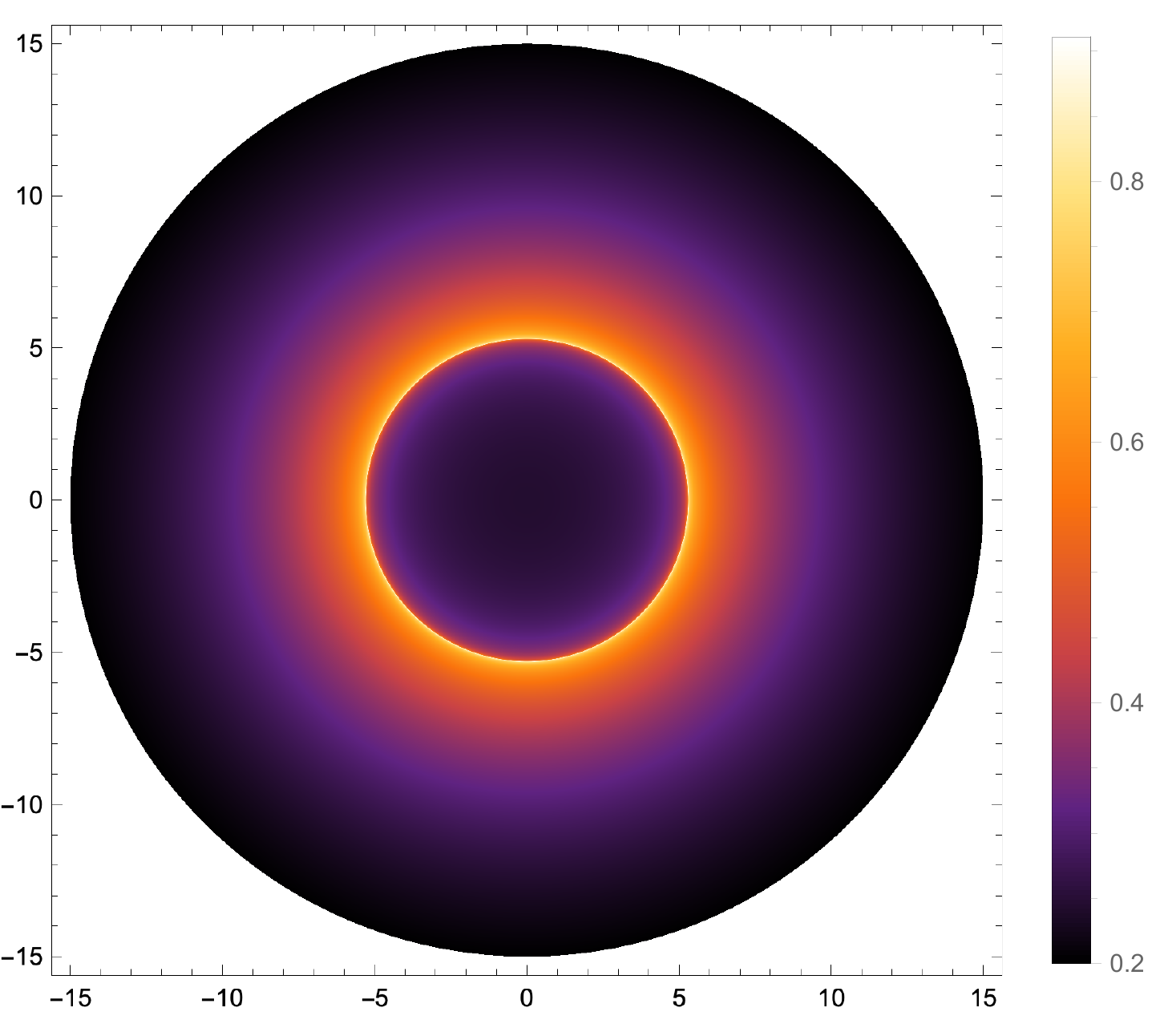}}
\caption{Specific intensity $I_{obs}(\nu_o)$ seen by a distant observer for a static spherical accretion. }\label{figjt}
\end{figure}

\begin{figure}[h]
\centering
\subfigure[]{\includegraphics[scale=0.25]{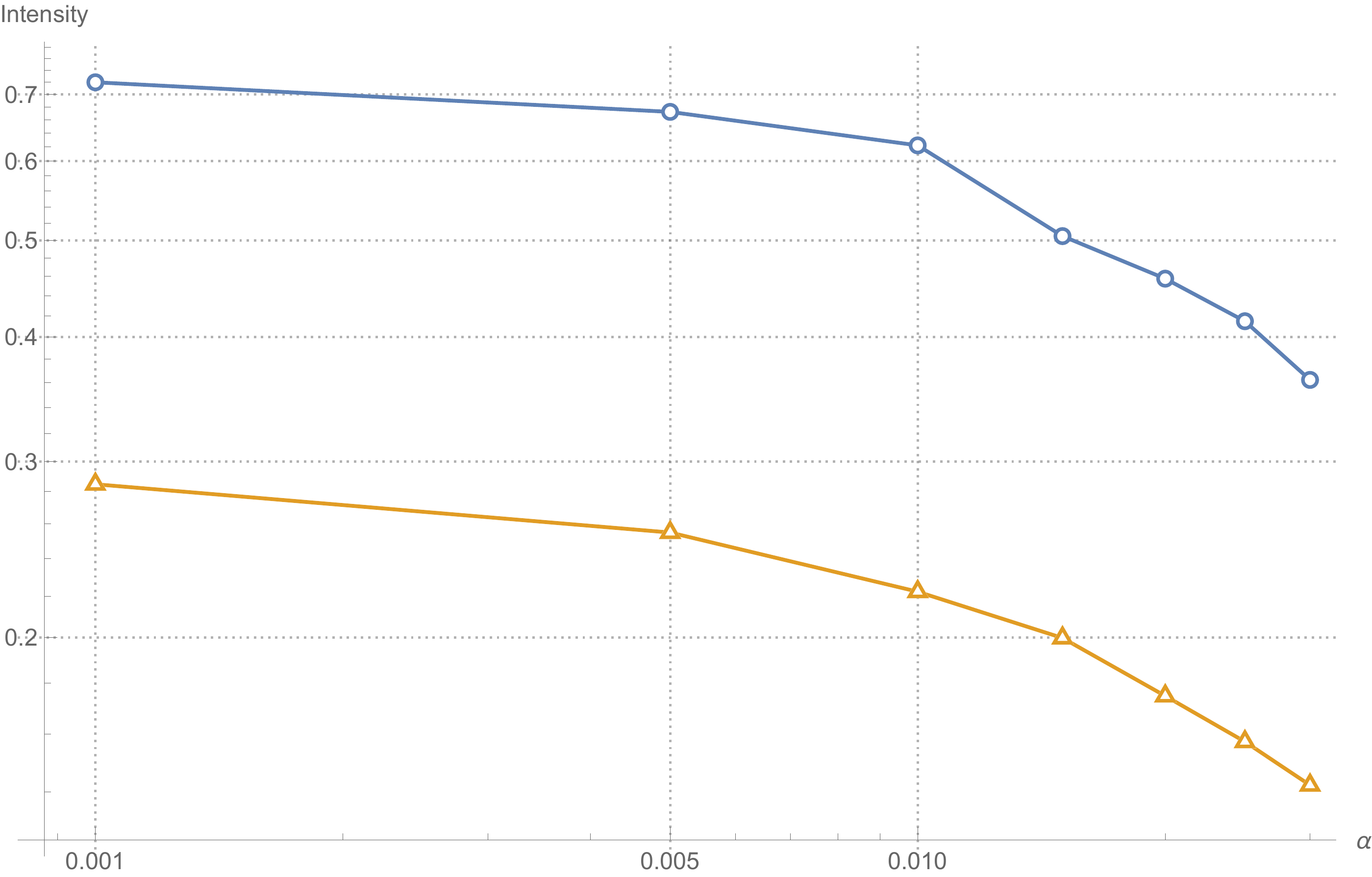}}\quad
\subfigure[]{\includegraphics[scale=0.25]{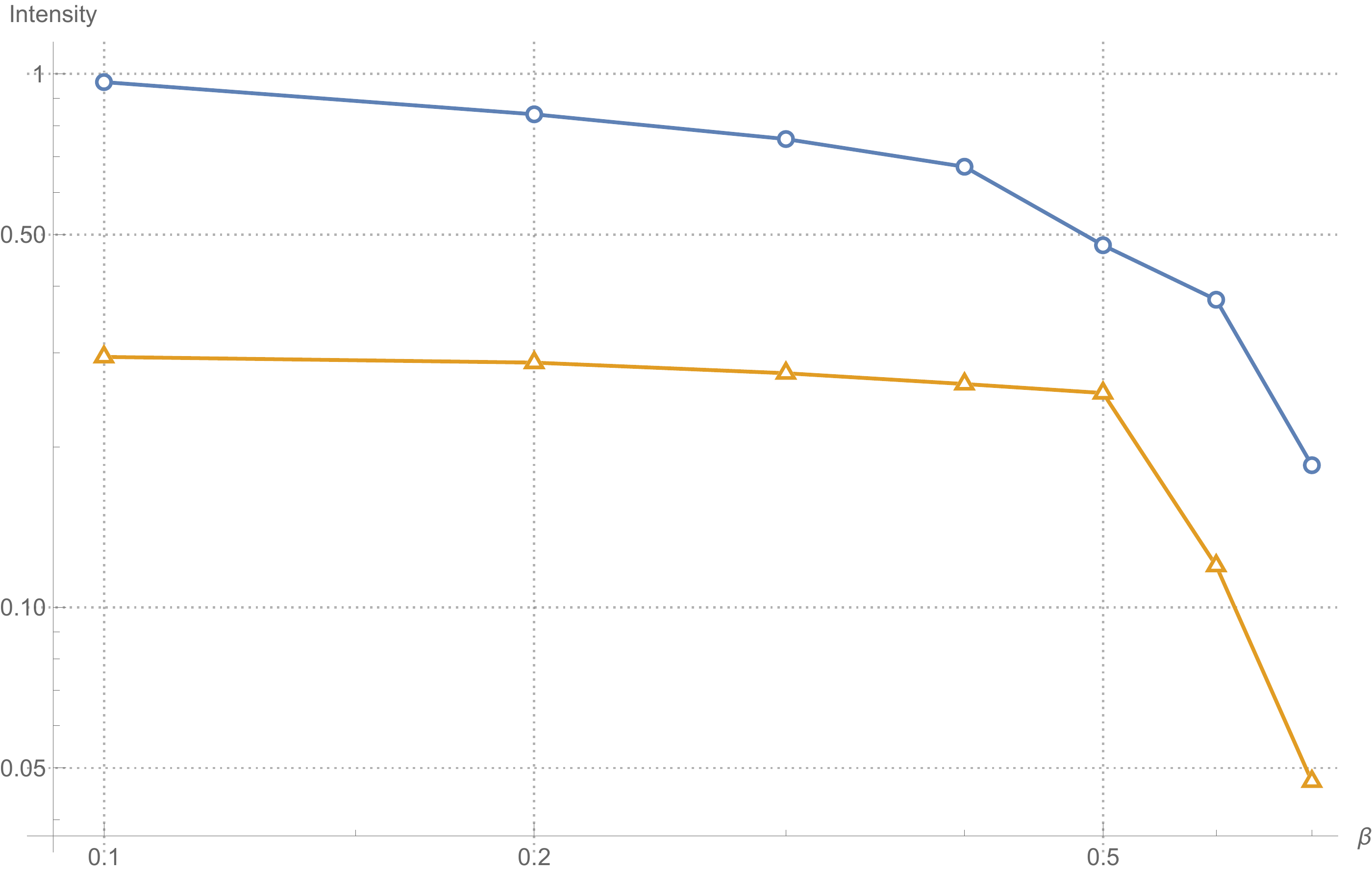}}
\caption{Maximal intensities and intensities at $b_c=10$ for the Brane-World black hole, where (a) and (b) represent the relationship of the intensities vs. $\alpha$ and $\beta$, respectively. The blue line corresponds to maximal intensities, and orange lines correspond to intensities at $b_c=10$, respectively.}\label{ad3fig2345}
\end{figure}

This subsection will focus on shadows and photon spheres when a static spherical accretion surrounds the Brane-World black hole. Based on \cite{ZXX1,ZXX2}, the observed specific intensity is
\begin{align}\label{eqq20}
    I(\nu_o) = \int_l g_{rs}^3 \cdot j(\nu_e) \cdot d\ell_{prop},
\end{align}
where $l$ is the trajectory of emitted photons, $\nu_o$ is the observed photon frequency, and $\nu_e$ is assumed to be the emitted frequency when the emitter radiates photons monochromatically. And,
\begin{align}
     &g_{rs} = A(r)^{1/2},\label{eqq21}\\
     &j(\nu_e)\propto \frac{\delta(\nu_e - \nu_m)}{r^2},\label{eqqq22}\\
     &d\ell_{prop} = \sqrt{ \frac{1}{A(r)} +r^2\left(\frac{d\phi}{dr}\right)^2}dr.\label{eqqq23}
\end{align}
Here, $g_{rs}$ and $d\ell_{prop}$ are the redshift factor and the infinitesimal proper length, respectively. $j(\nu_e)$ represents the emissivity per unit volume in the rest of the frame, which is assumed to be proportional to $1/r^2$. In the current context, by using the motion equation of photons and Eqs.(\ref{eqq21}), (\ref{eqqq22}), (\ref{eqqq23}), the observed intensity for a static observer can be finally expressed as:
\begin{align}\label{eqq22}
    I_{obs}(\nu_o) = \int_l \frac{A(r)^{3/2}}{r^2} \sqrt{\frac{1}{A(r)} + r^2 \left(\frac{d\phi}{dr} \right)}dr.
\end{align}
Considering the relationship between $r$ and $b_c$, we can plot the observed luminosity with respect to the impact parameter $b$ for the static accretion in Figure \ref{figjt}, clearly showing shadows and photon spheres of the Brane-World black hole.
From Figure \ref{figjt}, we can see that the observed luminosity increases with the impact parameter $b_c$, peaks at the photon sphere $b_p$, and finally, decreases to a certain value. Since the photons surround the black hole many times at the photon sphere, the maximum value of the observed luminosity naturally appears at $b_p$. Meanwhile, the intensities $I_{obs}(\nu_o)$, seen by a distant observer, decreased with the X-clod dark matter $\beta$ and cosmological parameter $\alpha$. The corresponding shadows and photon spheres cast by this black hole in the $(x, y)$ plane are shown in the second row. It should be noted that there is always a small value of $I_{obs}(\nu_o)$ in the inner region of the shadow. This is caused by the existence of a radiation field. {In addition, we further plot Figure \ref{ad3fig2345} to intuitively show the influence of $\alpha$ and $\beta$ on the observed intensity using several numerical data points.}

{The X-clod dark matter parameter $\beta=0.3$ shows that the maximal intensity and intensity at $b_c=10$ decrease with the cosmological parameter $\alpha$. When the cosmological parameter $\alpha=0.001$, the intensities decrease with the X-clod dark matter parameter $\beta$. It is true that the X-clod dark matter $\beta$ and cosmological constant $\alpha$ significantly affect shadows and the photon sphere.}

\subsection{The infalling spherical accretion } \label{sec42}
\begin{figure}[!h]
\centering
\leftline { \hspace{1.5mm}
\includegraphics[scale=0.37]{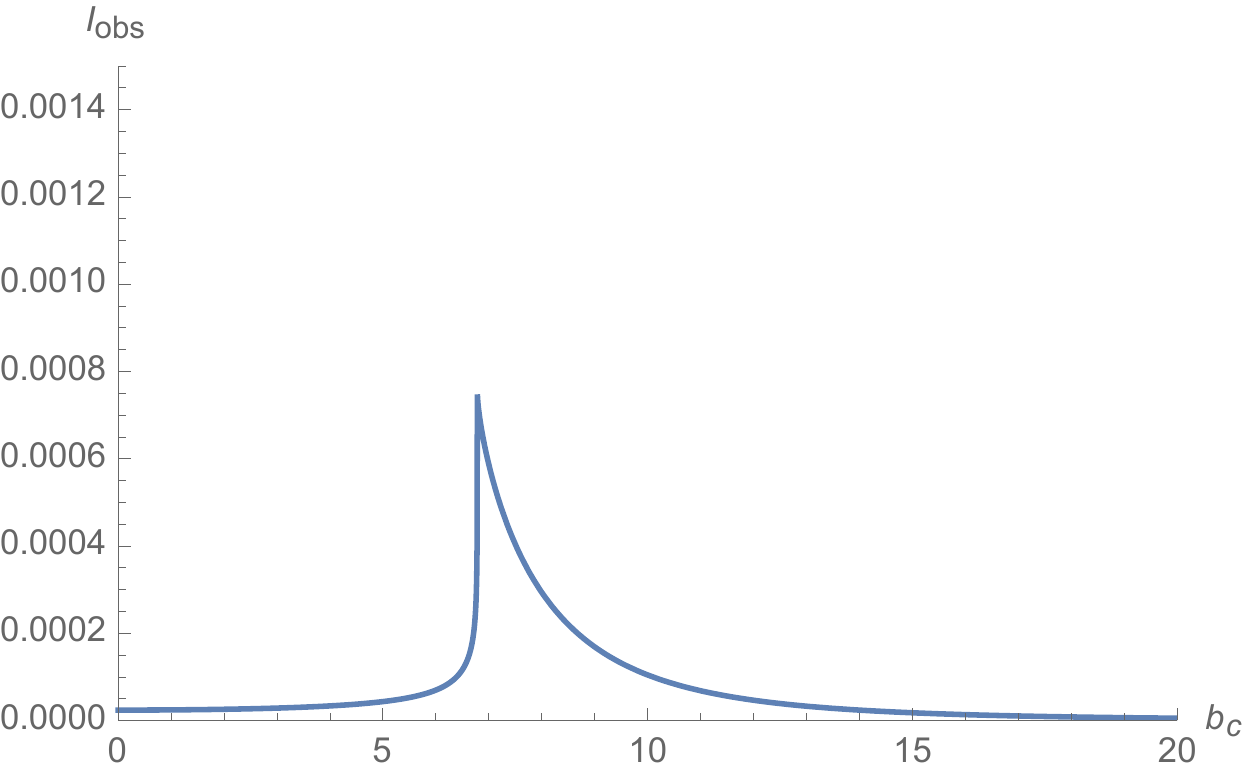} \hspace{6.7mm}
\includegraphics[scale=0.37]{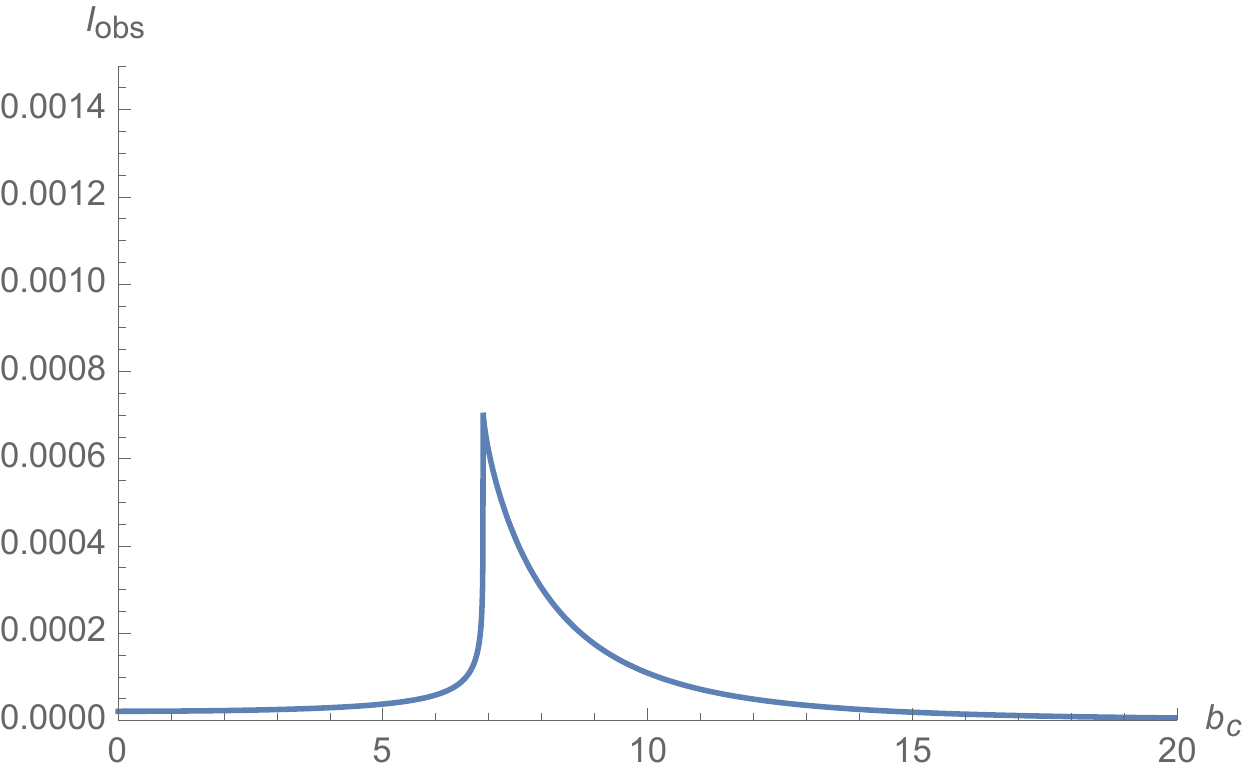} \hspace{6.9mm}
\includegraphics[scale=0.37]{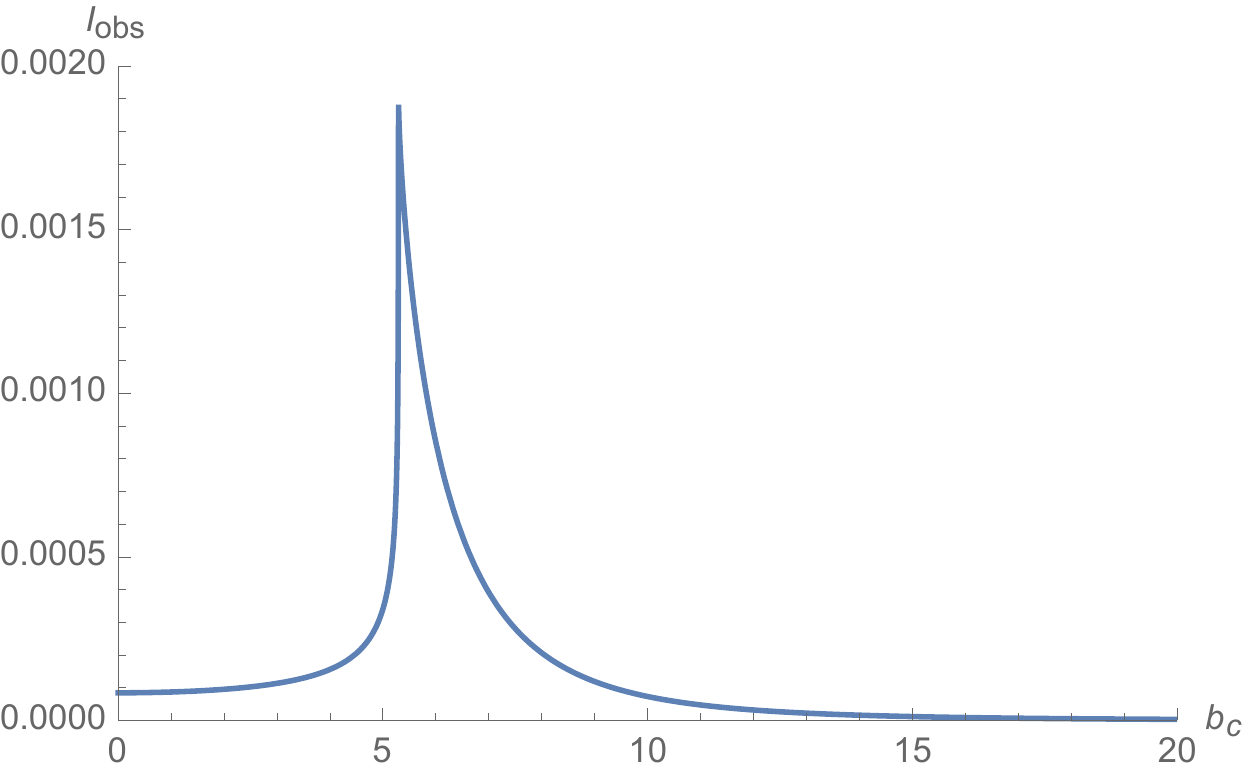}}
\leftline { \hspace{0mm}
\subfigure[$\alpha = 0.001, \beta =0.4$]{
\includegraphics[scale=0.35]{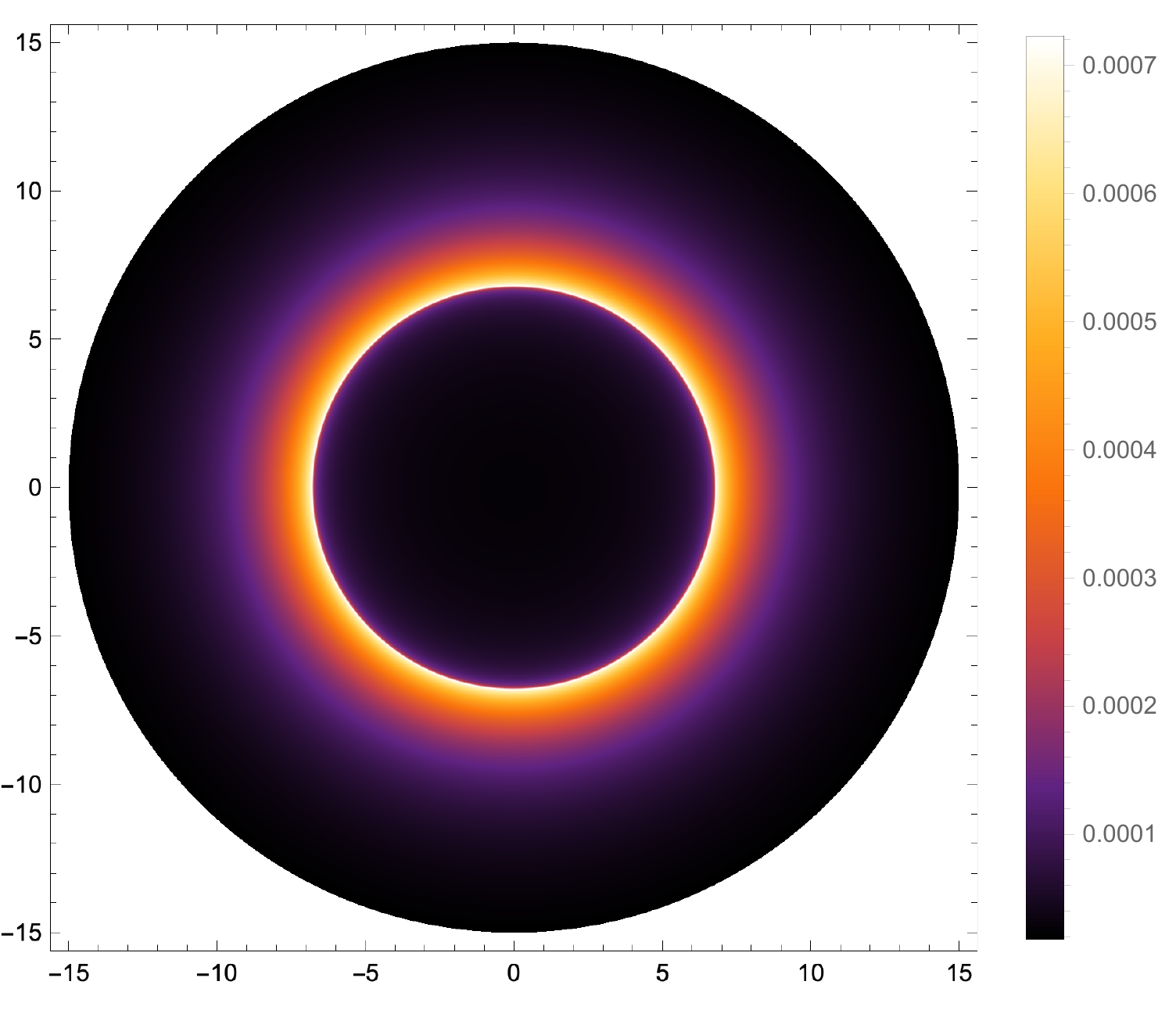}}
\subfigure[$\alpha = = 0.004, \beta = 0.4$]{
\includegraphics[scale=0.35]{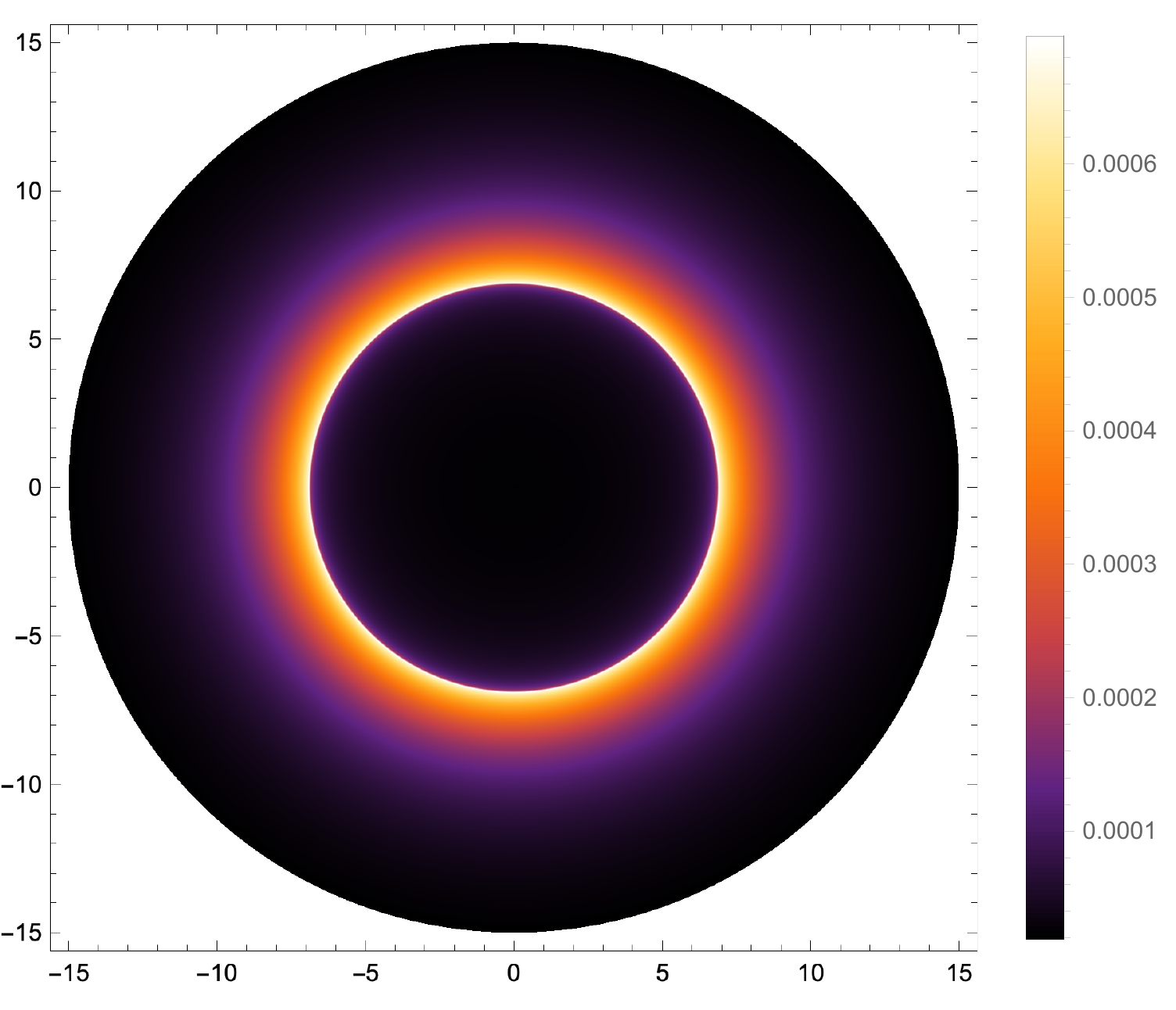}}
\subfigure[$\alpha = 0.004, \beta = 0.1$]{
\includegraphics[scale=0.35]{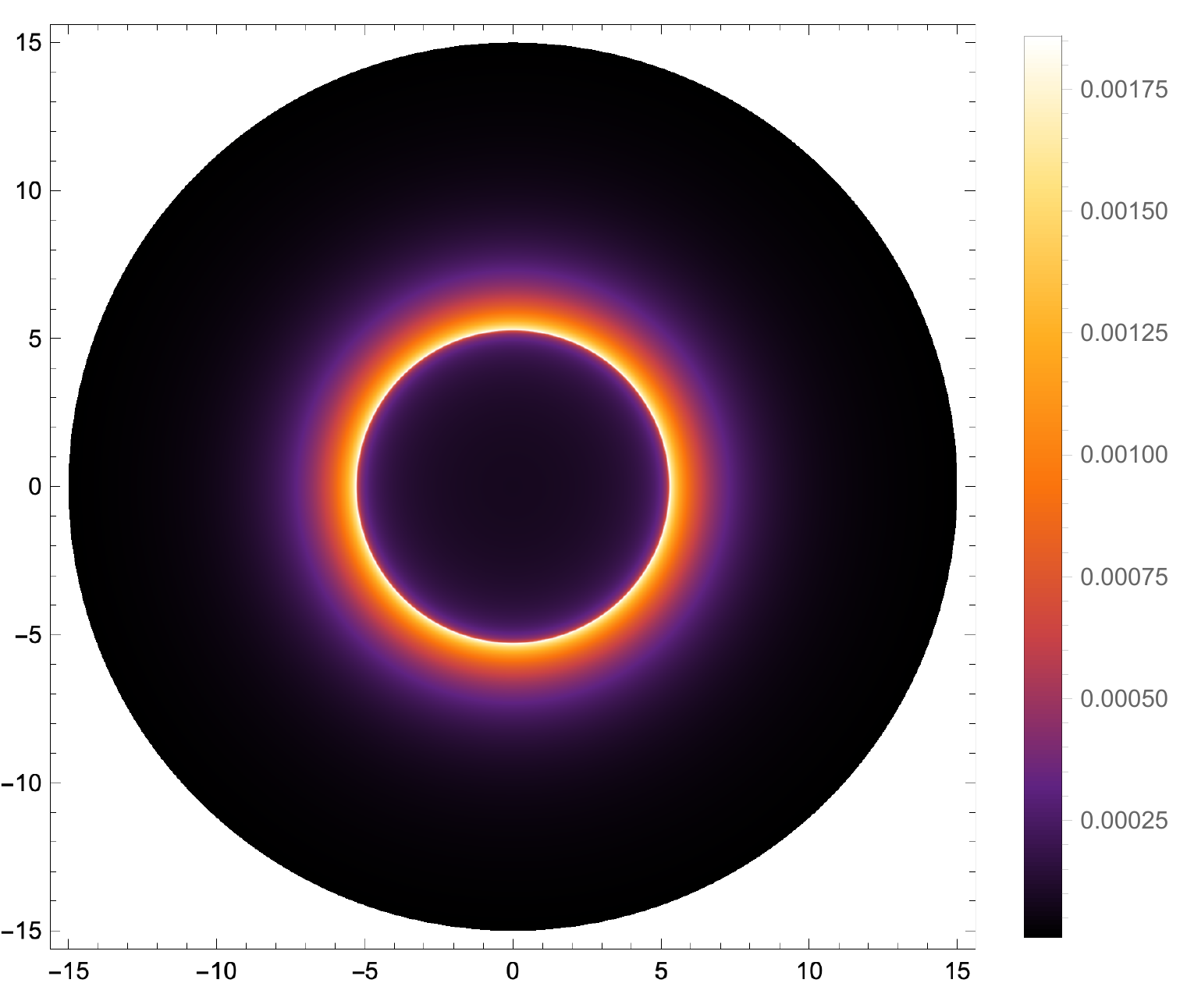}}}
\caption{Specific intensity $I_{obs}(\nu_o)$ seen by a distant observer for an infalling accretion. }\label{figdt}
\end{figure}

In our real universe, the accretion surrounding the black holes is always dynamic. Hence, the study of shadows and photon spheres in the context of infalling spherical accretion is interesting. Similarly, the observed intensity for a distant observer can be expressed as Eq.(\ref{eqq20}) for the infalling spherical accretion. However, as described in \cite{ZXX1,ZXX2}, the concrete form of the redshift factor, the proper distance, and the emissivity per unit volume should be reconsidered. In the current case, they can be written as:
\begin{align}\label{eqq23}
    &g_{rs} = \frac{\kappa_\alpha \cdot v_{obs}^\alpha}{\kappa_\beta \cdot v_{emi}^\beta},\\
    &d \ell_{prop} = \frac{1}{g_{rs}} \frac{\kappa_t}{|\kappa_r|} dr,\\
    &j(\nu_e)\propto \frac{\delta(\nu_e - \nu_m)}{r^5},
\end{align}
where the four velocities of the photon $\kappa^\mu \equiv \dot{x}_\mu$ can been obtained in Eqs.(1), (2), and (3). $v_{obs}^\mu$ $\equiv$ $(1,0,0,0)$ is defined as the four velocities for a static observer. The four velocities of the infalling accretion $v_{emi}^\mu$ read
\begin{align}\label{eqq24}
    v_{emi}^t =\frac{1}{A(r)}, \quad v_{emi}^r=-\sqrt{1-A(r)}, \quad v_{emi}^\theta=0, \quad v_{emi}^\phi=0.
\end{align}
In this consideration, using the above equations, the observed intensity for a static observer of the Brane-World black hole in the background of the infalling spherical accretion can be finally expressed as:
\begin{align}\label{eqq26}
    I_{obs}(\nu_o) \propto \int_l \frac{g^3}{r^5} \frac{\kappa_t}{|\kappa_r|} dr,
\end{align}
Similarly, we plotted shadows and photon spheres of the Brane-World black hole cast by the infalling accretion in Figure \ref{figdt}.

\begin{figure}[h]
\centering
\subfigure[]{\includegraphics[scale=0.17]{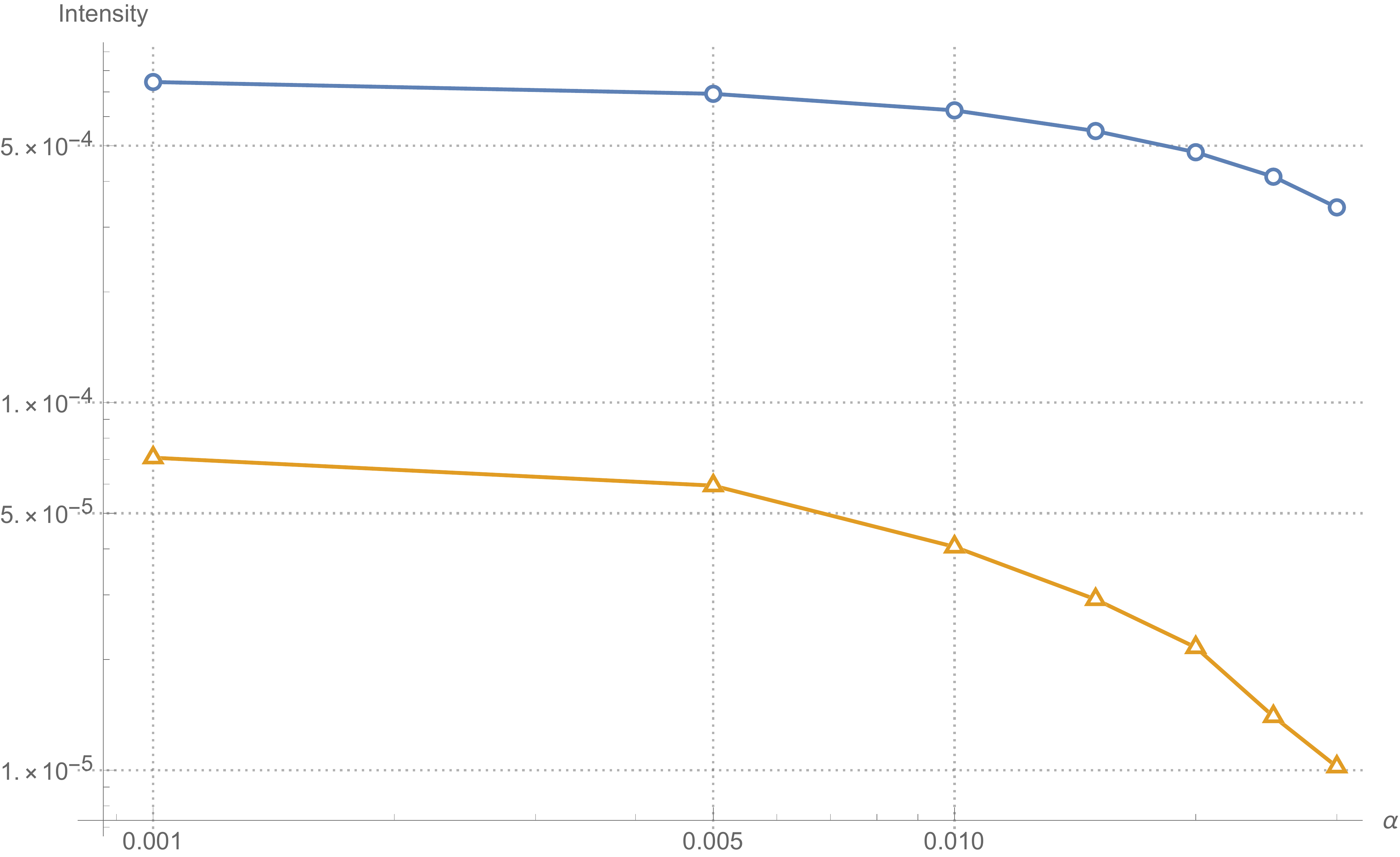}}\quad
\subfigure[]{\includegraphics[scale=0.17]{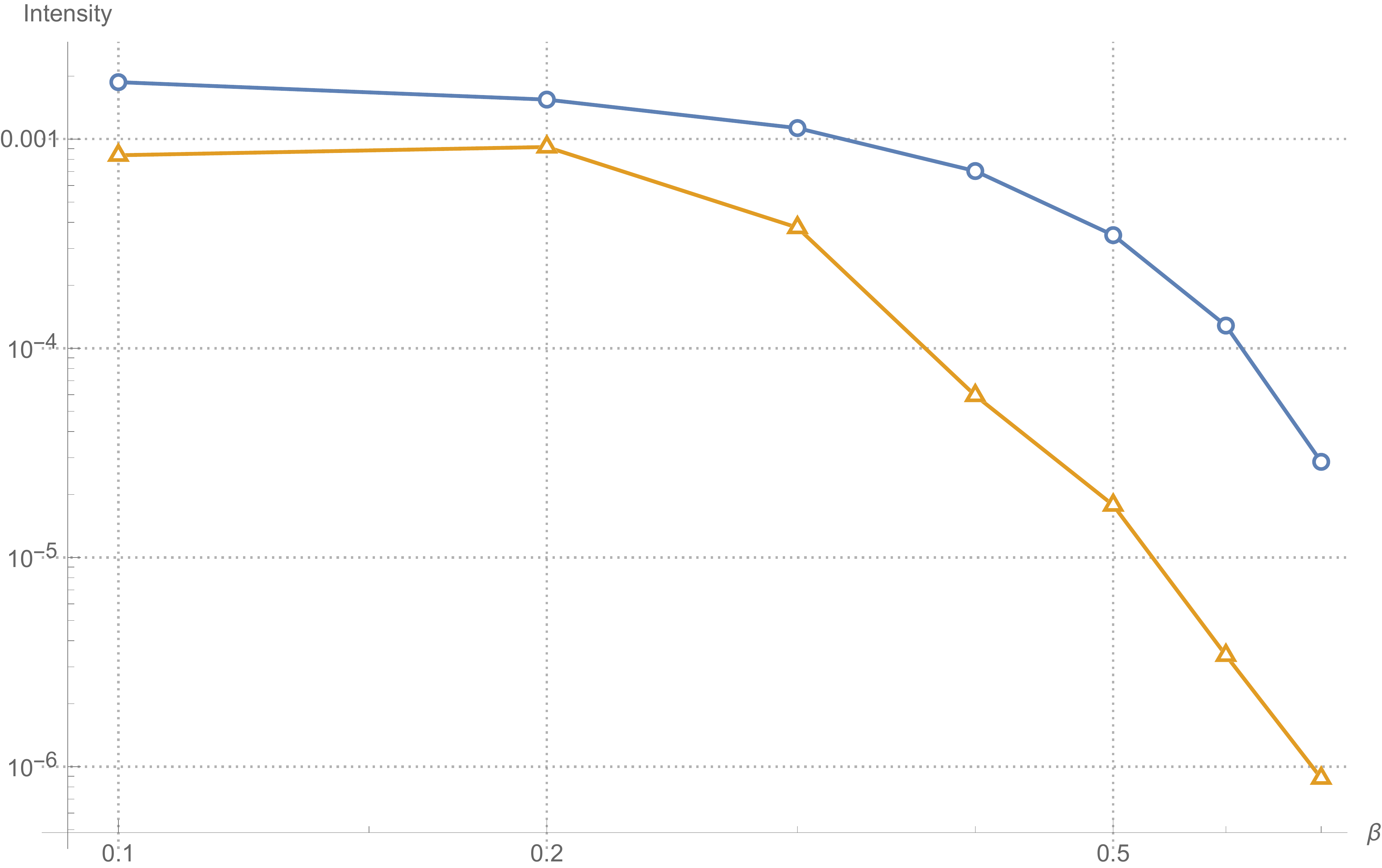}}
\caption{Maximal intensities and intensities at $b_c=10$ for the Brane-World black hole,  where (a) and (b) represent the relationship of the intensities vs. $\alpha$ and $\beta$, respectively. The blue line corresponds to maximal intensities, and orange lines correspond to intensities at $b_c=10$.}\label{add3fig2345}
\end{figure}

The observed luminosity also peaks at the photon sphere $b_p$, similar to that of the static accretion. In addition, the intensities $I_{obs}(\nu_o)$ also decrease with the X-clod dark matter $\beta$ and cosmological parameter $\alpha$. This means that the impact of dark matter and cosmological parameters on shadows is always significant and cannot be ignored.
In the second row of Figure \ref{figdt}, the corresponding shadows and photon spheres cast by this black hole and surrounded by the infalling spherical accretion can be seen. The effect of $\alpha$ and $\beta$ on the observed intensity is shown in Figure \ref{add3fig2345}.

{Similarly, as the X-clod dark matter parameter is fixed, the maximal intensity and intensity at $b_c=10$ decrease with $\alpha$. They also decrease with the parameter $\beta$ since the parameter $\alpha$ is fixed.}

{By comparing Figures \ref{7fig7}, \ref{figjt}, and \ref{figdt}, the shadows exhibit different features for different accretion models. {\bf i)}: The intensity for the infalling accretion is darker at the central region of shadow by comparing with that obtained in Sec.\ref{sec41}. This is caused by the Doppler effect here. {\bf ii)}: As the emissivity per unit volume $j(\nu_e)$ decays faster in this case, the $I_{obs}(\nu_o)$ also decays faster than that of the static accretion\footnote{Obviously, if the emissivity per unit volume $j(\nu_e)$ is chosen as Eq.(\ref{eqqq22}), the decay of the $I_{obs}(\nu_o)$ will be slower in this case.}. {\bf iii)}: The spherical accretion does not affect the radius of the shadow, and the observed luminosity always peaks at the photon sphere $b_p$ for both static and infalling accretions. {\bf IV)}: Different from spherical accretions, the radius of the shadow closely depends on the location of the accretion. The closer the accretion to the black hole, the smaller the size of the shadow. {\bf V)}: The thin disk gives rise to a more refined bright ring outside of the shadow, which includes direct emission, lensing ring, and photon ring. In a nutshell, accretion surrounding a black hole is too important to be ignored.
 }

\section{Conclusions and discussion}\label{Sec5}

In this paper, we numerically investigated the shadows and rings of a Brane-World black hole when the black holes are surrounded by various accretion models, including the effect of dark matter and the cosmological constant.
First, we studied the effective potential and photon orbits of a Brane-World black hole by employing the null geodesic. We discussed the effects of the parameters $(\alpha,\beta)$ on the trajectories of light rays emitted from the North Pole direction.
Then, as an optically and geometrically thin disk surrounding a black hole, we used the first three transfer functions to investigate the observed appearances of direct emissions and rings when the disk is located at different positions out of the black hole.
Finally, shadows and photon spheres were carefully addressed by considering the static and infalling spherical accretions.

The results show that the event horizon $r_h$ and the radius $r_p$ of photon sphere increase with the dark matter parameter $\beta$ and cosmological parameter $\alpha$, while the cosmological horizon $r_c$ decreases. The impact of parameter $b_p$ decreases with $\alpha$ and increases with $\beta$. The incident rays of photon trajectories slope to the abscissa axis since the observer is located at $r_c/5$.
Based on the definition of total orbits of $n = \phi/2\pi$, the thin disk shows that the trajectories of light rays emitted in the North Pole direction are classified as the direct emission, lensing ring, and photon ring. The corresponding trajectories are presented with different values of $\alpha$ and $\beta$ in Figure \ref{fig2}. Using transfer functions, we can see in Figures \ref{5fig2} - \ref{7fig7} that the direct emission and rings are distinguishable, the peaked observed luminosity of which increases with $\alpha$ and $\beta$ for Model I $I^1_{{em}}(r)$. For Model II $I^2_{{em}}(r)$, the peaked luminosity of the overlapped rings increases with $\alpha$ but remains almost unchanged with $\beta$.
For Model III $I^3_{{em}}(r)$, the direct emission, lensing ring, and photon ring are overlapped for the case($\alpha=0.001, \beta=0.3$), while the photon ring can be distinguished from the bright ring for the cases ($\alpha=0.003, \beta=0.1$) and ($\alpha=0.003, \beta=0.3$). The peaked luminosity of the photon ring decreases with $\alpha$ and $\beta$. In contrast, it increases with $\alpha$ and decreases with $\beta$ for the overlapped bright ring. More importantly, the total observed intensity mainly consists of the direct emission. At the same time, the lensing ring and photon ring have small and negligible contributions, respectively. {We also studied the total number of orbits $n$ and the trajectory of direct emission, lensing ring, and photon ring as the disk is located at the position with different inclination angles.}
In addition, by simulating the nominal resolution of the EHT, we blurred the middle column of Figure \ref{6fig6} and obtained the images as Figure \ref{concl123}. The following photos show that the direct emission and rings in all three models are indistinguishable after blurring, although some are distinguishable before blurring. It appears that the size of the dark area is closely related to the location of the disk and almost irrelevant to the critical curve.
\begin{figure}[h]
\centering
\subfigure[$I^1_{{em}}(r)$]{\includegraphics[scale=0.47]{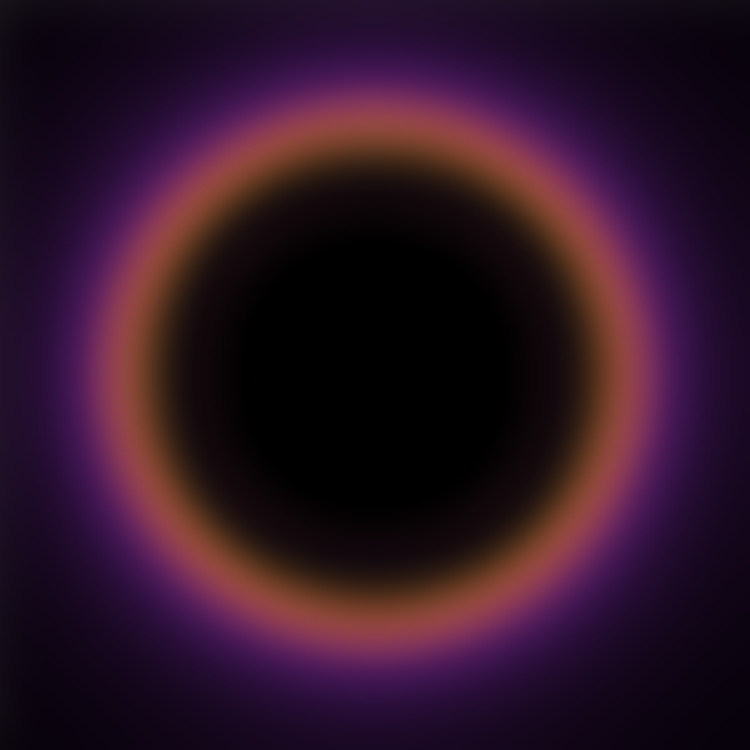}} \qquad
\subfigure[$I^2_{{em}}(r)$]{\includegraphics[scale=0.47]{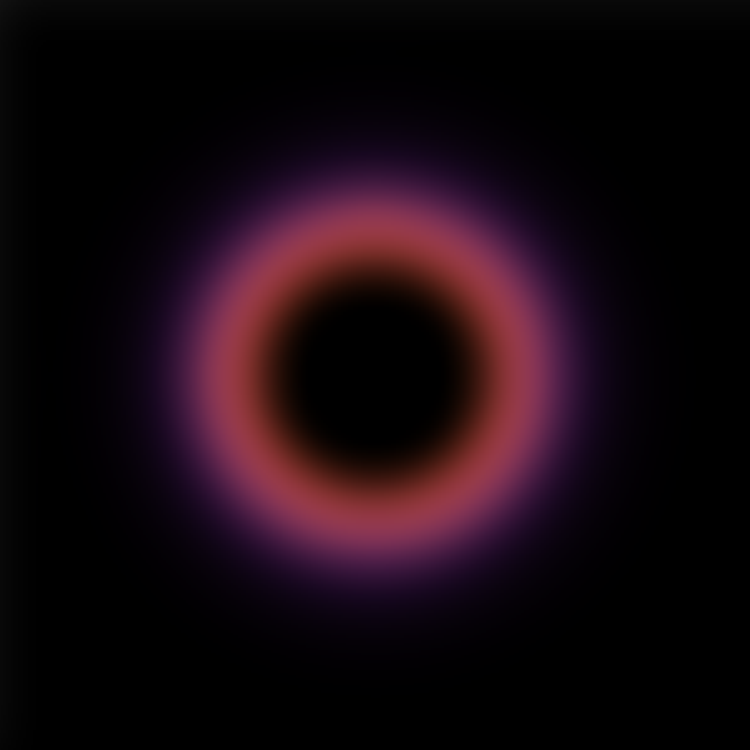}} \qquad
\subfigure[$I^3_{{em}}(r)$]{\includegraphics[scale=0.47]{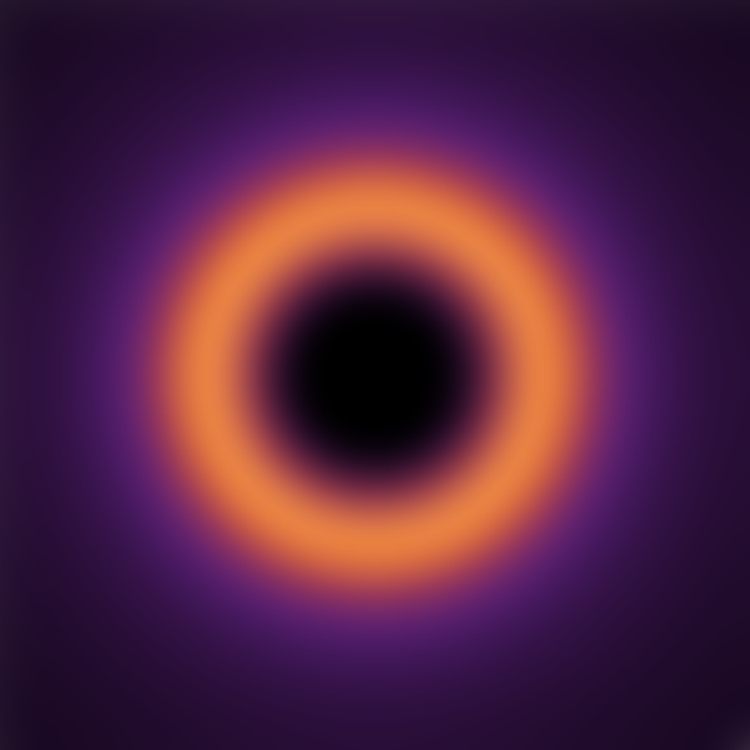}}
\caption{The blurred images for the middle column of Figure \ref{6fig6} with a Gaussian filter with a standard deviation equal to 1/12 of the field of view.}
\end{figure}\label{concl123}
For the static and infalling spherical accretions, according to shadow and photon sphere cast by the Brane-World black hole in Figure \ref{figjt}, we can see that the maximum value of observed luminosity always appears at the photon sphere $b_p$. The intensities $I_{obs}(\nu_o)$ seen by a distant observer decrease with the X-clod dark matter $\beta$ and cosmological parameter $\alpha$ for both accretions. In addition, the emissivity per unit volume $j(\nu_e)$ is closely related to shadow and photon sphere intensities. The Doppler effect will give rise to the shadow¡¯s brightness for the infalling case being darker than the static case.
Combined with the above mentioned facts, the impact of dark matter on shadows and rings of a Brane-World black hole illuminated by various accretions provides many prominent features that differ from that obtained for the Schwarzschild black hole. Hence, the shadows and rings of the black holes can be used to characterize the black holes that include the dark matter and cosmological constant.

Moreover, the EHT provides the angular size of the shadow of M87, which is $\delta = (42\pm3)\mu as$. Based on these facts, we can infer the diameter of the shadow in units of mass $M$, which is $d_{M87} = \frac{D \cdot \delta}{M} \approx 11 \pm 1.5$, in which $D$ is the distance to M87\cite{con1,con2,con3}. This implies that the diameter of the shadow must be located at the regions $9.5 \sim 12.5$ for 1$\delta$ uncertainties and $8\sim14$ for 2$\delta$ uncertainties. Using this condition, we can roughly constrain the X-clod dark matter parameter $\beta$ using the shadow, i.e., when $\alpha = 0.001$, the upper limits of the parameter $\beta$ are $0.33M$ for 1$\delta$ uncertainties and $0.42M$ for 2$\delta$ uncertainties. Hence, the shadow of a black hole can produce some upper limits on the X-clod dark matter parameter from this new exploratory research.
It is also interesting to continue the study of black hole shadows and rings, including the different models of accretions in the future, such as optically thin but geometrically thick accretions.

\vspace{10pt}

\noindent {\bf Acknowledgments}

\noindent
{The authors would like to thank the anonymous reviewers for their helpful comments and suggestions, which helped to improve the quality of this paper.} This work is supported by the National Natural Science Foundation of China (Grant Nos. 11875095 and 11903025), and by the starting fund of China West Normal University (Grant No.18Q062), and by the Sichuan Youth Science and Technology Innovation Research Team(21CXTD0038), and by the Central Government Funds of Guiding Local Scientific and Technological Development for Sichuan Province (No. 2021ZYD0032).\\

\end{document}